\documentclass[twosided,letterpaper,numberedappendix]{emulateapj}
\pdfoutput=1
\usepackage{thumbpdf}
\usepackage{amsmath,amssymb}
\usepackage{graphicx,mathptm}
\usepackage{natbib}
\usepackage[usenames]{color}
\usepackage{ifpdf}

\newcommand{\beq}{
\begin{equation}
}
\newcommand{\eeq}{
\end{equation}
}
\newcommand{\beqa}{
\begin{eqnarray}
}
\newcommand{\eeqa}{
\end{eqnarray}
}

\newcommand{\msun}     {\ensuremath{{\rm M}_{\scriptscriptstyle \odot}}}
\newcommand{\kms}      {\ensuremath{~\mathrm{km~s^{-1}}}}

\newcommand{\msigma}   {\ensuremath{M}{--}\ensuremath{\sigma}}
\newcommand{\ml}       {\ensuremath{M}{--}\ensuremath{L}}
\newcommand{\mbh}      {\ensuremath{M_{\mathrm{BH}}}}

\newcommand{\sigmaconf}   {\ensuremath{\sigma_{68}}}
\newcommand{\msint} {\ensuremath{8.12}}
\newcommand{\msinterr} {\ensuremath{8.12 \pm 0.08}}
\newcommand{\msslope} {\ensuremath{4.24}}
\newcommand{\msslopeerr} {\ensuremath{4.24 \pm 0.41}}
\newcommand{\msscat} {\ensuremath{0.44}}
\newcommand{\msscaterr} {\ensuremath{0.44 \pm 0.06}}

\newcommand{\mlint} {\ensuremath{8.95}}
\newcommand{\mlinterr} {\ensuremath{8.95 \pm 0.11}}
\newcommand{\mlslope} {\ensuremath{1.11}}
\newcommand{\mlslopeerr} {\ensuremath{1.11 \pm 0.18}}

\newcommand{\mlscaterr} {\ensuremath{0.38 \pm 0.09}}

\newcommand{\rinfres} {\ensuremath{R_{\mathrm{infl}} / d_{\mathrm{res}}}}

\def\spose#1{\hbox to 0pt{#1\hss}}
\newcommand{\lta}{\mathrel{\spose{\lower 3pt\hbox{$\mathchar"218$}}
      \raise 2.0pt\hbox{$\mathchar"13C$}}}
\newcommand{\gta}{\mathrel{\spose{\lower 3pt\hbox{$\mathchar"218$}}
      \raise 2.0pt\hbox{$\mathchar"13E$}}}
\def\simlt{\mathrel{\rlap{\lower 3pt\hbox{$\sim$}}\raise 2.0pt\hbox{$<$}}}
\def\simgt{\mathrel{\rlap{\lower 3pt\hbox{$\sim$}} \raise 2.0pt\hbox{$>$}}}

\ifpdf
\fi

\definecolor{KayhanCiteColor}{rgb}{0,0.0,0.0}
\definecolor{KayhanURLColor}{rgb}{0,0.08,0.35}
\definecolor{KayhanLinkColor}{rgb}{0,0.0,0.0}
\definecolor{KayhanPageColor}{rgb}{0,0.0,0.0}

\shorttitle{Intrinsic Scatter of the \texorpdfstring{$M$--$\sigma$}{M-Sigma} 
and \texorpdfstring{$M$--$L$}{M-L} Relations} 
\shortauthors{G\"{u}ltekin et al.}

\journalinfo{The Astrophysical Journal}
\submitted{Received 2008 December 27; accepted 2009 March 26}

\ifx\undefined\hyperref
\usepackage[pdftitle={The M-sigma and M-L Relations in Galactic Bulges and Determinations of their Intrinsic Scatter},pdfauthor={Kayhan Gultekin},pdfsubject={Astrophysics},pdfkeywords={black hole physics --- galaxies: general --- galaxies: nuclei --- stellar dynamics},letterpaper,linktocpage,colorlinks=true,linkcolor={KayhanLinkColor},urlcolor={KayhanURLColor},citecolor={KayhanCiteColor}]{hyperref}

\else\relax\fi
\makeatother


\begin{document}

\label{firstpage}

\title{The
\texorpdfstring{$M$--$\sigma$}{M-Sigma} and
\texorpdfstring{$M$--$L$}{M-L} Relations in Galactic Bulges,
and Determinations of their Intrinsic Scatter\footnotemark[]}\footnotetext[]{}

\author{Kayhan G\"{u}ltekin\altaffilmark{1}\altaffiltext{1}{Department
of Astronomy, University of Michigan, Ann Arbor, MI, 48109.  Send
correspondence to \href{mailto:kayhan@umich.edu}{kayhan@umich.edu}.},
Douglas O. Richstone\altaffilmark{1},
Karl Gebhardt\altaffilmark{2}\altaffiltext{2}{Department of Astronomy, University of Texas, Austin,
TX, 78712.},
Tod R. Lauer\altaffilmark{3}\altaffiltext{3}{National Optical
Astronomy Observatory, Tucson, AZ 85726.},
Scott Tremaine\altaffilmark{4}\altaffiltext{4}{School of
Natural Sciences, Institute for Advanced Study, Princeton, NJ 08540.},
M.~C. Aller\altaffilmark{5}\altaffiltext{5}{Department of
Physics, Institute of Astronomy, ETH Zurich, CH-8093 Zurich,
Switzerland.},
Ralf
Bender\altaffilmark{6}\altaffiltext{6}{Universit\"ats-Sternwarte
M\"unchen, Ludwig-Maximilians-Universit\"at, Scheinerstr. 1, D-81679
M\"unchen, Germany.},
Alan Dressler\altaffilmark{7}\altaffiltext{7}{Observatories of
the Carnegie Institution of Washington, Pasadena, CA 91101.},
S.~M. Faber\altaffilmark{8}\altaffiltext{8}{University of
California Observatories/Lick Observatory, Board of Studies in
Astronomy and Astrophysics, University of California, Santa Cruz, CA
95064.},
Alexei
V. Filippenko\altaffilmark{9}\altaffiltext{9}{Department of
Astronomy, University of California, Berkeley, CA 94720-3411.},
Richard Green\altaffilmark{10}\altaffiltext{10}{LBT
Observatory, University of Arizona, Tucson, AZ 85721.},
Luis C. Ho\altaffilmark{7},
%
John Kormendy\altaffilmark{2},
%
John Magorrian\altaffilmark{11}\altaffiltext{11}{Department of 
Physics, University of Durham, Durham DH1 3LE, UK.},
Jason Pinkney\altaffilmark{12}\altaffiltext{12}{Department of
Physics and Astronomy, Ohio Northern University, Ada, OH 45810.}, and
Christos Siopis\altaffilmark{13}\altaffiltext{13}{Institut
d'Astronomie et d'Astrophysique, Universit\'{e} Libre de Bruxelles,
B-1050 Bruxelles, Belgium.}}
\affil{}

\begin{abstract}
We derive improved versions of the relations between supermassive
black hole mass (\mbh) and host-galaxy bulge velocity dispersion
($\sigma$) and luminosity ($L$) (the \msigma\ and \ml\ relations),
based on 49 $M_{\mathrm{BH}}$ measurements and 19 upper limits.
Particular attention is paid to recovery of the intrinsic scatter
($\epsilon_0$) in both relations.  We find
$\log(M_{\mathrm{BH}}/\msun) = \alpha + \beta \log(\sigma/200~\kms)$
with ($\alpha$, $\beta$, $\epsilon_0$) = (\msinterr, \msslopeerr,
\msscaterr) for all galaxies and ($\alpha$, $\beta$, $\epsilon_0$) =
($8.23 \pm 0.08$, $3.96 \pm 0.42$, $0.31 \pm 0.06$) for ellipticals.
The results for ellipticals are consistent with previous studies, but
the intrinsic scatter recovered for spirals is significantly larger.
The scatter inferred reinforces the need for its consideration when
calculating local black hole mass function based on the \msigma\
relation, and further implies that there may be substantial selection
bias in studies of the evolution of the \msigma\ relation.  We
estimate the \ml\ relationship as $\log(M_{\mathrm{BH}}/\msun) =
\alpha + \beta \log(L_V/10^{11} {\rm L}_{{\scriptscriptstyle
\odot},V})$ of ($\alpha$, $\beta$, $\epsilon_0$) = (\mlinterr,
\mlslopeerr, \mlscaterr); using only early-type galaxies.  These
results appear to be insensitive to a wide range of assumptions about
the measurement errors and the distribution of intrinsic scatter.  We
show that culling the sample according to the resolution of the black
hole's sphere of influence biases the relations to larger mean masses,
larger slopes, and incorrect intrinsic residuals.
\end{abstract}

\section{Introduction to Black Hole Mass Relations}

\setcounter{footnote}{14}
\label{intro}

Studies of elliptical galaxies and spiral bulges (``hot'' galaxies)
have led to the discovery that most such galaxies contain massive dark
objects at their centers, presumably black holes
\citep[BHs;][]{kr95,richstoneetal98,kg01}. Moreover there is a
remarkably tight correlation between the BH mass and the slit-averaged
velocity dispersion of the hot component of the galaxy
\citep{fm00,gebhardtetal00a}.  This \msigma\ relation suggests a
strong link between BH formation, galaxy formation, and active
galactic nuclei (AGNs).

The slope of the \msigma\ relation has been estimated several times in
the last ten years: originally $3.75\pm 0.3$ \citep{gebhardtetal00}
and $4.8\pm 0.5$ \citep{fm00}, then $4.72\pm 0.36$ \citep{mf01a} and
$4.02\pm 0.32$ \citep{tremaineetal02}, and more recently $4.86\pm
0.43$ \citep{ff05} and $3.68 \pm 0.42$ \citep[for barless
galaxies;][]{graham08}.  The reasons for the differences among the
different measures are discussed by \citet{tremaineetal02} and
\citet{ff05}.  

There is also a relation between the BH mass
and the bulge or spheroid luminosity of the galaxy \cite[e.g.,][and the
ratio of BH mass to bulge mass was found to be
$2.2^{+1.6}_{-0.9} \times 10^{-3}$ by
\citealt{kr95}]{dressler89,kormendy93a,magorrianetal98}, but the scatter in the
relation is larger than in \msigma.  No other single parameter or
combination of parameters of the host galaxy has been found to predict
the BH mass with less scatter than stellar velocity dispersion
\citep[][but see also \citealt{mh03}, who find a comparable scatter
for the relation between BH mass and host bulge
mass.]{2003ApJ...583...92G}.

Fundamental to the understanding of the \msigma\ relation is the
measurement of the relation's \emph{intrinsic} or \emph{cosmic
scatter}, as distinct from scatter due to measurement errors.  Indeed,
the fact that there is a relation between BH mass and stellar
velocity dispersion is not surprising, but the scatter is remarkably
small, estimated by \citet{tremaineetal02} to be no larger than 
0.25--0.3 dex.  \citet{nfd06} carried out an extensive investigation
of the residuals from proposed \msigma\ relations and variants.  Their
work highlights the critical role of understanding measurement errors
in assessing the scatter of the various relations.

The magnitude of the intrinsic scatter is extremely important for
several reasons.  First, the range of BH masses in galaxies of a given
velocity dispersion or bulge luminosity constrains BH formation and
evolution theories.  For the past several years, many theories of BH
formation and galaxy evolution have used the \msigma\ relation either
as a starting point for further work or as a prediction of the theory
\citep[e.g.,][]{sr98,bs01,agr01,adamsetal03}; for a review, see
\citet{richstone04}.  A further test of such theories is whether they
can reproduce the observed cosmic scatter in the relation.  Some
predictions that are testable in principle already exist; for example,
\citet{volonteri07} predicts that there should be an increased
intrinsic scatter in low-mass galaxies because BHs are ejected by
asymmetric gravitational wave emission and low-mass spheroids have
lower escape velocities.

Understanding the scatter in the \msigma\ relation is also essential
for estimating the space density of the most massive BHs in the local
universe.  One of the most useful aspects of the \msigma\ relation is
that it allows one to estimate a galaxy's central BH mass from the
more easily measured velocity dispersion.  Because of the steep
decline in number density of galaxies having high velocity dispersion
\citep{shethetal03, bernardietal06,laueretal07}, the majority of the
extremely large BHs will reside in galaxies with moderate velocity
dispersions that happen to contain BHs that are overmassive for the
given velocity dispersion \citep{yt02,marconietal04,laueretal07}.
Knowing the magnitude of the intrinsic scatter is thus required to
find the density of the most massive BHs.  For example, the number
density of BHs with $M > 10^{10}~\msun$ is $\sim 3~\mathrm{Gpc}^{-3}$
if the intrinsic scatter is 0.15 dex and $\sim 30~\mathrm{Gpc}^{-3}$
if the intrinsic scatter is 0.30 dex \citep{laueretal07}.

Both the magnitude of the intrinsic scatter and its distribution
(e.g., normal or log-normal in mass) are also important to know for
studies of the evolution of the \msigma\ relation
\citep[e.g.,][]{tmb04, treuetal07, hopkinsetal06, pengetal06,
shenetal07, shenetal08, vestergaardetal08}.  \citet{laueretal07c}
showed that there is a bias when comparing BH masses derived from
observations of inactive galaxies at low redshifts to BH masses from
active galaxies at higher redshift.  The bias arises because the
sample of nearby galaxies measures the distribution of BH masses for a
given host velocity dispersion or luminosity, whereas the sample from
high-redshift galaxies tends to measure the distribution of the host
luminosity or host velocity dispersion for a given BH mass.
\citet{laueretal07c} found that the bias in the inferred logarithmic
mass scales as the square of the intrinsic scatter in logarithmic
mass.  In order to account for this bias correctly, not only the
magnitude but also the distribution of the deviations from the
\msigma\ relation is needed.

Given the importance of the intrinsic scatter, we focus on a detailed
examination of the scatter in the \msigma\ and \ml\ relations.  This
paper addresses the two most fundamental questions regarding the
scatter: (1) \emph{What is the magnitude of the intrinsic scatter?}
and (2) \emph{What is the shape of the distribution of the intrinsic
residuals?}  A central part of this paper is examination of the
intrinsic scatter from galaxies lying in a narrow range of velocity
dispersion.  This sample of galaxies includes measurements from a
companion paper \citep{Gultekin_etal_2008} that presents BH mass
measurements for five galaxies selected to fall within a narrow range
of velocity dispersion ($180~\kms < \sigma < 220~\kms$).  By focusing
on these galaxies and the others in this narrow range of velocity
dispersion, we may study the distribution of BH masses at a given
value of $\sigma$.

We also combine these new mass measurements with BH mass
measurements and upper limits from the literature to provide new
estimates of the parameters of the \msigma\ and $M$--$L$ relations and
their intrinsic scatter.  We discuss our sample selection in
\S~\ref{sample}.  Results of fits to \msigma\ and \ml\ as well as
an analysis of their scatter are presented in \S~\ref{results}.  We
also demonstrate in \S~\ref{influence} that a bias is incurred by 
selecting a sample of BHs based on the resolution of the sphere of 
influence; this is our reason for including all reliable
BH masses, regardless of resolution.  The implications of our
results are discussed in \S~\ref{discuss} and summarized in
\S~\ref{concl}.  The details of our likelihood method are described in
Appendix~\ref{analysis}, and tests of various error distributions and
subsample selections are provided in Appendix~\ref{difffits}.

\section{Sample of Black Hole Masses}
\label{sample}
  
Our analysis of the \msigma\ and \ml\ relations uses a
$M_{\mathrm{BH}}$ sample comprising the entire corpus of measurements
published before 20 Nov 2008, augmented with the four new measurements
(and one additional upper limit) presented by
\citet{Gultekin_etal_2008}.  We surveyed the literature to make a list
of dynamically detected central BHs, starting with the
compilations of \citet{tremaineetal02}, \citet{mh03}, \citet{ff05}, and 
\citet{graham08}.    We include only direct dynamical measurements,
and since reverberation-mapping measurements
\citep[e.g.,][]{petersonetal04} are normalized to the \msigma\
relation \citep{onkenetal04}, they are not included here.
Tables~\ref{t:bigtable}, \ref{t:ultable}, and~\ref{t:insecure} list
all galaxies with dynamical BH measurements, with masses scaled from
the original publications to our preferred distances (also listed)
assuming $\mbh \propto D$ and $H_0 = 70~\mathrm{km~s^{-1} Mpc^{-1}}$.
We also include upper limits to BH masses, listed in
Table~\ref{t:ultable}.  Most upper limits are not restrictive, in the
sense that they are consistent with a wide range of values for
\msigma\ intercept, slope and intrinsic scatter.  Our fits only use BH
masses from elliptical galaxies or from galaxies with classical bulges
or pseudobulges \citep{kk04}.  Masses from some previous papers by our
``Nuker team'' \citep{2000AJ....119.1157G,2003ApJ...583...92G} were
9\% smaller than the correct values used here due to an error in units
conversion.  To correct this, the published values of $\mbh$,
$M_\mathrm{low}$, and $M_\mathrm{high}$ should be multiplied by 1.099.
This correction is small compared to the measurement uncertainties,
and it is indicated in Table~\ref{t:bigtable} by a superscript beside
the galaxies to which it has been applied.

It has been argued \citep[e.g.,][]{ff05} that BH mass determinations
are unreliable if the kinematic observations do not resolve the sphere
of influence of radius $R_{\mathrm{infl}} = G \mbh \sigma^{-2}$.  In
particular the argument generally requires that the full width at
half-maximum intensity (FWHM) spatial resolution of the kinematic
observations, $d_{\mathrm{res}}$, satisfy $R_{\mathrm{infl}} /
d_{\mathrm{res}} > 1$ to avoid bias in the BH mass determination.
We do not agree with this argument: our tests
\citep{2003ApJ...583...92G,kormendy04} show that the smaller values of
\rinfres\ lead to large error bars but not systematic bias.  We expand
on the misconception that under-resolved spheres of influence lead to
biased BH mass determinations in \S~\ref{nomoresoi}.  On the other
hand, we show in \S~\ref{influence} that excluding measurements on the
basis of \rinfres\ leads to systematic bias, not in the individual BH
masses, but in the estimates of the parameters of the \msigma\ and
\ml\ relations.  For these reasons we do not adopt any
resolution-based cutoff in measurements included in our sample.  We
also investigate each BH mass detection individually and decide
whether any systematic uncertainties suggest that it should be omitted
from our sample.  In Table~\ref{t:insecure} we list published BH
masses that we omit from our fitting sample as well as the reasons.
In general, galaxies are omitted because 
(1) the authors of the study, themselves, expressed doubts of their
models' ability to securely determine the mass,
(2) there is no quantitative analysis of how well their model fits the
data (e.g., the value of reduced $\chi^2$ for the best fit), or
(3) the provided quantitative analysis of goodness of fit is poor.
Note that one of the major conclusions of this paper is that the
intrinsic scatter of the \msigma\ relation is larger than most
previous studies have been found.  If we did not omit the galaxies in
Table~\ref{t:insecure}, the scatter would increase further.
After removing these BH masses, we are left with a sample including
upper limits (SU, see Table~\ref{t:abbrev} for abbreviations used to
denote our samples) and a sample excluding upper limits (S).

In Appendix~\ref{difffits} we also consider an additional sample.
Following the suggestion of \citet{ff05} and others, we create a
``restricted sample'' (RS) of only 20 galaxies in which we (1) require
$\rinfres \ge 1.0$, (2) use only detected BH masses (i.e., no upper
limits), (3) exclude masses deemed suspicious by \citet{ff05} or by
\citet{tremaineetal02} in their final ``culled sample,'' (4) exclude
any galaxy in which multiple measurements are inconsistent, and (5)
make a subjective judgment that the quality of the mass determination
is adequate.  Because of the bias introduced by restrictions based on
resolution, we do not recommend this approach, but we present fits to
Sample RS in Appendix~\ref{difffits}.

\begin{figure*}[h]
\centering
\includegraphics[width=0.95\textwidth]{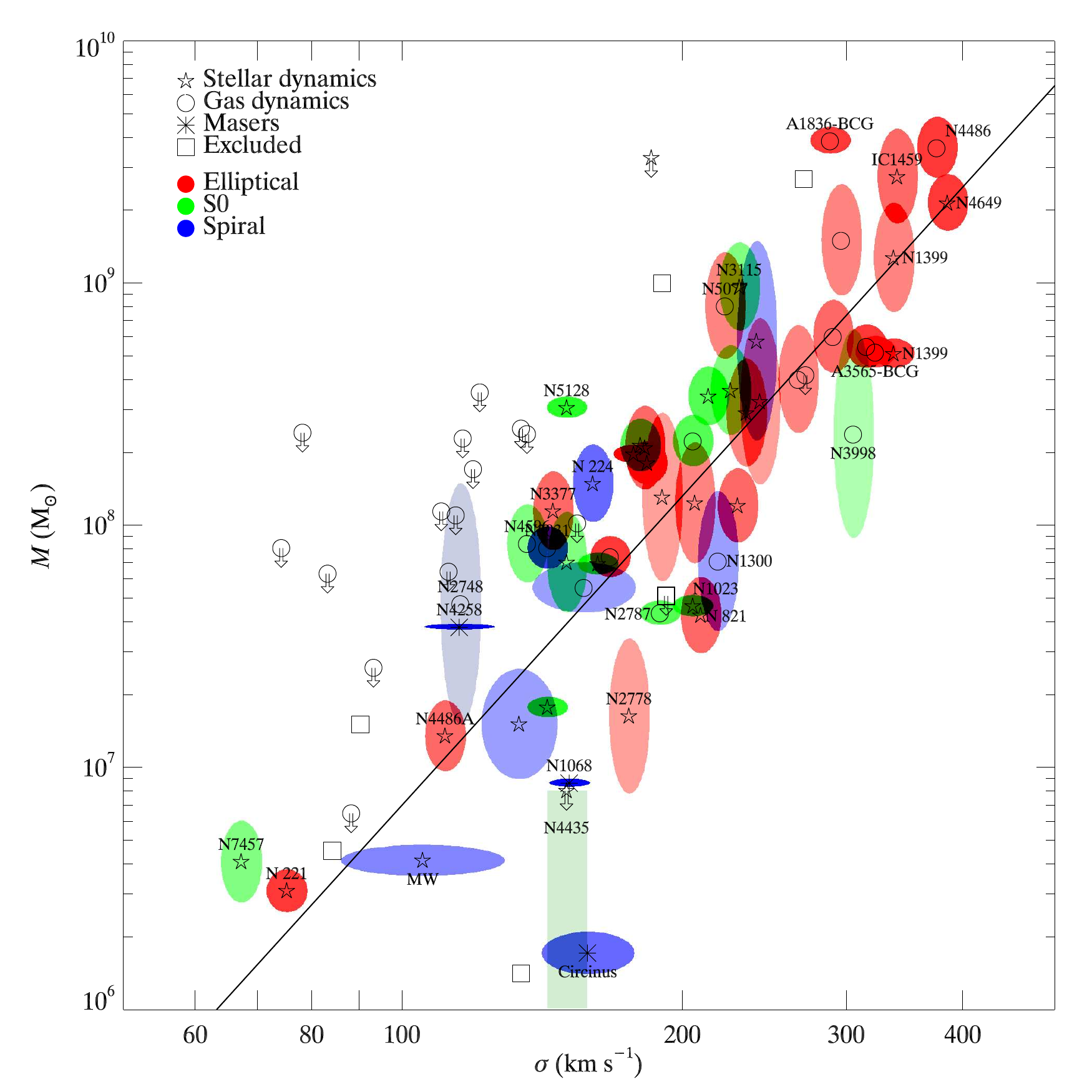}
\caption{\footnotesize The \msigma\ relation for galaxies with
dynamical measurements.  The symbol indicates the method of BH mass
measurement: stellar dynamical (\emph{pentagrams}), gas dynamical
(\emph{circles)}, masers (\emph{asterisks}).  Arrows indicate
3$\sigmaconf$ upper limits to BH mass.  If the 3$\sigmaconf$ limit is
not available, we plot it at 3 times the 1$\sigmaconf$ or at 1.5 times
the 2$\sigmaconf$ limits.  For clarity, we only plot error boxes for
upper limits that are close to or below the best-fit relation.  The
color of the error ellipse indicates the Hubble type of the host
galaxy: elliptical (\emph{red}), S0 (\emph{green}), and spiral
(\emph{blue}).  The saturation of the colors in the error ellipses or
boxes is inversely proportional to the area of the ellipse or box.
Squares are galaxies that we do not include in our fit.  The line is
the best fit relation to the full sample:
$M_{\mathrm{BH}} = 10^{\msint}~\msun(\sigma/200~\kms)^{\msslope}$.
The mass uncertainty for NGC~4258 has been plotted much larger than
its actual value so that it will show on this plot.  For clarity, we
omit labels of some galaxies in crowded regions.}
\label{f:msigma}
\end{figure*}

Membership in all samples (SU, S, RS) is given in
Tables~\ref{t:bigtable} and~\ref{t:ultable}.  We plot the masses of
the BHs as a function of velocity dispersion in
Figure~\ref{f:msigma}.

\subsection{Velocity Dispersion}
\begin{figure}[htb]
\centering
\includegraphics[width=1.05\columnwidth]{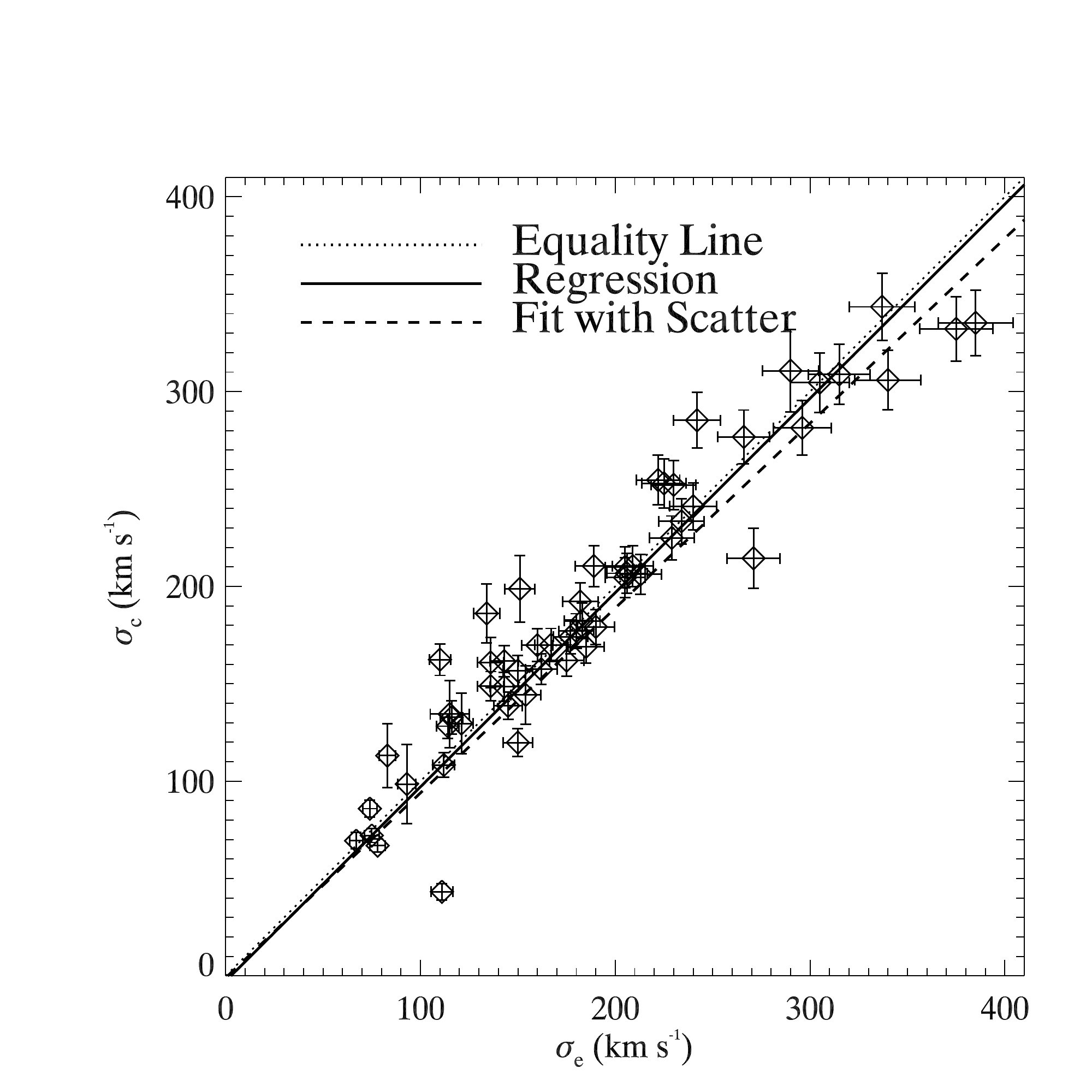}
\caption{Central stellar velocity dispersion $\sigma_c$ from HyperLEDA
as a function of effective stellar velocity dispersion $\sigma_e$
(eq.~\ref{e:sigmae}).  There is no systematic bias to high or low
values.  The dotted line shows $\sigma_c = \sigma_e$.  The solid line
is the best-fit regression, $\sigma_c = (-2.7 \pm 1.4~\kms) + (1.00
\pm 0.02)\sigma_e$.  The dashed line is a fit that includes
uncertainties in both variables as well as an intrinsic scatter:
$\sigma_c = (-0.9 \pm 7.2~\kms) + (0.95 \pm 0.09)\sigma_e$ with
intrinsic scatter of $22 \pm 5~\kms$.  Both fits are consistent with
$\sigma_c = \sigma_e$ within their uncertainties.  }
\label{f:sigmasigma}
\end{figure}

Whenever possible we use the effective velocity dispersion $\sigma_e$
as defined by
\beq
\sigma^2_e \equiv \frac{\int_{0}^{R_e} \left({\sigma^2 + V^2}\right) I\left(r\right) dr}{{\int_{0}^{R_e} I\left(r\right) dr}},
\label{e:sigmae}
\eeq
where $R_e$ is the effective radius of the galaxy and $V$ is the
rotational component of the spheroid.  If this is not available, we
use the central stellar velocity dispersion ($\sigma_c$) found in
HyperLeda.  In galaxies for which both are available, we compare
$\sigma_e$ to $\sigma_c$ in Figure~\ref{f:sigmasigma} and find no
systematic bias to high or low values.  Because of stellar template
mismatches and possible errors in the determination of $R_e$, we
impose a minimum error in $\sigma_e$ of 5\%.  The contribution to
velocity dispersion arising from the relatively small rotational
component of disks in spiral galaxies is unlikely to lead to large
systematic errors in velocity dispersion.  For this reason, we include
spirals in our main sample.

\subsection{Luminosities}
We use extinction-corrected, bulge (which we obviously interpret as
total luminosity for ellipticals), $V$-band luminosities--calculated
from the extinction-corrected magnitudes, $M_{V,{\mathrm{bulge}}}^0$,
using $\log(L_V / {\rm L}_{{\scriptscriptstyle \odot}, V}) = 0.4 (4.83
- M_{V,{\mathrm{bulge}}}^0)$.  The choice of $V$ band is a compromise
between more widely available luminosities in the $B$ and $V$ bands
and less extinguished $K$-band luminosities.  Because bulge-disk
decomposition of spiral galaxies is fraught with difficulties and
possible systematic errors, we do not include spirals in the $M_{\rm
BH}$--$L$ fits.  We \emph{do} include bulge luminosities for S0
galaxies for which we are confident in the bulge-disk decomposition
because the disk is faint and contains little dust and few young
stars.  Errors in bulge-disk decomposition were estimated by examining
the range of values in the literature.  Thus the galaxies included in
\ml\ fits are those which have a $M_{V,{\mathrm{bulge}}}^0$ value
listed in Table~\ref{t:bigtable} or \ref{t:ultable}.  Most of our
$V$-band absolute magnitudes (total and bulge) come directly from
\citet{rc3} as compiled by \citet{laueretal05} and
\citet{laueretal07b}.

\section{Measurement of the \texorpdfstring{\msigma\ and 
$M$--$L$}{M-sigma and M-L} Relations, and their Intrinsic Scatter}

\label{results}
\subsection{Best Fit \texorpdfstring{$M$-$\sigma$}{M-Sigma} Relation}
\label{msigmaresults}
We are principally interested in predictors of the form
\beq
\log{(M/\msun)} = \alpha + \beta \log{\left(\sigma_e/200\kms\right)},
\label{e:msigmaform}
\eeq
with an intrinsic or cosmic scatter $\epsilon_0$ that is the root-mean 
square (rms) deviation in $\log(\mbh/\msun)$ from this relation (for 
zero measurement error).  We assume for simplicity that $\epsilon_0$ is
independent of $\sigma_e$.  The details of the fitting method are
discussed in Appendix~\ref{analysis}.  Based on tests
in \S~\ref{msigmascatter}, a log-normal distribution of intrinsic
scatter $\epsilon_0$ is an adequate description of the scatter.  Based
on tests in Appendix~\ref{bootstrap}, a generalized maximum-likelihood
method is appropriate to use here.  Given these, the
best-fit relation for the full sample (SU) is
\beq
\log{\left(\frac{M}{\msun}\right)} = (\msinterr) + (\msslopeerr) 
\log{\left(\frac{\sigma_e}{200\kms}\right)},
\label{e:noulmsigmafit}
\eeq
which has an intrinsic rms scatter $\epsilon_0
= \msscaterr$.  These values are very close to those obtained by
\citet{tremaineetal02}.  We show in Appendix~\ref{difffits} that the
choice of objects to include in the sample can have a significant
impact on the slope, intercept, and residual scatter, but that the error
distribution assumed (among those we considered) has little impact.

The full sample may be split into subsamples: early-type (elliptical
and S0) and late-type (spiral) galaxies, ellipticals and
non-ellipticals, BH mass measurements by gas dynamics and
other methods, low $\sigma_e$ and high $\sigma_e$, and barred and
non-barred galaxies.  The fits are summarized in
Table~\ref{t:subsamplefits}.  In most cases the fits to different
subsamples are consistent with each other.  The intrinsic scatter for
the ellipticals-only sample, however, is $\epsilon_0 = 0.31 \pm 0.06$,
which is $\sim$2$\sigmaconf$%
\footnote{We use $\sigmaconf$ to mean 68\%-confidence level so as to
distinguish it from velocity dispersion.}
 smaller than either the full sample
($\epsilon_0 = \msscaterr$) or the sample of non-ellipticals
($\epsilon_0 = 0.53 \pm 0.10$).  This may reflect either greater
unaccounted errors in BH mass and $\sigma_e$ measurements in
later-type galaxies, or that ellipticals lie closer to the
ridge line of the \msigma\ relation than later-type galaxies.

\begin{deluxetable*}{lrrrrrr}
  \footnotesize
  \tablecaption{$M$--$\sigma$ Relation for Subsamples}
  \tablehead{
     \colhead{Subsample} & 
     \colhead{$N_m$} & 
     \colhead{$N_u$} & 
     \colhead{$\alpha$} & 
     \colhead{$\beta$} & 
     \colhead{$\epsilon_0$} &
     \colhead{$P_\emptyset$}
  }
  \startdata

Full sample         & 49 & 18 & \msinterr & \msslopeerr & \msscaterr & $0.0004 \pm 0.018$\\
\\
Early type          & 38 &  6 & $8.22 \pm 0.073$ & $3.86 \pm 0.380$ & $0.35 \pm 0.031$ & $0.0145 \pm 0.031$\\
Late type           & 11 & 12 & $7.95 \pm 0.286$ & $4.58 \pm 1.583$ & $0.56 \pm 0.141$ & $0.0006 \pm 0.040$\\
\\
Ellipticals         & 25 &  2 & $8.23 \pm 0.084$ & $3.96 \pm 0.421$ & $0.31 \pm 0.063$ & $0.0006 \pm 0.018$\\
Non-ellipticals     & 24 & 16 & $8.01 \pm 0.156$ & $4.05 \pm 0.831$ & $0.53 \pm 0.097$ & $0.0010 \pm 0.031$\\
\\
Stars and masers    & 32 &  2 & $8.11 \pm 0.107$ & $4.05 \pm 0.554$ & $0.49 \pm 0.075$ & $0.0002 \pm 0.021$\\
Gas dynamics        & 17 & 16 & $8.16 \pm 0.122$ & $4.58 \pm 0.652$ & $0.35 \pm 0.096$ & $0.0036 \pm 0.040$\\
\\
$\sigma_e < 200~\kms$ & 25 & 16 & $8.07 \pm 0.172$ & $3.97 \pm 0.869$ & $0.50 \pm 0.091$ & $0.0013 \pm 0.031$\\
$\sigma_e > 200~\kms$ & 24 &  2 & $8.12 \pm 0.158$ & $4.47 \pm 0.921$ & $0.35 \pm 0.079$ & $0.0026 \pm 0.024$\\
\\
Non-barred & 41 &  7 & $8.19 \pm 0.087$ & $4.21 \pm 0.446$ & $0.43 \pm 0.064$ & $0.0006 \pm 0.017$\\
Barred     &  8 & 11 & $7.67 \pm 0.115$ & $1.08 \pm 0.751$ & $0.17 \pm 0.078$ & $0.1809 \pm 0.147$\\
\\
Classical bulges    & 39 & 16 & $8.17 \pm 0.086$ & $4.13 \pm 0.434$ & $0.45 \pm 0.066$ & $0.0009 \pm 0.024$\\
Pseudobulges       & 10 &  2 & $7.98 \pm 0.156$ & $4.49 \pm 0.903$ & $0.28 \pm 0.096$ & $0.0034 \pm 0.037$
  \enddata
  \label{t:subsamplefits}
  \tablecomments{Results from fits to subsamples of our full sample,
  based on morphological type, BH mass-measurement method.  $N_m$ and
  $N_u$ are the number of galaxies in each group with BH mass
  measurements and upper limits, respectively.}
  \end{deluxetable*}

\subsection{Examination of the Intrinsic Scatter in 
\texorpdfstring{$M$--$\sigma$}{M-Sigma}}
\label{msigmascatter}

\begin{figure}
\includegraphics[width=0.4\textwidth,angle=90]{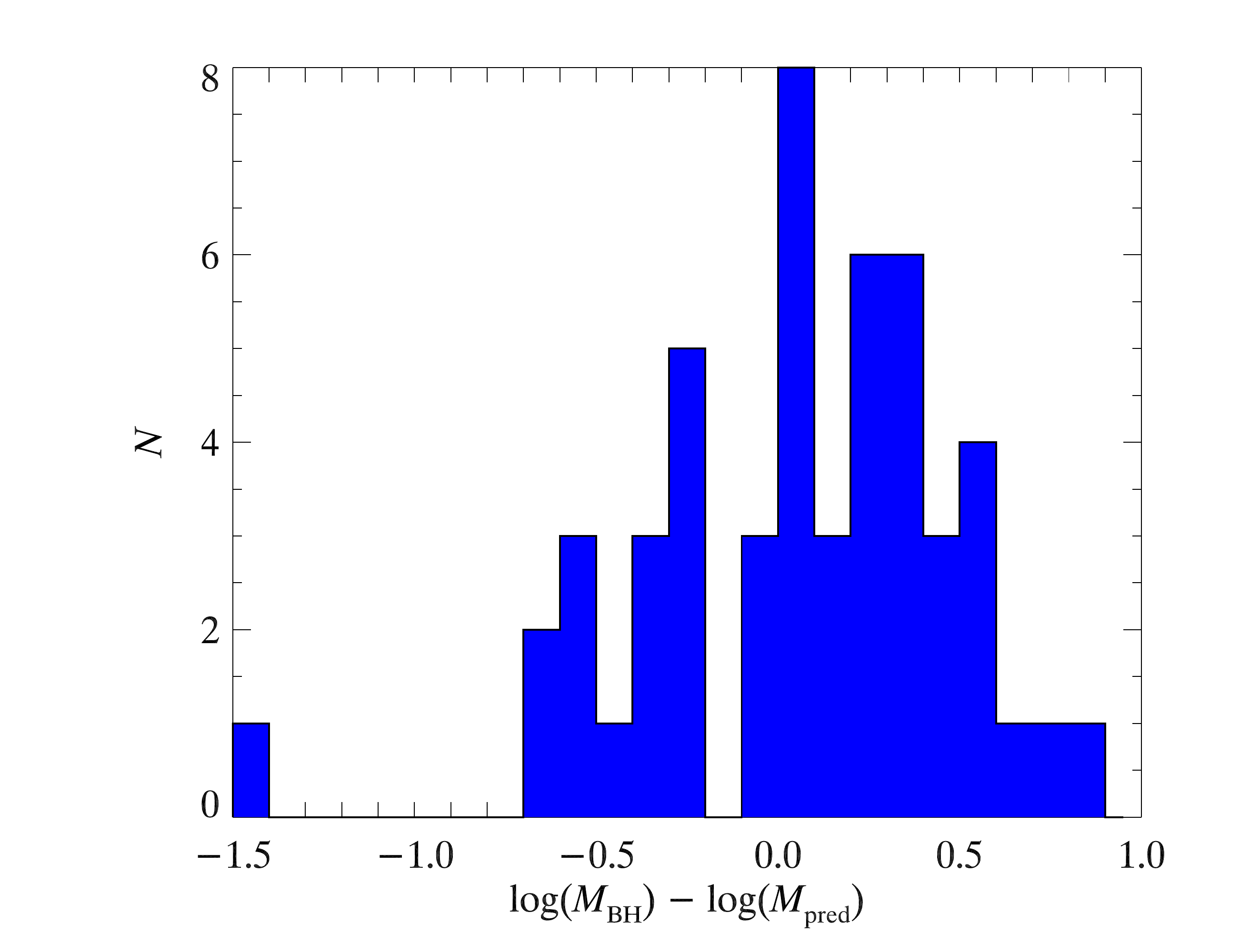}
\caption{Histogram of residuals from the best-fit \msigma\ relation in
sample~S.}
\label{f:resids}
\end{figure}

The distribution of residuals from the \msigma\ relation is of
practical interest as discussed in \S~\ref{intro}.
Figure~\ref{f:resids} is a histogram of the residuals in
$\log{(\mbh)}$ in the best-fit \msigma\ relation from sample S.  The
distribution of the residuals appears consistent with a normal or
Gaussian distribution in logarithmic mass, although the distribution
is noisy because of the small numbers.  For a more direct test of
normality we look at $\log(\mbh)$ in galaxies with $\sigma_e$ between
$165$ and $235\kms$, corresponding to a range in
$\log(\sigma_e/200\kms)$ from approximately $-0.075$ to $0.075$.  The
predicted masses for the 19 galaxies in this narrow range differ by at
most a factor of 4.3, given our best-fit relation.  The power of
having a large number of galaxies in a narrow range in velocity
dispersion is evident here, as there is no need to assume a value for
the slope of \msigma\ or even that a power-law form is the right
model.  The only assumption required is that the ridge line of any
\msigma\ relation that may exist does not change substantially across
the range of velocity dispersion.  The mean of the logarithmic mass in
solar units is 8.16, and the standard deviation is 0.45.  The expected
standard deviation in mass is 0.19, based on the rms dispersion of
$\log(\sigma_e/200\kms)$ (0.046) in this range times the \msigma\
slope $\beta$; thus the variation in the ridge line of the \msigma\
relation in this sample is negligible compared to the intrinsic
scatter.  We perform an Anderson-Darling test for normality with
unknown center and variance on this sample of logarithmic masses
\citep{stephens74,numericalrecipes} and find that the distribution is
consistent with normality at better than the 15\% level.  The same
sample fails an Anderson-Darling test for normality in
$M_{\mathrm{BH}}$ --- as opposed to $\log(M_{\mathrm{BH}})$ --- at the
1\% level.

The Anderson-Darling tests show that log-normal (i.e., Gaussian in
logarithmic mass) is an acceptable description of the distribution of
the residuals.  The residuals may also be well represented by other
distributions, which may be compared to a log-normal description with
an odds ratio (Eq.~\ref{e:oddsratio}).  The calculated odds ratio of
Gaussian to Lorentzian in logarithmic mass is 14.49, of Gaussian to
double-sided exponential in logarithmic mass is 1.78, of Gaussian to
the sum of two Gaussians in logarithmic mass is 1.55, and of Gaussian
to a Gaussian with different standard deviations above and below the
mean is 2.30.  In these cases considered, the odds ratio (${\mathcal
R}_{ab}$) of the Gaussian distribution to the other distributions in
logarithmic mass is greater than unity.  Thus the Gaussian
distribution is favored.  While it should be noted that these tests
convolve the intrinsic dispersion with variance from imprecise
measurements, the residuals in the \msigma\ relation are well
described by a Gaussian distribution in logarithmic mass.

One component of the intrinsic scatter in both the \msigma\ and \ml\
relations is sure to be random errors in distance.  

All BH mass measurements in our sample scale other than the Milky Way
linearly with the distance.  For these galaxies random errors in
distance contribute directly to uncertainty in the mass and thus
contributes to the intrinsic scatter.  The Milky Way, which scales
approximately as $D^{1.8}$ \citep{ghezetal08}, has the uncertainty in
distance incorporated in its mass uncertainty in our tables.  The
contribution of random errors in distance to the intrinsic scatter,
however, is sure to be a small as the random errors in distance are
typically around 10\% or 0.04 dex \citep[e.g.,][]{tonryetal01}
compared to the scatter of \msscat~dex.
Another source of intrinsic scatter may be unaccounted systematic
errors in \mbh\ measurements.  In this paper, we assume that the
measurement errors accurately reflect the uncertainty, but we discuss
some possible cuases in \S~\ref{systematics}.  If systematics are
large, then they may be an important contribution to the inferred
intrinsic scatter.

\subsection{Log-Quadratic Fits to \texorpdfstring{$M$--$\sigma$}{M-Sigma}}
\label{logquad}
We may also fit a log-quadratic function to the data as suggested by
\citet{wyithe06}:
\beqa
\log{\left(\frac{\mbh}{\msun}\right)}= \alpha &+& \beta 
\log{\left(\frac{\sigma_e}{200\kms}\right)}\nonumber\\ &+& 
\gamma \left[\log{\left(\frac{\sigma_e}{200\kms}\right)}\right]^2.
\label{e:logquad}
\eeqa
For the full sample, we find $\alpha = 8.08 \pm 0.10$, $\beta = 4.47
\pm 0.50$, $\gamma = 1.72 \pm 1.71$, and $\epsilon_0 = 0.44 \pm 0.06$.
The results are consistent with log-linear ($\gamma = 0$) at the
1$\sigmaconf$ level, and the intrinsic scatter is not significantly
decreased from a log-linear model.  The value and significance for
$\gamma$ is similar to that found by \citet{wyithe06erratum}.  The
odds ratio (eq.~\ref{e:oddsratio}) of a log-linear model to a
log-quadratic model is ${\mathcal R}_{ab} = 4.68$, indicating the
log-linear model is favored.

\subsection{Best-Fit \texorpdfstring{$M$--$L$}{M-L} Relation}
\label{mlresults}

We also look at fits to the \ml\ relation.  Spiral galaxies, which
present problems in determining the bulge luminosity because of the
difficulty in getting a precise bulge-disk decomposition, are
excluded.  We therefore limit our fits to ellipticals and those S0
galaxies for which we have reliable bulge-disk decomposition.  The
sample of galaxies used may be discerned from Tables~\ref{t:bigtable}
and~\ref{t:ultable} by the presence of a value in the column for bulge
magnitude ($M^{0}_{V,\mathrm{bulge}}$), which in the case of
ellipticals is equal to the total magnitude.

Using the same fitting method as before on all galaxies meeting the
above criteria, we find
\beq 
\log{\left(\frac{M}{\msun}\right)} = (\mlinterr) + (\mlslopeerr) 
\log{\left(\frac{L_V}{10^{11} {\rm L}_{{\scriptscriptstyle \odot}, 
V}}\right)}
\label{e:noulmlfit}
\eeq
with an intrinsic scatter of $\epsilon_0 = \mlscaterr$.

\begin{figure*}[tb]
\centering
\includegraphics[width=1.0\textwidth]{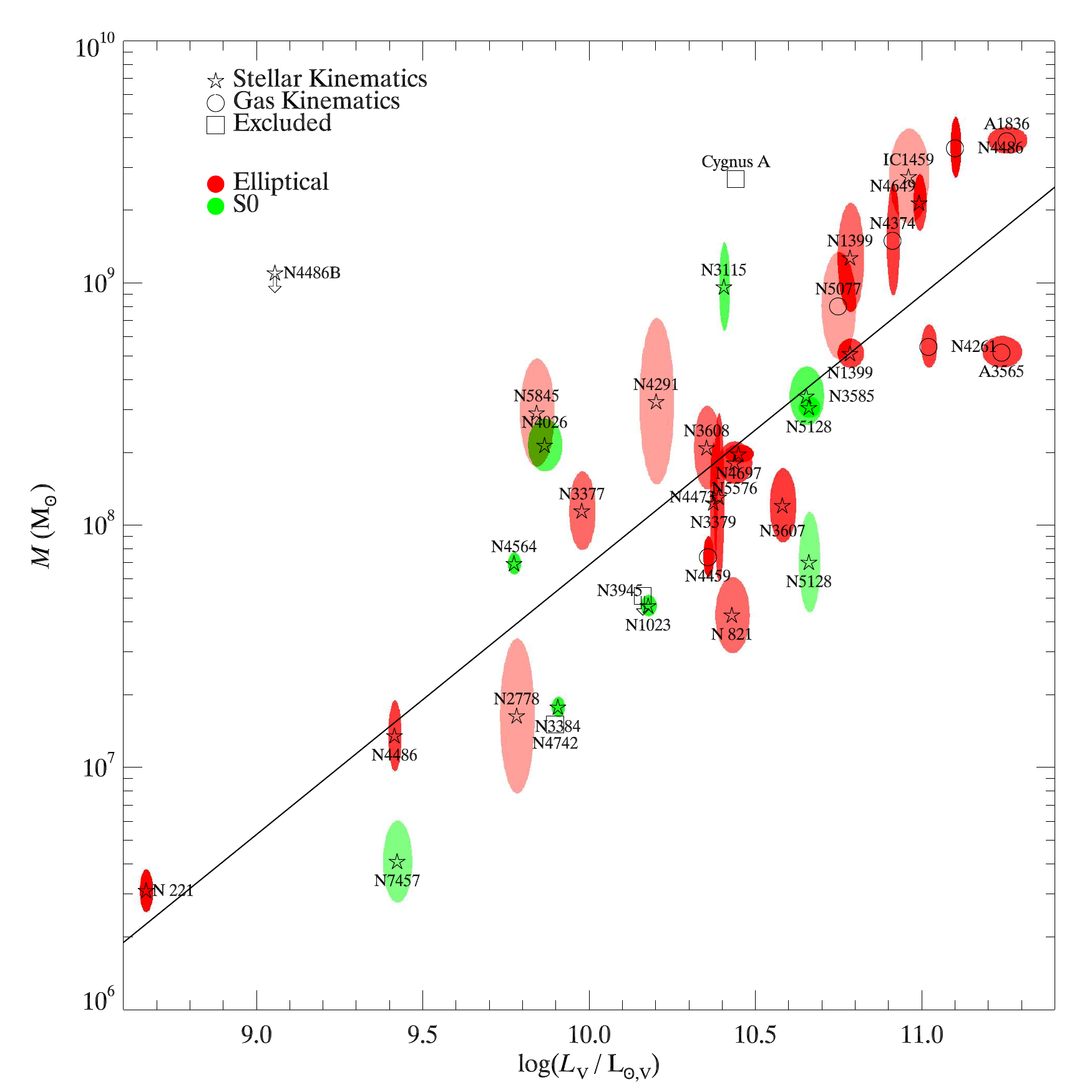}
\caption{The $M$--$L$ relation for galaxies with dynamical
measurements.  The symbol indicates the method of BH mass measurement:
stellar dynamical (\emph{pentagrams}) and gas dynamical
(\emph{circles)}.  Arrows indicate upper limits for BH mass. Squares
are galaxies that we omitted from the fit.  The color of the error
ellipse indicates the Hubble type of the host galaxy [elliptical
(\emph{red}) and S0 (\emph{green})] and the saturation of the color is
inversely proportional to the area of the ellipse.  The line is the
best-fit relation for the sample without upper limits:
$M_{\mathrm{BH}} = 10^{\mlint}~\msun (L_V/10^{11}~{\rm
L}_{{\scriptscriptstyle \odot},V})^{\mlslope}$.}
\label{f:ml}
\end{figure*}

\subsection{Examination of the Intrinsic Scatter in 
\texorpdfstring{$M$--$L$}{M-L}}

\begin{figure}
\includegraphics[width=0.4\textwidth,angle=90]{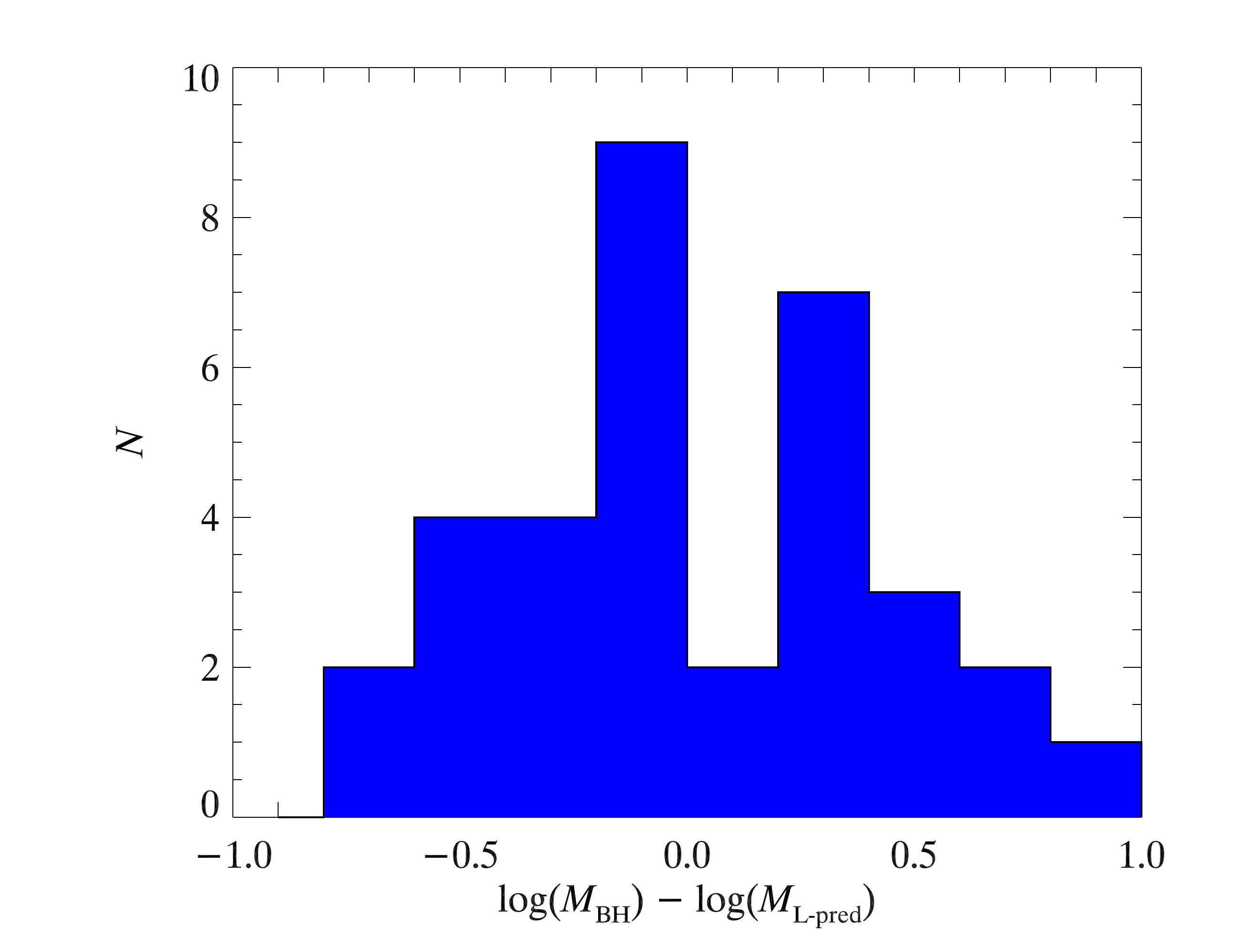}
\caption{Histogram of residuals from best-fit \ml\ relation.}
\label{f:mlresids}
\end{figure}

The scatter determined here for the \ml\ relation is notably smaller
than other studies have found, and it is consistent with the intrinsic
scatter in \msigma\ in early-type galaxies.  Other studies have found
similar scatter between the \ml\ and \msigma\ relations
\citep[e.g.,][]{mh03}.  Figure~\ref{f:mlresids} shows a histogram of the
residuals from the fit.  We test the distribution of 12 masses in a
narrow range in luminosity ($10.2 < \log{(L_{\mathrm{V}}/{\rm
L}_{{\scriptscriptstyle \odot}\mathrm{,V}}) < 10.7}$) for normality
and log-normality using an Anderson-Darling test with unknown mean and
variance.  The mean logarithmic BH mass in that range is 8.21 with
standard deviation 0.36.  The expected standard deviation in
logarithmic mass is 0.16, if all the galaxies lie exactly on the ridge
line of the \ml\ relation, based on standard deviation in logarithmic
luminosity of 0.14.  Thus the variation in the ridge line of the \ml\
relation in this sample is negligible compared to the intrinsic
scatter.  The distribution of masses is consistent with a log-normal
distribution, but is inconsistent with a normal distribution at the
1\% level.

Again, we consider several possible functional forms of the intrinsic
scatter in logarithmic mass at constant luminosity and calculate odds
ratios of Gaussian to Lorentzian (3.37), to double-sided exponential
(1.10), to double Gaussian (1.36), and to a Gaussian with different
standard deviations above and below the mean (1.99).  Thus, normal or
double-sided exponential in logarithmic mass are equally acceptable
descriptions of the intrinsic scatter.  There are, however, relatively
few mass measurements in this range, so this it is not as strong a
test as it is for \msigma.

\section{Biases Introduced When Culling the Sample by Black Hole 
Sphere-of-Influence Resolution}
\label{influence}

Our analysis has used all \mbh\ measurements in the sample without
regard to how well the sphere of influence is resolved.  In essence,
we trust the uncertainties returned by the dynamical models to reflect
the quality of the \mbh\ determination---given the assumptions---for
whatever resolution was realized.  This approach conflicts with the
caveats advanced by, e.g., \citet{ferrarese02}, who concluded that
$\rinfres > 1$ was a necessary (but not sufficient) condition to make
a precise \mbh\ determination.  Such concerns have motivated a number
of groups studying the BH relations \citep[e.g.,][]{mf01a,ff05} to
cull the more poorly resolved BHs from their samples.  We show in this
section that far from improving the accuracy of the estimated
relations, this procedure will actually produce biased estimates of
the intercept, slope, and intrinsic scatter of the relations.  

\subsection{What Resolution of the Sphere of Influence Does and Does Not Mean}
\label{nomoresoi}
We advocate use of the sphere-of-influence scale only as a rough guide
to the needed spatial resolution of the observations.  In the
arguments that follow, we will show that \mbh\ determinations from
datasets with $\rinfres < 1$ are unbiased when using three-integral
models.  This, however, contrasts with claims commonly made in other
works: that strict resolution of the sphere of influence is required
for credible \mbh\ determinations and more importantly, that \mbh\
determinations made from observations that do not resolve the sphere
of influence will be biased.  Given the strong prevalence of this
viewpoint, and prompted by comments from the referee, we review its
development and application in the literature.  We find, in fact, that
there is little or no support for the conclusion that \mbh\
determination become increasingly biased with decreasing resolution.
It appears that the common but uncritical application of
sphere-of-influence-resolution as a way to cull \mbh\ determinations
cannot be justified by careful reading of the very works often cited
in its support.

In their review article, \citet{ff05} write, ``All studies which have
addressed the issue [of BH mass determination and resolution level]\dots
have concluded that resolving the sphere of influence is an
important (although not sufficient) factor: not resolving
[$R_\mathrm{infl}$] can lead to systematic errors on \mbh\ or even
spurious detections,'' and cite the following: \cite{fm00, mf01a,
mf01b, grahametal01, ferrarese02, mh03}.  We consider each of
these in turn.

\citet{fm00} found that the ground-based \mbh\ measurements by
\citet{magorrianetal98} were higher for fixed velocity dispersion than
the predictions of their empirical \msigma\ relation and judged them
to be therefore biased.  The discrepancy with their \msigma\ relation
increased with increasing distance.  While discrepancy with the
\msigma\ relation is not a justifiable reason for excluding \mbh\
measurements from the relation (the argument is circular), the masses
from \citet{magorrianetal98} were, in fact, biased to high values by
roughly a factor of 3.  The reason for the bias, however, is that they
came from two-integral, isotropic, axisymmetric models, \emph{not}
because they were more poorly resolved \citep{mf01b,2003ApJ...583...92G}.

\citet{mf01a} present similar arguments as do \citet{mf01b} who also
go on to describe the reason two-integral models yield masses that are
biased somewhat high.  \citet{mf01b} do mention that three-integral
models provide an increased space of orbits that can lead to an
increased range in acceptable black hole masses but \emph{not} that
they are biased, resolved or not.

\citet{grahametal01} find decreased scatter in the relation between
\mbh\ and concentration index when removing more poorly resolved
galaxies in addition to another galaxy.  Again, using a provisional
relation to exclude potential data points from the relation is not the
same as the identification of a bias.

\citet{ferrarese02} does not present new arguments or studies but
repeats the arguments just described.

\citet{mh03} say nothing about a bias, only that they used $\rinfres$
as a criterion for determining reliable masses and that they find a
smaller scatter in the $K$-band \ml\ relation when excluding lower
resolution observations.

\citet{vme04} are frequently cited as having shown that resolving the
sphere of influence is necessary for accurate \mbh\ determination,
\emph{but, in fact, they do not make such a claim}.  The most
important result from their work was to show that when using too few
orbits, spurious black hole mass determinations may appear.  When
enough orbits are used, \citet{vme04} found that a wide range of
values for \mbh\ are acceptable with their synthetic data set, but
this turns out to be caused by the lack of real measurement noise in
the data set \citep{magorrian06}.  Even so, the input \mbh\ in their
simulations was always within their range of acceptable values, i.e.,
not biased.  The bias that \citet{vme04} do identify is what results
from over-regularizing (i.e., requiring smoothness in orbit solutions)
not from under-resolving the sphere of influence.

\begin{figure*}
\hspace{-0.75cm}\includegraphics[width=0.55\textwidth]{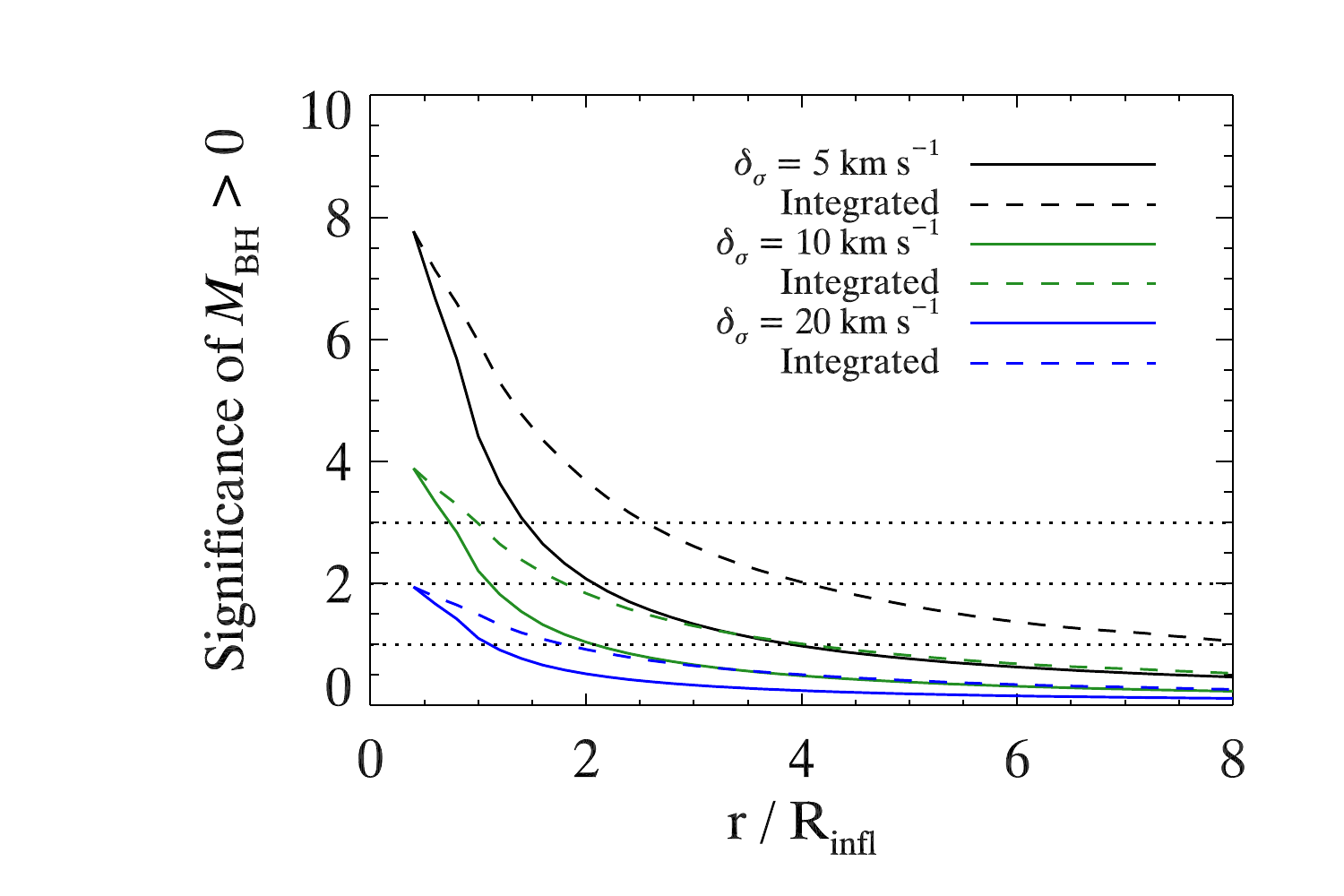}
\includegraphics[width=0.55\textwidth]{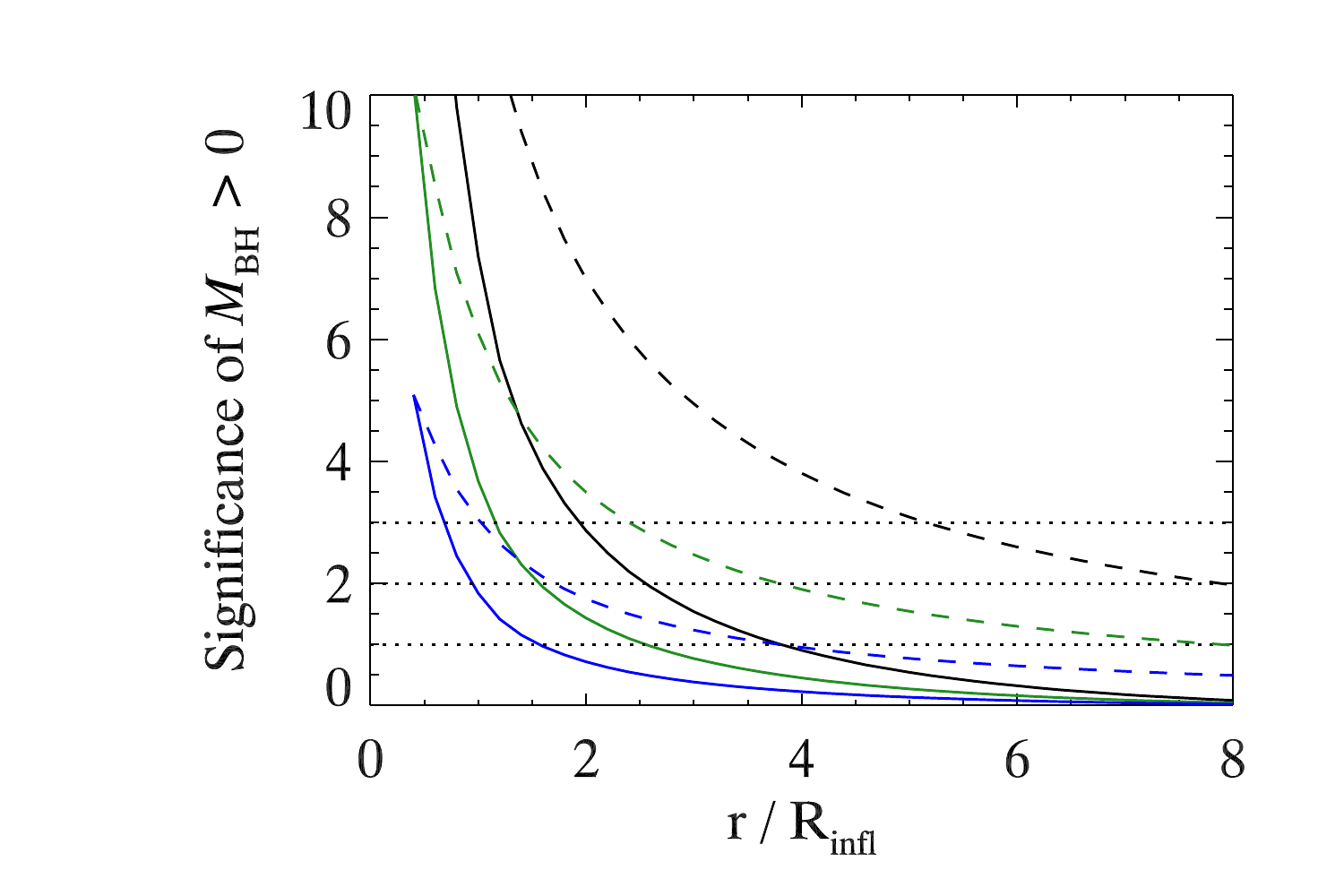}
\caption{Plots of significance of the difference between velocity
dispersions of a galaxy with a BH and one without.  The solid lines
show $(\sigma_\mathrm{BH} - \sigma_0) / \delta_\sigma$ as a function
of radial distance in units of $R_\mathrm{infl}$ for three different
values of measurement error in velocity dispersion.  The dashed curves
show the difference between light-weighted integrated velocity
dispersion profiles.  The left panel is derived from NGC~3607, a
galaxy with a core surface brightness profile, and the right panel is
derived from NGC~4026, a galaxy with a power-law surface brightness
profile.  It is clear that (1) it is possible to discern the presence
of a BH outside of the sphere of influence and (2) the ability to
discern the presence of a BH depends on the measurement error.}
\label{f:showrinfl}
\end{figure*}

The degree of resolution of the sphere of influence of the BH cannot
tell the complete story of the reliability of the BH mass
determination.  In the quantity $\rinfres$ there is encoded no
information about, e.g., the spectroscopic resolution of the data.  To
illustrate this, we perform a simple experiment.  We take surface
brightness profiles, $I(r)$, from two galaxies with velocity
dispersion $\sigma_e \approx 200 \kms$, one a core galaxy (NGC~3607)
and one a power-law galaxy (NGC~4026), and deproject to get luminosity
densities.  We assume a constant mass-to-light ratio of $\Upsilon = 4$
and calculate the projected velocity dispersion profile, $\sigma_0(r)$
from a spherical, isotropic model, assuming a seeing of 0\farcs1.  We
repeat this process but this time assuming there is a black hole with
mass $\mbh = 10^8 \msun$ at the center to get $\sigma_\mathrm{BH}(r)$.
We then calculate the difference in projected velocity dispersions
from the two models.  The ability to discriminate between these two
models depends on the measurement error of velocity dispersion,
$\delta_\sigma$.  We calculate the significance of the difference
between these models $[\sigma_\mathrm{BH}(r) - \sigma_0(r)] /
\delta_\sigma$ for three values of the measurement error
$\delta_\sigma = 5$, $10$, and $20 \kms$, which roughly correspond to
signal-to-noise ratios from \emph{HST} STIS data of more than $50$,
$40$ to $50$, and $35$, respectively, depending on the details of the
observation.  We plot these curves in Figure~\ref{f:showrinfl} as
solid lines.  Two conclusions are immediately obvious from these
curves: (1) It is possible to discern a significant difference between
the two profiles outside of $r / R_\mathrm{infl} = 2$ (corresponding
to $\rinfres < 1$), and (2) the significance of the difference depends
on the measurement errors.  An actual observation would amplify the
difference because, rather than comparing the two velocity dispersion
profiles at one location in radius, it would integrate the velocity
dispersion within a resolution bin.  To illustrate this we calculate
$[\Sigma_\mathrm{BH}(r) - \Sigma_0(r)] / \delta_\sigma$, where
\beq
\Sigma^2_\mathrm{BH}(r) = \frac{\int_0^r \sigma^2_\mathrm{BH}(r) I(r) r dr}{\int_0^r I(r) r dr}
\eeq
and similarly for $\Sigma_0$.  We plot these as dashed lines in
Figure~\ref{f:showrinfl} for each of the representative measurement
errors.  These curves show that it is possible to determine the
presence of a BH when your data do not resolve the sphere of
influence, depending on the quality of the data.

It is further evident from the figure that for a given measurement
error in velocity dispersion, the ability to discriminate between
different BH masses decreases further away from the center, i.e.,
errors in BH mass increase as $\rinfres$ decreases.  This fact has
been seen several times before in the literature.
\citet{2003ApJ...583...92G} modeled 12 galaxies using (1) ground-based
data only and (2) ground-based data combined with space-based data.
In every case, the models recover larger uncertainties for the data
sets with only ground-based data.  A separate study, by
\citet{kormendy04}, considered the many mass measurements of the BH in
M32 using a variety of modeling techniques and data that range in
level of resolution by over an order of magnitude.  The study revealed
that the error bars in the \mbh\ determination increased as \rinfres\
decreased but that the \mbh\ values were not biased at any resolution.
In fact, the value for \mbh\ found from the highest resolution data
was consistent with all of the lower resolution results.  This held
true despite the fact that the values considered were derived from
widely varying modeling methods.

\begin{figure}
\hspace{-0.75cm}\includegraphics[width=0.45\textwidth,angle=90]{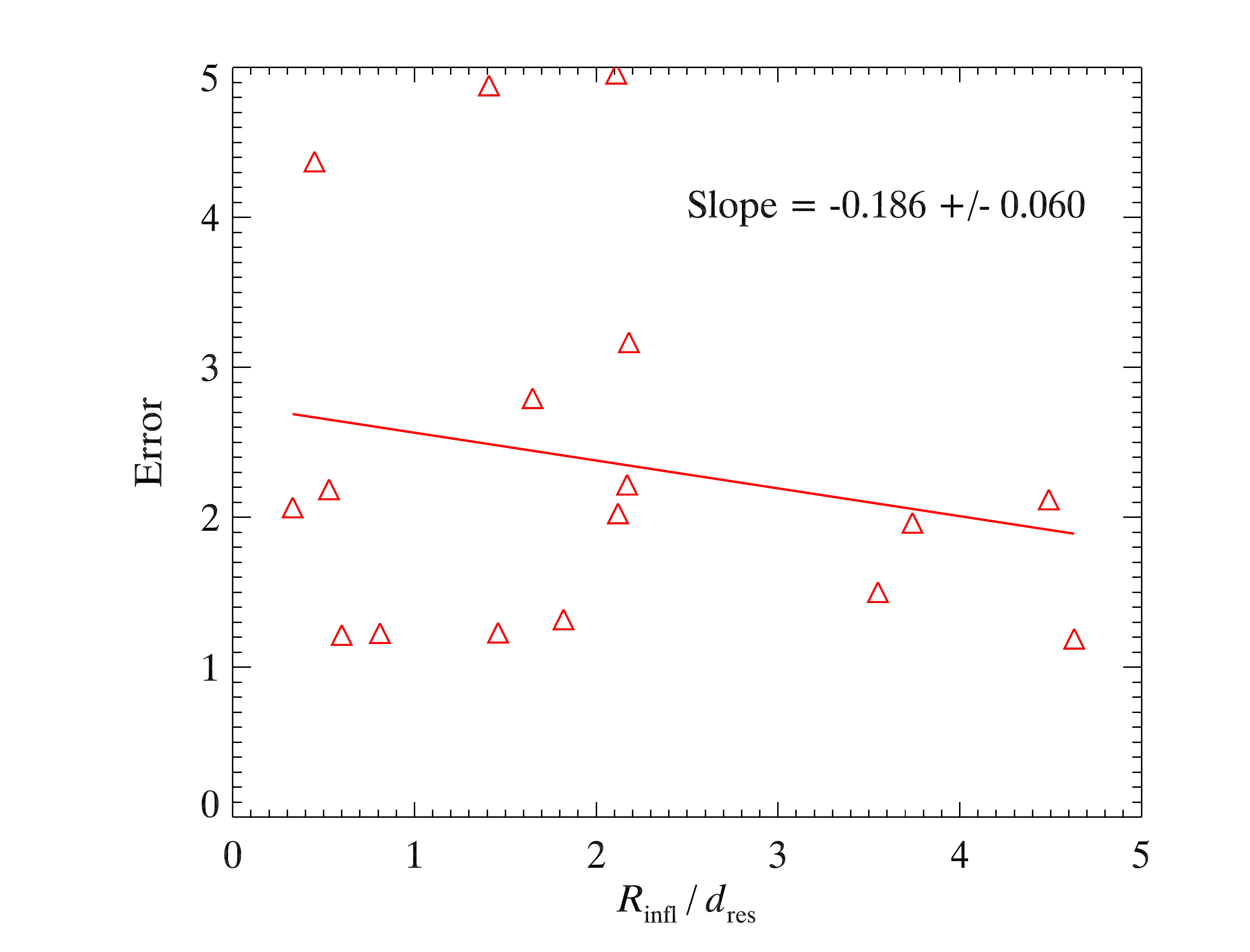}
\caption{Plot of size of errors in BH mass, calculated as
$M_\mathrm{high}/M_\mathrm{low}$, as a function of $\rinfres$ for
stellar dynamical measurements using the code described in
\protect{\citet{siopisetal08}}.  There is a weak but significant trend
indicating that lower resolution contributes to larger errors in BH
mass, but it is not the only factor in determining the uncertainty.}
\label{f:errrinfl}
\end{figure}

When looking at the data in Table~\ref{t:bigtable}, it is tempting to
look for a correlation between the size of the error in BH mass and
$\rinfres$, but the size of the error depends on many factors.  Among
the factors are differences in method of BH mass determination,
differences in codes within similar methods, and varying quality data
among the different galaxies.  When considering only stellar dynamical
mass determinations by the code described by \citet{siopisetal08},
there is a weak yet significant trend, as seen in
Figure~\ref{f:errrinfl}, which plots $M_\mathrm{high} /
M_\mathrm{low}$ as a function of $\rinfres$.  This trend is weak
because of other factors that determine the precision of the
measurement.

We stress again that larger error bars do not imply a bias in the
value of BH mass recovered from observations with smaller $\rinfres$.
For example, \citet{2003ApJ...583...92G} show that there is no
significant bias in their figure~9, which plots BH masses obtaimed
from ground-based data alone as a function of masses from the same
ground-based data combined with space-based data.  The line of
equality in their figure goes through 10 of the 12 error bars.  In 4
of the 12 galaxies, the difference between the best-fit values is less
than 25\%, and in 9 of the 12 it is less than 50\%.  While 9 of the 12
have smaller values of \mbh\ for ground-only datasets (actually the
\emph{opposite} bias to that seen in the \citet{magorrianetal98}
models, which is additional evidence that the limited resolution does
not produce a systematic bias), for 12 independent trials of an event
with 50\% chance of success, there will be 9 or more successes 7.3\%
of the time.  This is entirely consistent with no bias.  So while
increased spatial resolution will always improve the precision of BH
measurements, this trend appears to be correctly reflected in the larger error bars that emerge from the modeling procedure.

\subsection{Simulated Sample of Galaxies}
\label{mcsample}

We demonstrate the effects of such a culled sample selection with a
series of simple Monte Carlo experiments on a synthetic \msigma\ data
set.  Each synthetic data set consists of a sample of 40 galaxies,
uniformly distributed in volume out to a distance of 30~Mpc.  Each
galaxy is given a velocity dispersion from a normal distribution in
$\log{(\sigma/200~\kms)}$ centered at 0 with standard deviation 0.2.
Each galaxy is given a BH mass from an \msigma\ relation with $\alpha
= 8$, $\beta = 4.0$, and log-normal intrinsic scatter with $\epsilon_0
= 0.3$~dex.  The BH's logarithmic mass is measured with a normally
distributed measurement error of 0.2~dex and the velocity dispersion
has a 5\% error.  Since each galaxy has a distance, a BH mass, and a
velocity dispersion, we calculate $R_{\mathrm{infl}} = G \mbh
\sigma^{-2}$ and assume the galaxy to be observed with an instrument
with resolution of $d_{\mathrm{res}} = 0\farcs1$.  This process is
repeated for $10^5$ realizations for each of the choices in assembling
the sample described below.

We first tried fitting for the parameters of the \msigma\ relation
using these simulated data sets and the same fitting procedures that
we applied earlier to the actual data.  We successfully recovered the
input parameters with only a slight bias to low intrinsic scatter 
(recovering $\epsilon = 0.27$ instead of the input value of
$0.3~\mathrm{dex}$). The samples were then culled in two ways: (1) we
removed galaxies that fell below a given cutoff in \rinfres, where
$R_\mathrm{infl}$ is computed from the measured BH mass and
velocity dispersion; and (2) we removed galaxies that fell below a given
cutoff in \rinfres, where $R_\mathrm{infl}$ is computed from the
BH mass predicted by the \msigma\ relation.  Since the Galaxy
plays an important role in the observed sample (because of its
position near the low end of the range of observed $\sigma_e$ and the
small uncertainty in the measured mass), we also augmented some of the
samples by adding one galaxy with the actual measured values of the
Galaxy's BH mass and velocity dispersion with their
corresponding uncertainties.

\subsection{Culling the Simulated Sample Based on Traditional Sphere 
of Influence}
\label{mcthresh}
\begin{figure*}
\centering
\hbox{\hspace{-1.5cm}\includegraphics[width=0.575\textwidth]{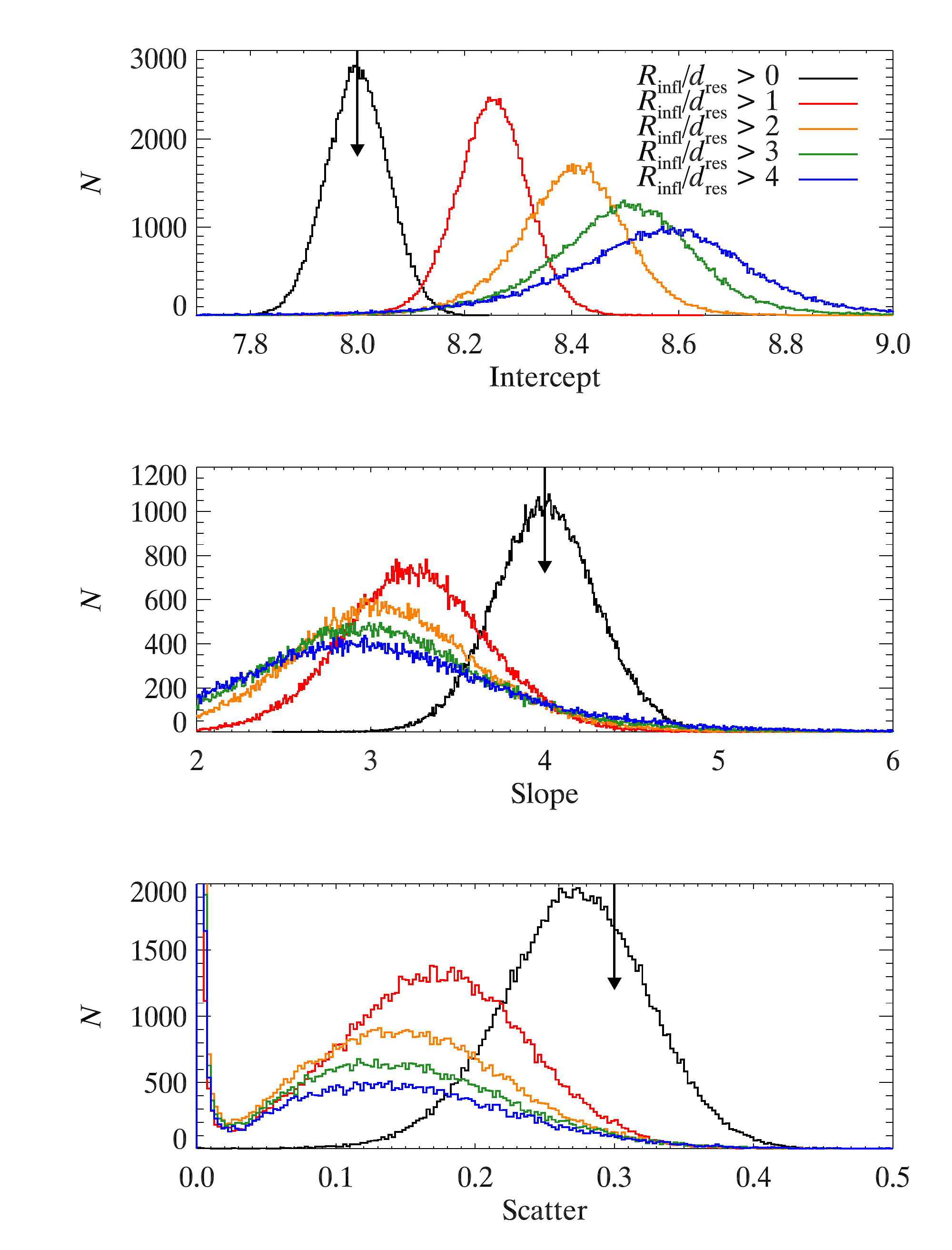}\hspace{-1cm}\includegraphics[width=0.575\textwidth]{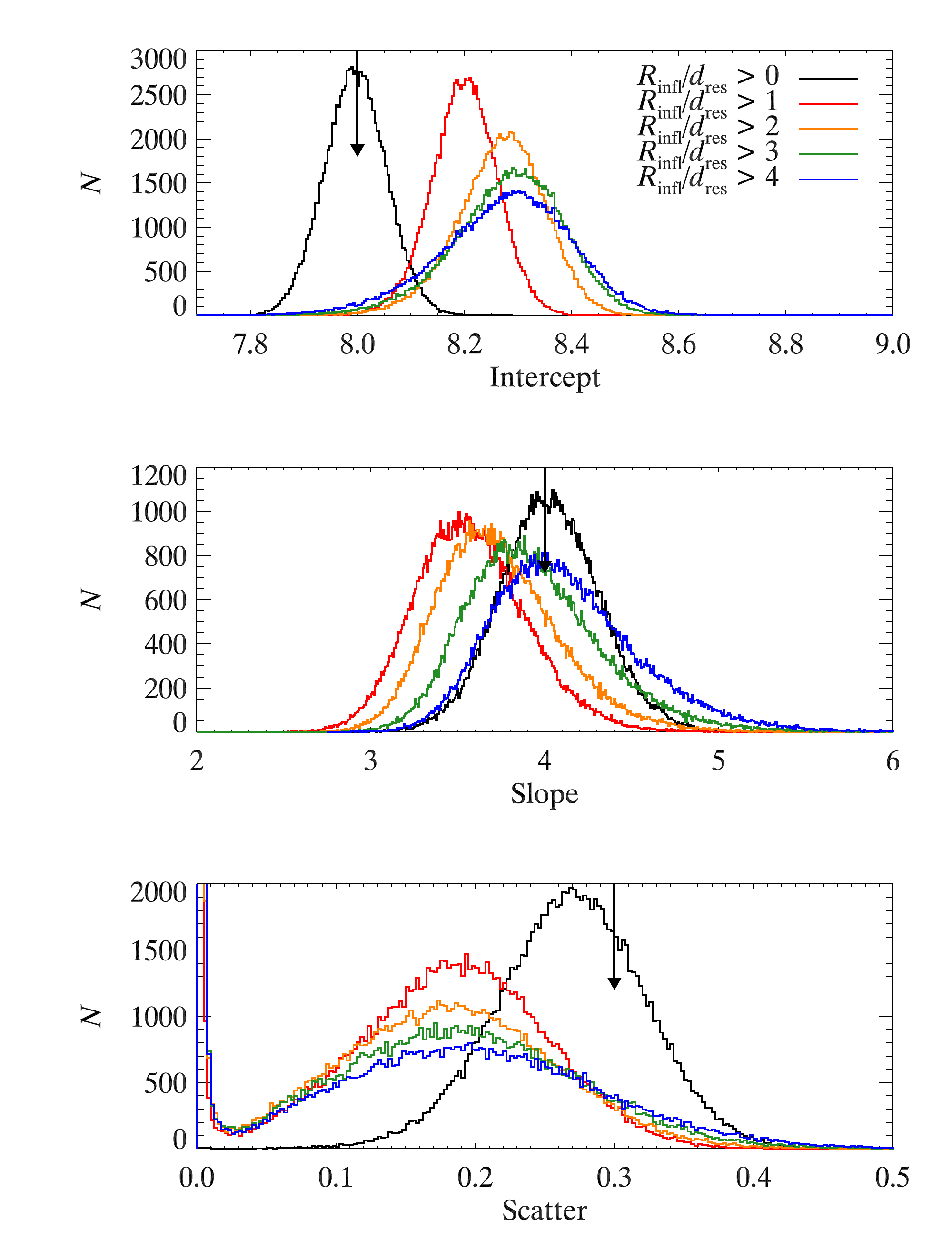}}
\caption{\footnotesize Histograms of fit parameters from Monte Carlo simulations of
$10^5$ realizations of 40 galaxies with BH masses derived from
a model with $\alpha = 8$, $\beta = 4.0$, and $\epsilon_0 = 0.3$ with
rms measurement errors in log mass of $0.2$.  The right-hand panels differ
from the left panels in that one of the 40 galaxies generated by the
Monte Carlo procedure is replaced by a galaxy having the mass and
velocity dispersion of the Milky Way.  The black curves show results
when all galaxies are included in the fits.  These curves show that
our fit method is able to recover the original parameters with a
slight bias to lower intrinsic scatter.  The remaining curves come
from imposing a cutoff in \rinfres\ and fitting the remaining galaxies
from the original 40.  The colors correspond to a cutoff value of 1.0
(\emph{red}), 2.0 (\emph{orange}), 3.0 (\emph{green}), and 4.0
(\emph{blue}).  The biases in the left panels can be summarized as
follows: (1) the intercept $\alpha$ increases monotonically with
increasing cutoff in \rinfres; (2) the slope $\beta$ decreases
monotonically with increasing cutoff; and (3) the intrinsic scatter
$\epsilon_0$ decreases monotonically with increasing cutoff, with a
large number of samples consistent with $\epsilon_0 = 0$.  The orange
curve ($\rinfres > 1$) may be compared with our fits to sample RS,
which does not contain the Galaxy.  If our Galaxy is included (right
panels), the biases are (1) the intercept increases monotonically
with increasing cutoff in \rinfres, though the presence of the Galaxy
mitigates this somewhat; (2) the slope is biased to slightly lower
values for the first three cutoffs and to a slightly higher value for
the highest cutoff; and (3) the inferred intrinsic scatter is biased
to a lower value, with a significant number of runs consistent with
$\epsilon_0 = 0$.}
\label{f:threshnomw}
\label{f:threshmw}
\end{figure*}

First, we use various cuts in \rinfres\ to eliminate galaxies from the
simulated sample.  The results from this simulation are shown in
Figure~\ref{f:threshnomw}.  Without the Galaxy, the trends are simple
and monotonic: increasing the \rinfres\ cutoff increases the value of
the intercept and decreases the values of the slope and scatter.
Because our sample RS (see Appendix~\ref{difffits}) omits the Galaxy,
comparison of the orange curve to the black curve in
Figure~\ref{f:threshnomw} approximates comparison of the fit values in
our RS sample to our S sample.  They show the same trends in
intercept, slope, and scatter.

\begin{figure*}
\centering
\hbox{\hspace{-1.5cm}\includegraphics[width=0.45\textwidth,angle=90]{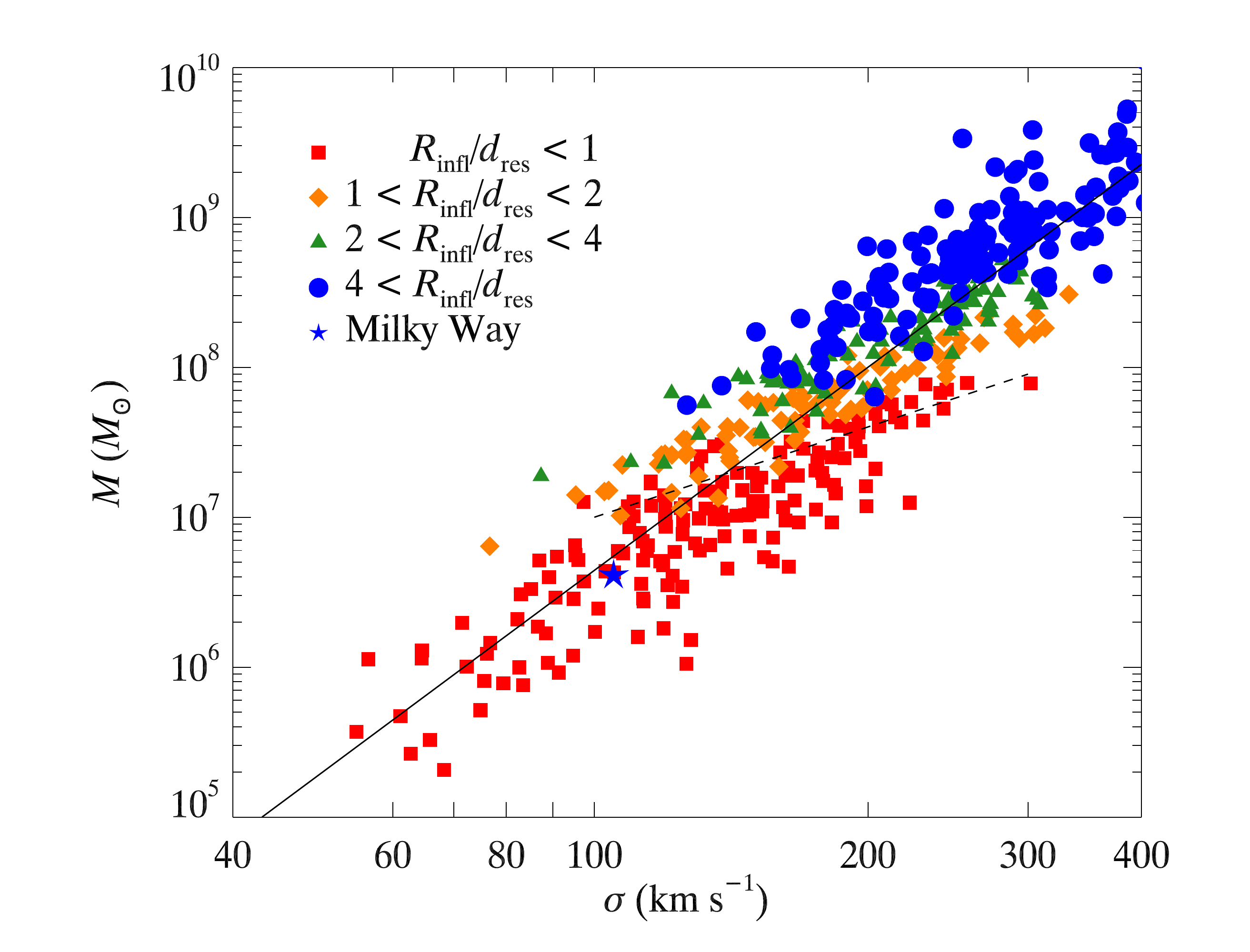}\hspace{-1cm}\includegraphics[width=0.45\textwidth,angle=90]{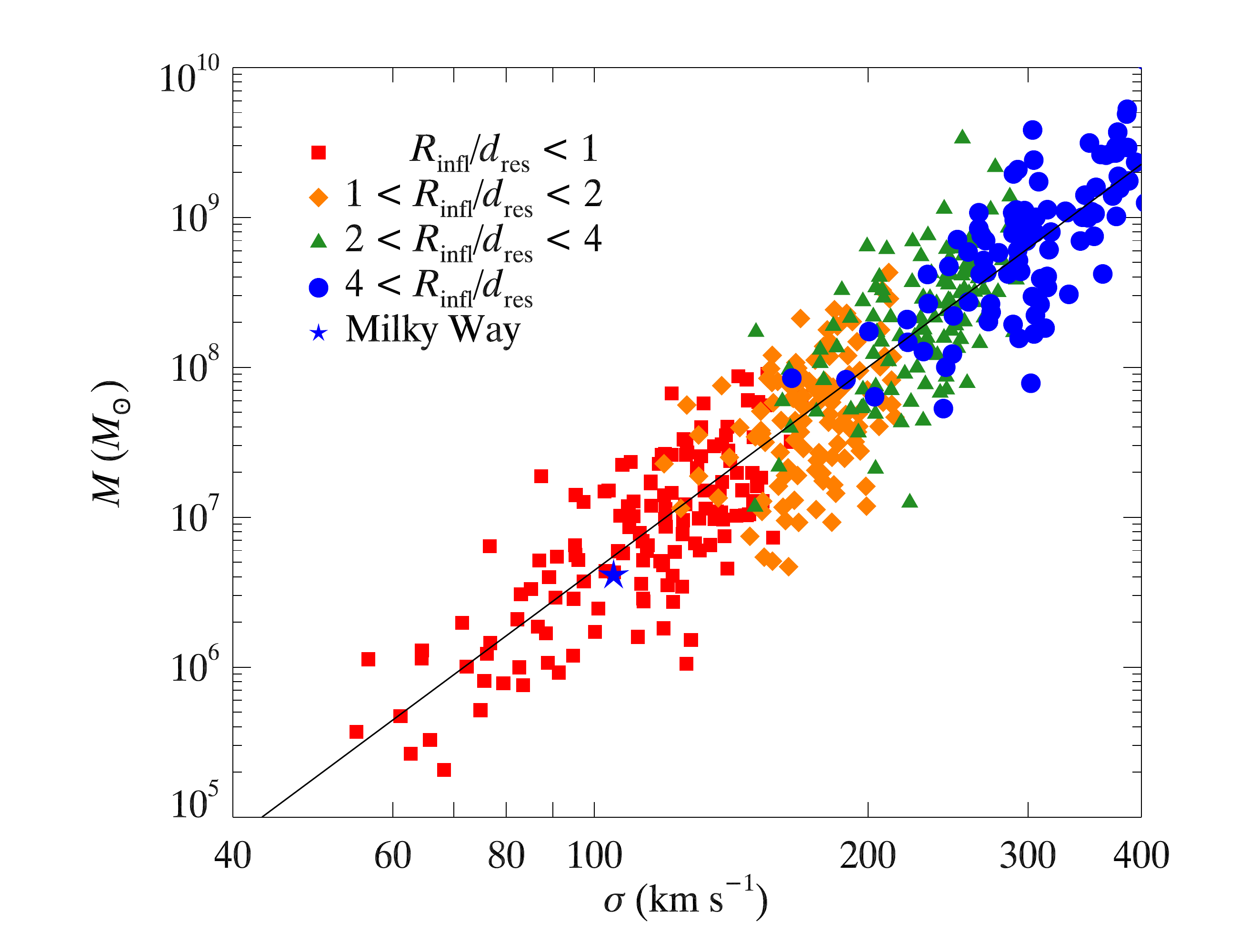}}
\caption{Monte Carlo sample of 500 galaxies with BH mass generated
from an \msigma\ relation with $\alpha = 8$ and $\beta = 4.0$ and a
log-normal scatter of 0.3 dex with measurement errors of 0.2 dex.  The
ridge line of the \msigma\ relation is drawn as a black line.  The
Galaxy is plotted as a pentagram.  Both panels use different colors
and symbols for different resolution levels.  The \emph{left} panel
defines $R_{\mathrm{infl}} = G\mbh/\sigma^2$ with \mbh\ equal to the
measured BH mass; and the \emph{right} panel uses our
alternate definition using the expected mass of the BH based on the
\msigma\ relation.  For each definition, the symbols are $\rinfres <
1.0$ (\emph{red squares}), $1.0 < \rinfres < 2.0$ (\emph{orange
diamonds}), $2.0 < \rinfres < 4.0$ (\emph{green triangles}), and
$\rinfres > 4.0$ (\emph{blue circles}).  In the left panel, fitting
the combined subsamples with \rinfres\ exceeding a given value yields
a shallower slope and a smaller intercept than the underlying \msigma\
relation..  The reason for the bias in slope is that cuts in
$R_{\mathrm{infl}}$ tend to fall along lines of $\mbh \propto
\sigma^{\beta - 2}$ (since $R_{\mathrm{infl}} \propto \mbh
\sigma^{-2}$ and $\mbh \propto \sigma^\beta$).  This is illustrated by
the dashed line of slope 2.0.  Cuts using the expected
$R_{\mathrm{infl}}$ (\emph{right panel}) do not suffer from this bias
because the cuts in \rinfres\ with this definition tend to run at
constant $\sigma$.  The bias they introduce is far less than with the
traditional definition.  The uncertainties in the parameters when
fitting, however, are much larger due to the decreased dynamic range
in $\sigma$.}

\label{f:rinfmc}
\label{f:altrinfmc}
\end{figure*}

The reasons for these biases can be easily seen in
Figure~\ref{f:rinfmc}a, which illustrates the effects by plotting a
sample of 500 galaxies randomly generated in the same way and plotting
different ranges in \rinfres\ with different symbols.  Because
$R_{\mathrm{infl}} \approx \mbh \sigma^{-2}$ and $\mbh \approx
\sigma^\beta$, lines in constant \rinfres\ tend to fall on lines of
$\mbh \approx \sigma^{\beta - 2}$.  (This argument neglects the
different distances of the galaxies, but since the Monte Carlo
simulation is uniformly distributed in volume, most galaxies are near
the outer edge of the volume.)  Since the synthesized data set uses
$\beta = 4.0$, each subsample has a decreased scatter because the cuts
fall along lines with slope $\beta - 2 = 2.0$ and take out the bottom
portion of the scatter.  The slopes are similarly biased to low values
because the points have been removed from systematically smaller
masses and systematically smaller velocity dispersions.  The intercept
increases as each subsample's mean increases.

When the Galaxy is included (Fig.~\ref{f:threshmw}b), the magnitude of
the biases decreases.  The intercepts are still biased to high values.
The slopes are biased to low values except those from the most
restrictive cutoff, which are biased to a slightly higher value.  The
scatters are biased to a smaller value; a significant number are
consistent with $\epsilon_0 = 0$.  The bias when using older, lower
values for our Galaxy's BH mass \citep{2005ApJ...620..744G} would have
produced more extreme changes in slopes.

\subsection{Culling the Simulated Sample Based on Sphere of Influence 
from Expected Black Hole Mass}
\label{mcfaberthresh}
\begin{figure*}
\centering
\hbox{\hspace{-1.5cm}\includegraphics[width=0.575\textwidth]{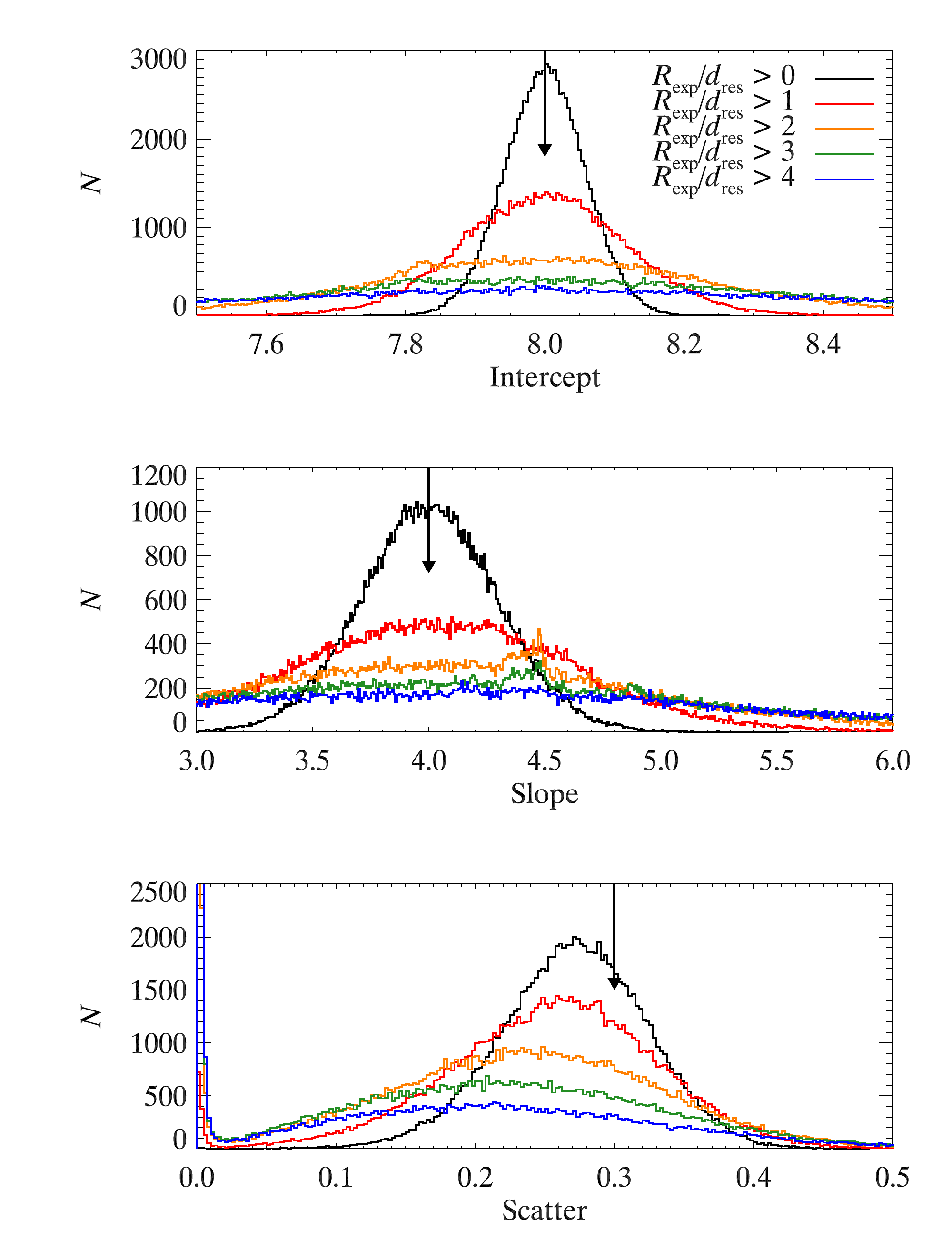}\hspace{-1cm}\includegraphics[width=0.575\textwidth]{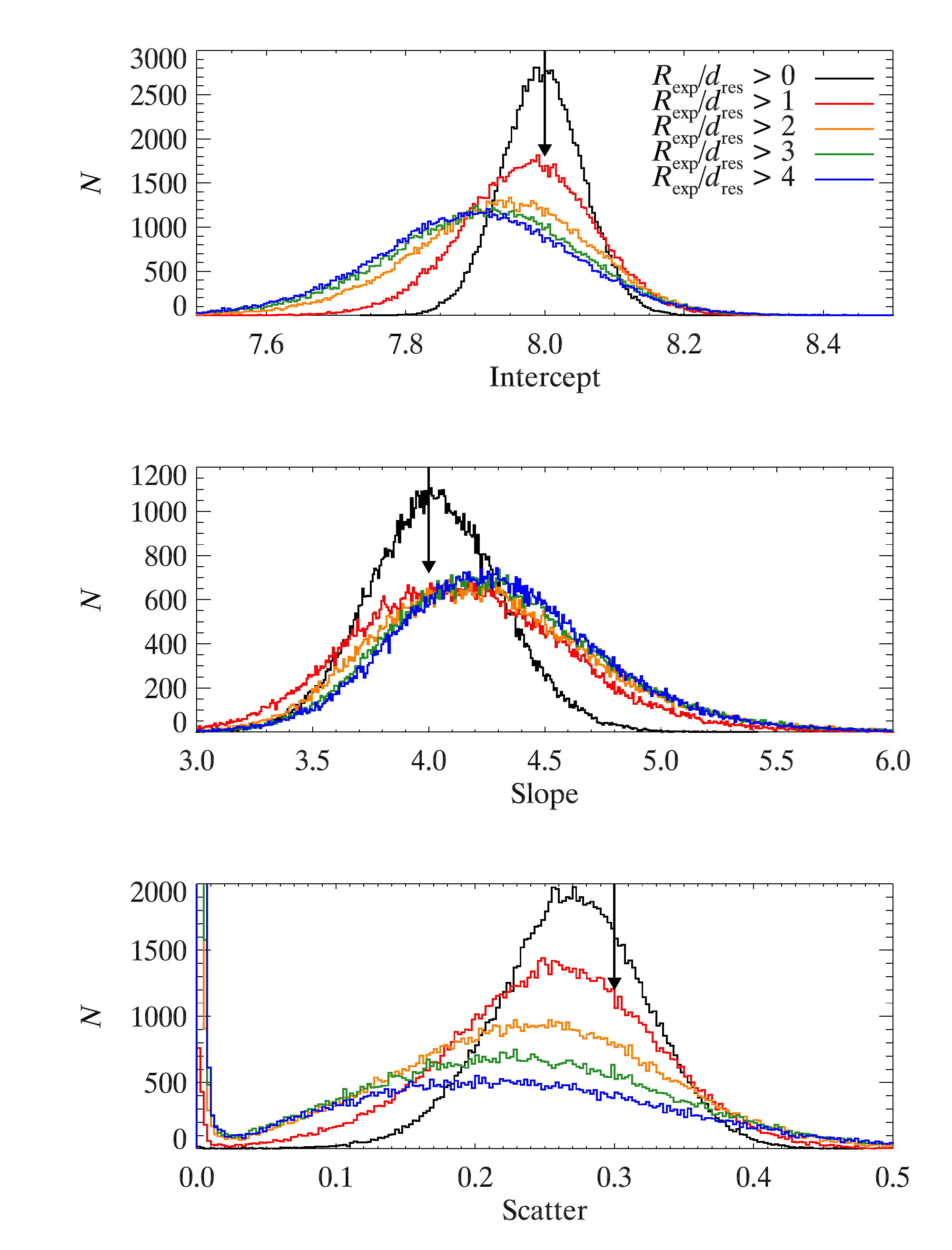}}
\caption{Histograms of fit parameters from Monte Carlo simulations as
in Figure~\ref{f:threshnomw}, except the cuts are based on
$R_{\mathrm{exp}}/d_{\mathrm{res}}$ as defined in
\S~\ref{mcfaberthresh}.  The colors correspond to a cutoff value of
1.0 (\emph{red}), 2.0 (\emph{orange}), 3.0 (\emph{green}), and 4.0
(\emph{blue}).  The biases in the left panels are minimal, but the
efficiency of recovering the original parameters is greatly reduced
since the dynamic range in $\sigma$ is greatly reduced.  In the right
panels, the biases incurred from these cutoffs are small, and the
efficiency of recovering the original parameters is greatly improved
compared to the left panels since the dynamic range in $\sigma$ always
includes the Galaxy.  There is, however, still a tendency to infer an
intrinsic scatter consistent with $\epsilon_0 = 0$ when making a large
cut.
\bigskip
}
\label{f:altthreshnomw}
\label{f:altthreshmw}
\end{figure*}

For the second culling strategy, we adopt a different definition of
$R_\mathrm{infl}$ that does not lead to the extreme biases seen above.
Because the traditional definition of $R_\mathrm{infl}$ depends on the
measured BH mass, it is sensitive to whether the BH is overmassive or
undermassive for a given velocity dispersion.  Thus, using
$R_\mathrm{infl}$ as a sample criterion necessarily biases any
measurement in scatter.  Instead, we offer an alternative definition
based on the \emph{expected} value of $M_\mathrm{BH}$ from the
\msigma\ relation.  That is, we replace culling based on \rinfres\ by
culling based on $R_{\mathrm{exp}}/d_{\mathrm{res}}$, where
$R_\mathrm{exp} = G M_\mathrm{pred} \sigma^{-2}$ and $M_\mathrm{pred}
= 10^\alpha (\sigma / 200~\kms)^\beta~\msun$.  Using this as a
selection criterion for finding the \msigma\ relation requires only a
simple and rapidly convergent iterative procedure.  It is possible
that this procedure will not always converge to a unique answer, but
this is unlikely to happen if the initial guess is chosen close to
the actual relation.  Adopting this strategy substantially decreases
the bias in the fitting parameters (Fig.~\ref{f:altthreshnomw}) but
dramatically increases their uncertainties since the range of $\sigma$
used in the fit is reduced.  The reason for this is seen in the right
panel of Figure~\ref{f:altrinfmc}.  Instead of falling along lines of
slope 2.0, the cuts tend to fall on lines of constant $\sigma$.

\subsection{Culling Based on Traditional Sphere of Influence Using Observed Sample with Simulated Masses}
\begin{figure}
\centering
\hbox{\hspace{-1.5cm}\includegraphics[width=0.575\textwidth]{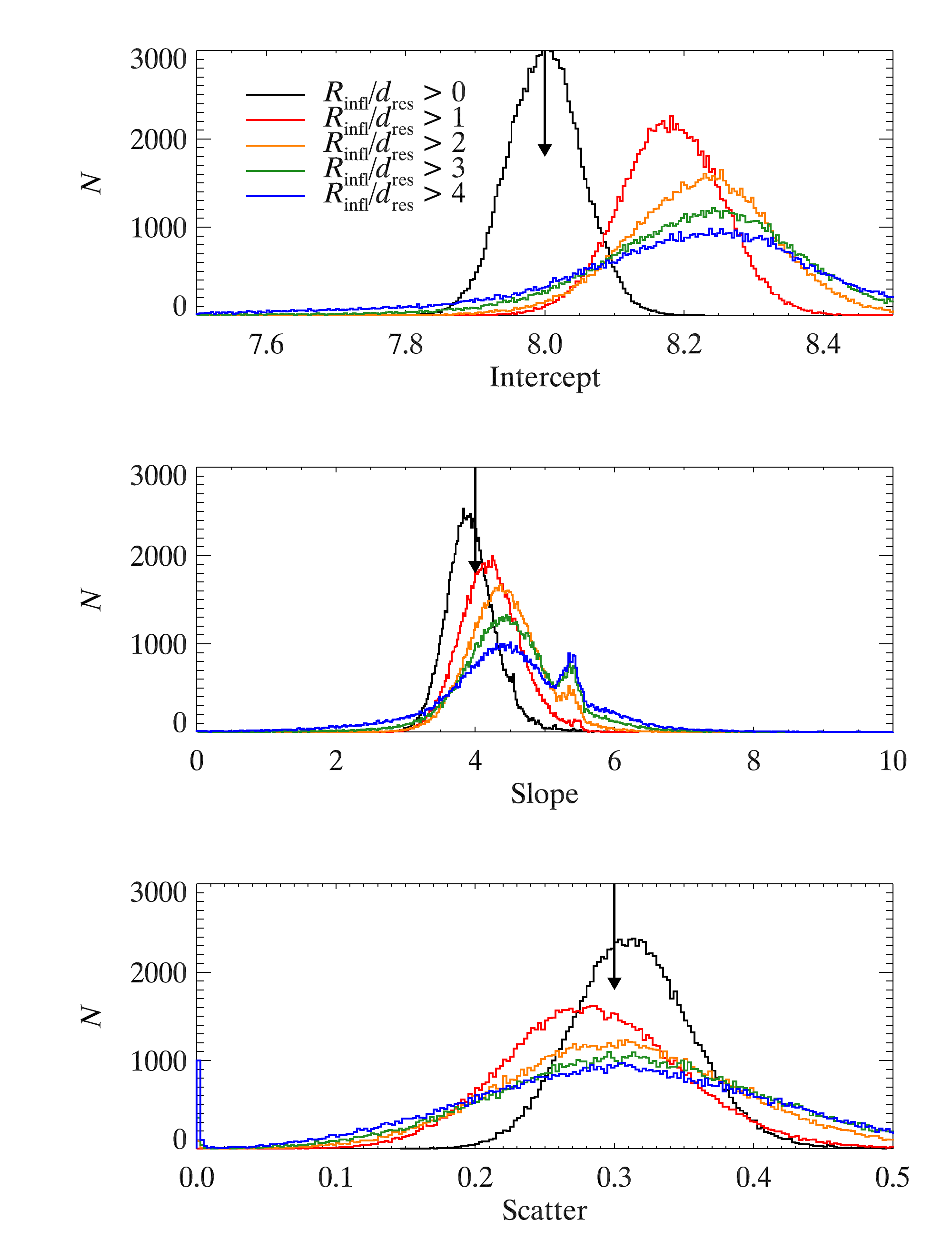}}
\caption{Histograms of fit parameters from Monte Carlo simulations as
in Figure~\ref{f:threshnomw}, except that the velocity dispersions,
distances, and errors in mass measurements are all derived from the
actual observed sample of galaxies.  The masses are synthesized from
an \msigma\ model.
\bigskip
}
\label{f:realnomw}
\end{figure}

\begin{figure}
\centering
\hbox{\hspace{-1.5cm}\includegraphics[width=0.45\textwidth,angle=90]{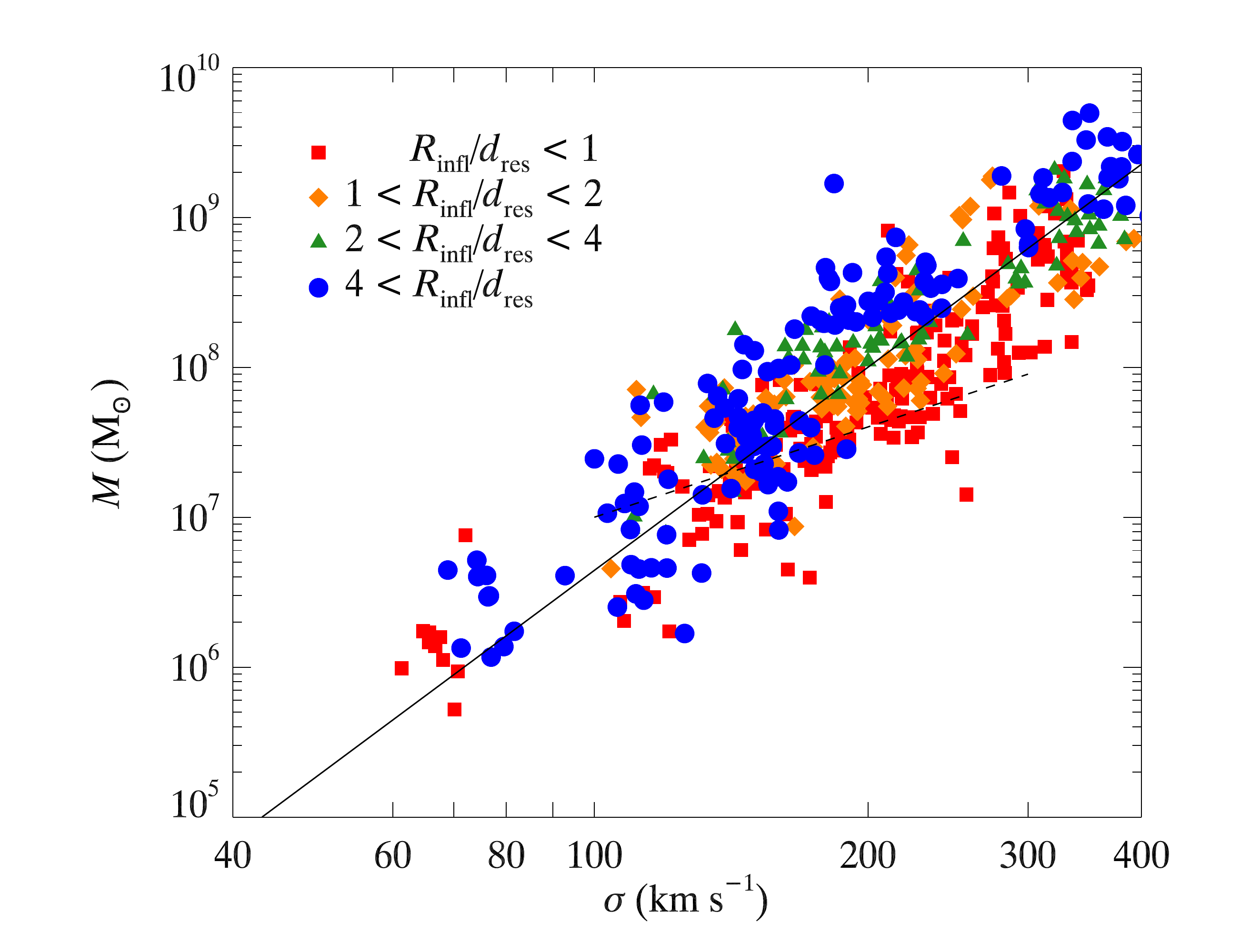}
}
\caption{Same as Figure~\ref{f:rinfmc}, except that the velocity dispersions,
distances, and errors in mass measurements are all derived from the
actual observed sample of galaxies.  The masses are synthesized from
an \msigma\ model.}

\label{f:realrinfmc}
\end{figure}

The above synthetic datasets represent an idealized scenario in which
the galaxies come from a volume-limited sample.  The real sample of BH
mass measurements, however, is not uniformly distributed in volume.
As a final experiment on a synthetic data set, we use the observed
sample of velocity dispersions, errors in velocity dispersion,
distances, errors in mass measurements, and instrumental resolution.
The masses are synthesized from an \msigma\ model as before and
censored according to \rinfres\ as calculated from the synthesized BH
mass.  There is no need to artificially include the Galaxy in these
simulations since it is used along with all of the other data.  The
resulting distributions of fit parameters are presented in
Figure~\ref{f:realnomw}.  Again, using the entire sample recovers the
input parameters.  Cuts in \rinfres\ show the same bias to high values
of intercept as well as reducing the efficiency of the estimates.  The
slope is biased to high values.  The intrinsic scatter is similarly
biased to low values, with a noticeable number of realizations
recovering zero intrinsic scatter.  We plot 5 realizations and
indicate their level of resolution in Figure~\ref{f:realrinfmc}.  The
same trends as seen in the idealized simulations are present.  The
trends in parameter estimation from culled samples seen here may
explain why \citet{ff05} find a higher intercept ($\alpha = 8.22 \pm
0.06$), a much higher slope ($\beta = 4.86 \pm 0.43$), and a
``negligible'' scatter with their sample restricted to $\rinfres > 1$.

\begin{figure}
\hspace{-0.75cm}\includegraphics[width=0.55\textwidth]{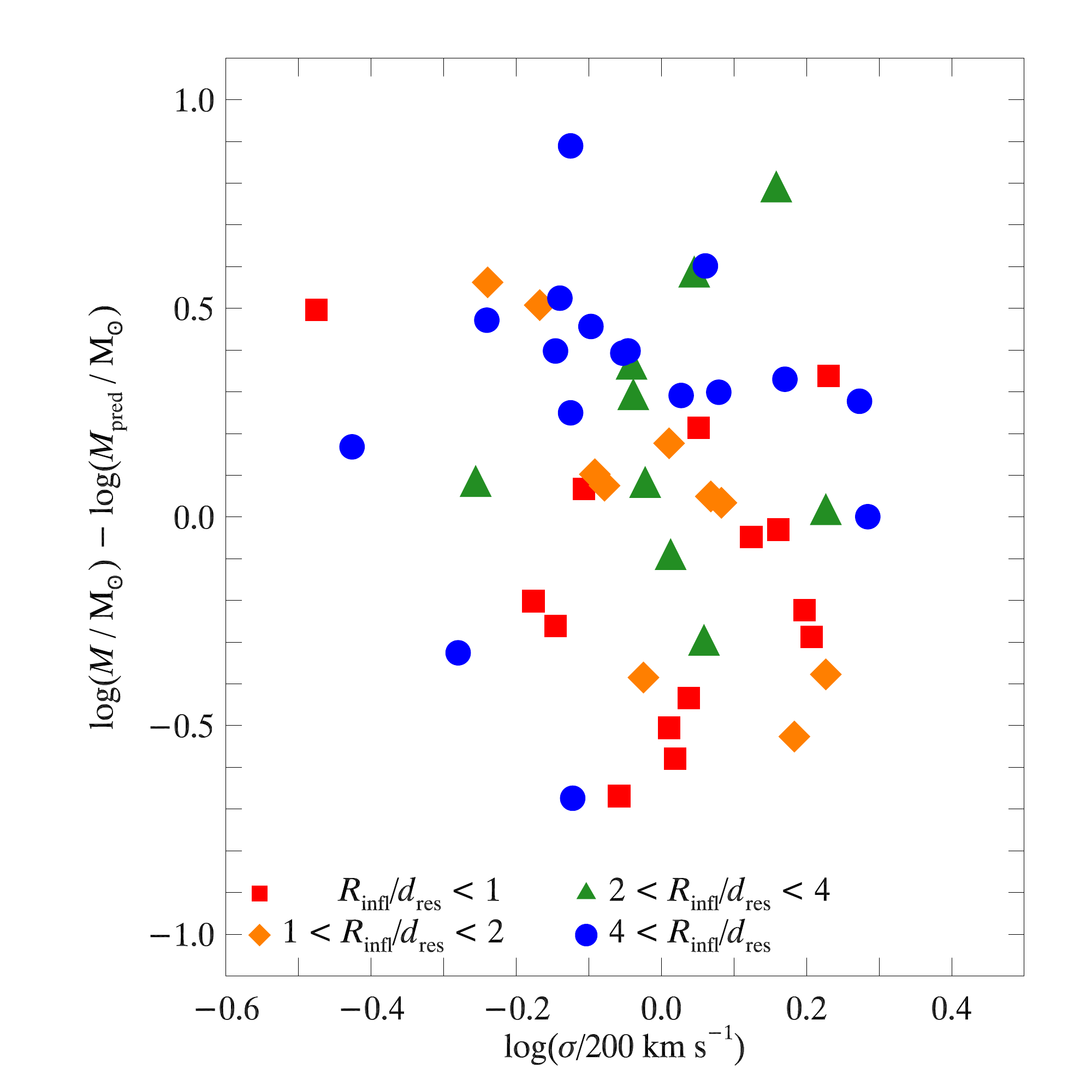}
\caption{Residuals in mass from our fit to \msigma\ for the full
sample.  The colors and shapes of the subsamples correspond to how
well resolved the BH's sphere of influence is (from low to
high \rinfres: red squares, orange diamonds, green triangles, and blue
circles).  Removing the more poorly resolved objects from the sample
leaves a residual trend, even though there is no such trend for the
sample as a whole.
\bigskip
}
\label{f:msigmarinf}
\end{figure}

The systematic effects of $\rinfres$ cutoffs can also be seen in
Figure~\ref{f:msigmarinf}, which plots the residuals to our \msigma\
fit as a function of $\sigma_e$.  The lower resolution objects tend to
be found toward the bottom and right of the plot, and the most highly
resolved galaxies clearly show a residual trend in $\sigma_e$ even
though the sample as a whole shows no such trend. Hence, eliminating
reliable BH mass measurements (in the sense that the error
bars accurately reflect the uncertainty in the measurement, even if
they are large) by selecting on \rinfres\ will bias the resulting
fits.  This bias could be mitigated by using $R_{\mathrm{exp}}$ rather than
$R_{\mathrm{infl}}$ in the culling criterion but then the
uncertainties in the fit become much larger.  Thus, our recommendation
is to include BH masses without regard to the value of \rinfres.

\section{Discussion}
\label{discuss}
\subsection{Systematic Uncertainties in Measurements of BH Mass}
\label{systematics}
Our measurements of the intrinsic scatter in the \msigma\ and \ml\
relations assume that the stated errors in \mbh\ accurately reflect the
measurement uncertainties.  If unaccounted systematic errors are
large, however, they will significantly increase the 
inferred intrinsic scatter.  We list some potential sources of
systematic errors in \mbh\ measurements.

\begin{itemize}
\item{Models in which the value of $\chi^2$ i smuch larger than the
number of parameters are inadequate.  We have removed \mbh\
measurements from our sample where this is a problem.}
\item{For more massive galaxies, the derived stellar mass-to-light
ratio ($\Upsilon_*$) may be wrong due to an un-modeled contribution of
the dark matter halo to the stellar kinematics at the large radii used
to constrain the stellar population.  If the dark matter halo is less
centrally concentrated than the stellar mass distribution, as is
expected from cosmological simulations of adiabatic collapse
\citep{gnedinetal04}, then the intrinsic $\Upsilon_*$ would be lower
than inferred, implying larger than estimated \mbh\ values
\citep{gt09}.}
\item{If the BH mass is measured by gas kinematics, uncertainty in
  inclination and non-gravitational forces may cause systematic errors
  \citep{hoetal02}.}
\item{Galaxies that have recently merged such as Cen~A may not be in
  the equilibrium state that is assumed by stellar kinematic models.}
\item{There are galaxies with AGN, whose central kinematics are hard
  to measure. This is systematic noise since it depends on how one
  handles the AGN subtraction.}
\item{Uncertain deprojection of the gas density profiles may be a
source of systematics.}
\end{itemize}

\subsection{The Value of the Slope in the \texorpdfstring{\msigma}{M-sigma} 
Relation}
\label{flatness}
The slope we report here ($\beta = \msslopeerr$) is noticeably,
though not significantly, steeper than the slope obtained by
\citet{tremaineetal02} ($\beta = 4.00 \pm 0.31$).  For comparable assumptions, however, this
difference disappears.  The most direct comparison is to apply the
symmetric least-squares fitting method of \citet[][see
Table~\ref{t:fits}]{tremaineetal02} to our sample without upper
limits, which yields $\beta = 4.02 \pm 0.37$, compared to $\beta =
4.00 \pm 0.31$ found by \citet{tremaineetal02}.  The two slopes are
consistent with each other.

\subsection{The Value of the Scatter in the \texorpdfstring{\msigma}{M-sigma} 
Relation}
\label{sizeofscatter}
The value of the intrinsic scatter obtained here ($\epsilon_0 =
\msscaterr$) is significantly larger than the value estimated by
\citet{tremaineetal02}, which was no more than $\epsilon_0 = 0.25$ to
$0.3$.  Since the primary interest of the \msigma\ relation derives
from its tightness, a conclusion that the scatter is larger than
previously believed warrants attention.  \citet{tremaineetal02} used
both a different fitting method and a different sample.  We examine
both of these, finding that the sample differences, rather than the
fitting method or updated \mbh\ measurements for a given galaxy, are
the cause of the different scatter measurements.

We fitted several different samples: (1) the original \citet{tremaineetal02}
sample, (2) the \citet{tremaineetal02} sample updated to the values
used in this paper, (3) sample S from this paper, and (4) galaxies in
sample S that are not in \citet{tremaineetal02}.  Fits were done using
both the maximum likelihood method developed for this paper and the
symmetric least-squares method used by \citet{tremaineetal02}.
We use our sample
without upper limits to make the most direct comparison between the
methods.  The intrinsic scatter that we find in sample S is almost
identical to that in sample SU.  When used on the same sample, there
was never a significant difference between the two methods, which
leads us to quote only the results from the maximum likelihood method
of this paper.  The original \citet{tremaineetal02} sample yields an
intrinsic scatter of $\epsilon = 0.31 \pm 0.06$.  Updating the values
in the same sample increases the intrinsic scatter to $\epsilon = 0.35
\pm 0.06$.  The intrinsic scatter in sample S is $\epsilon = 0.43 \pm
0.06$, and the intrinsic scatter from just the 19 galaxies in sample S
not found in \citet{tremaineetal02} is $\epsilon = 0.51 \pm 0.11$.
Since the two methods used do give consistent results, they cannot
be the source of the difference.  

Next we examine the difference between the original
\citet{tremaineetal02} sample and the new galaxies.
Kolmogorov-Smirnov tests on the distributions of mass, velocity
dispersion, and error in logarithmic mass cannot rule out the null
hypothesis that the two samples came from the same parent distribution
at better than 50\% confidence.

A chi-square test on the distribution of morphological types, however,
reveals that the probability that the two samples in galaxy type come
from the same parent distribution is only approximately 20\%.  This is
far from conclusive, but it is noteworthy that the more recent \mbh\
measurements consist of a relatively larger number of spirals.  In
\S~\ref{msigmaresults}, we find that the intrinsic scatter in the
population of late-type galaxies may be larger than the scatter in
early-type galaxies.  This may be a result of unrecognized systematic
effects in \mbh\ measurement of spirals or because the scatter is
actually larger.  In contrast, we find consistent intrinsic scatter
estimates, at about 0.3, when fitting only early-type galaxies from
the two samples.

In this context, \citet{hu08} has noted that pseudobulges appear to
host relatively smaller black holes than do ``classical'' bulges.  We
have included both bulge types in the analysis, and the differences in
the relative contributions of the two types between various samples
may enhance the intrinsic scatter.

One of the spiral galaxies that was not included in the
\citet{tremaineetal02} samples is Circinus, the largest outlier in our
sample.  When we exclude Circinus from sample S, the scatter reduces
to $\epsilon = 0.36 \pm 0.05$, in near agreement with
\citet{tremaineetal02}.  While it is disturbing that a single galaxy
can cause such a large change in our scatter estimates, Circinus is an
extreme outlier, a factor of $\sim30$ below the \msigma\ ridge line.
In order for any \msigma\ model to explain such an outlier that does
not have large measurement errors, a large intrinsic scatter is
needed.

We summarize the reasons for the difference in intrinsic scatter
measurements as follows.
\begin{itemize}
\item{The difference in fitting methodology is not the source of the
difference in intrinsic scatter estimates.}
\item{The difference in samples is the source of the difference.}
\item{Updating the \mbh\ measurements of galaxies in
\citet{tremaineetal02} slightly increases the intrinsic scatter
estimate.}
\item{The galaxies in the \citet{tremaineetal02} sample and the new
galaxies in this sample possibly have different distributions of
Hubble types.  If they do, the increased fraction of spirals in the
newly added galaxies may be a source of the increased scatter. The
underlying cause may be either because (1) spiral galaxies are
susceptible to more systematic error or (2) there is a larger
intrinsic scatter in \mbh\ in spiral galaxies, as might occur if there
is a significant difference in the population of black holes hosted by
pseudobulge versus classical bulges.}
\item{When considering only early-type galaxies, the scatter values of
the samples are consistent with each other and are close to the value
quoted by \citet{tremaineetal02}.}
\item{The inclusion of Circinus makes a large difference in the
scatter estimates.}
\end{itemize}

\subsection{The Role of Upper Limits}
\label{upperlimits}
Upper limits provide information on the population of BH masses.  Weak
upper limits provide little information, but when the limit is
comparable to detected BH masses at a similar velocity dispersion or
luminosity, it can provide a valuable constraint on the fit.  Thus,
excluding upper limits from the fit excludes potentially useful
information.  Since restrictive upper limits will lie at lower masses,
excluding them may also introduce a bias, especially in estimates of
the intrinsic scatter and intercept.  For these reasons, we view our
fit including upper limits as the most complete answer.  To avoid
being biased by upper limits measured in galaxies with no BH, which do
not belong in either the \msigma\ or \ml\ relations, we include the
probability of a galaxy's having no BH ($P_\emptyset$) in our fit.  As
the simplest possible form, we assume $P_\emptyset$ is constant for
all galaxies.  Given that smaller galaxies may be less likely to form
central BHs and that their smaller escape velocities mean ejection of
their BHs is more likely, this assumption may not hold.  Tests using a
simple linear dependence on velocity dispersion, $P_\emptyset = \zeta
+ \eta \log{(\sigma / 200 \kms)}$, found $\eta = 0.001 \pm
0.047$.  So assuming a constant form is unlikely to introduce a
strong bias, and in any event the data are not strong enough to
support a more elaborate model.
Excluding upper limits reduces the slope, although not by a
statistically significant amount.

\subsection{Implications for Space Density of Black Holes and Studies 
of \texorpdfstring{\msigma}{M-sigma} Evolution}
\label{bhdf}
\begin{figure*}[hbt]
\centering
\includegraphics[width=0.49\textwidth]{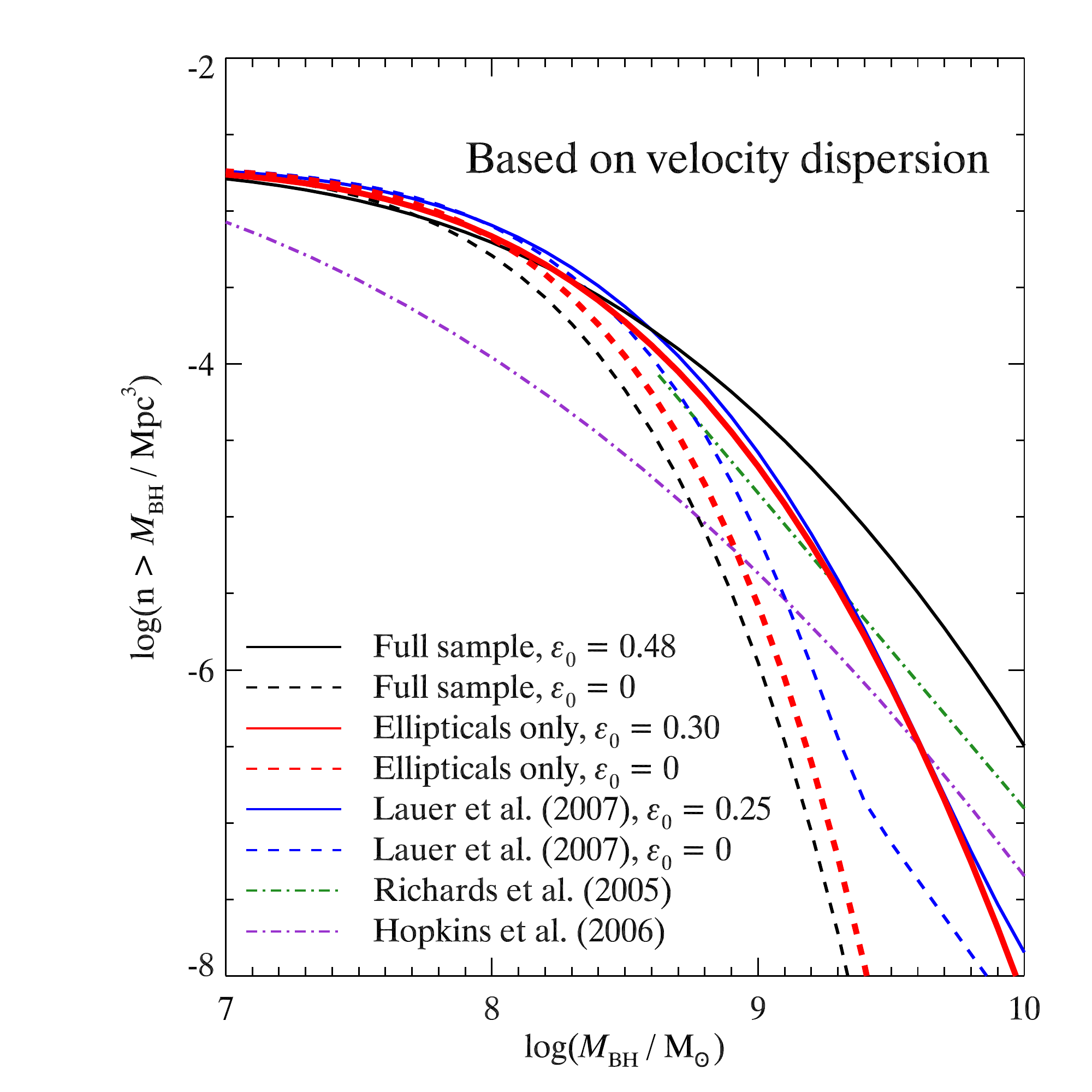}
\includegraphics[width=0.49\textwidth]{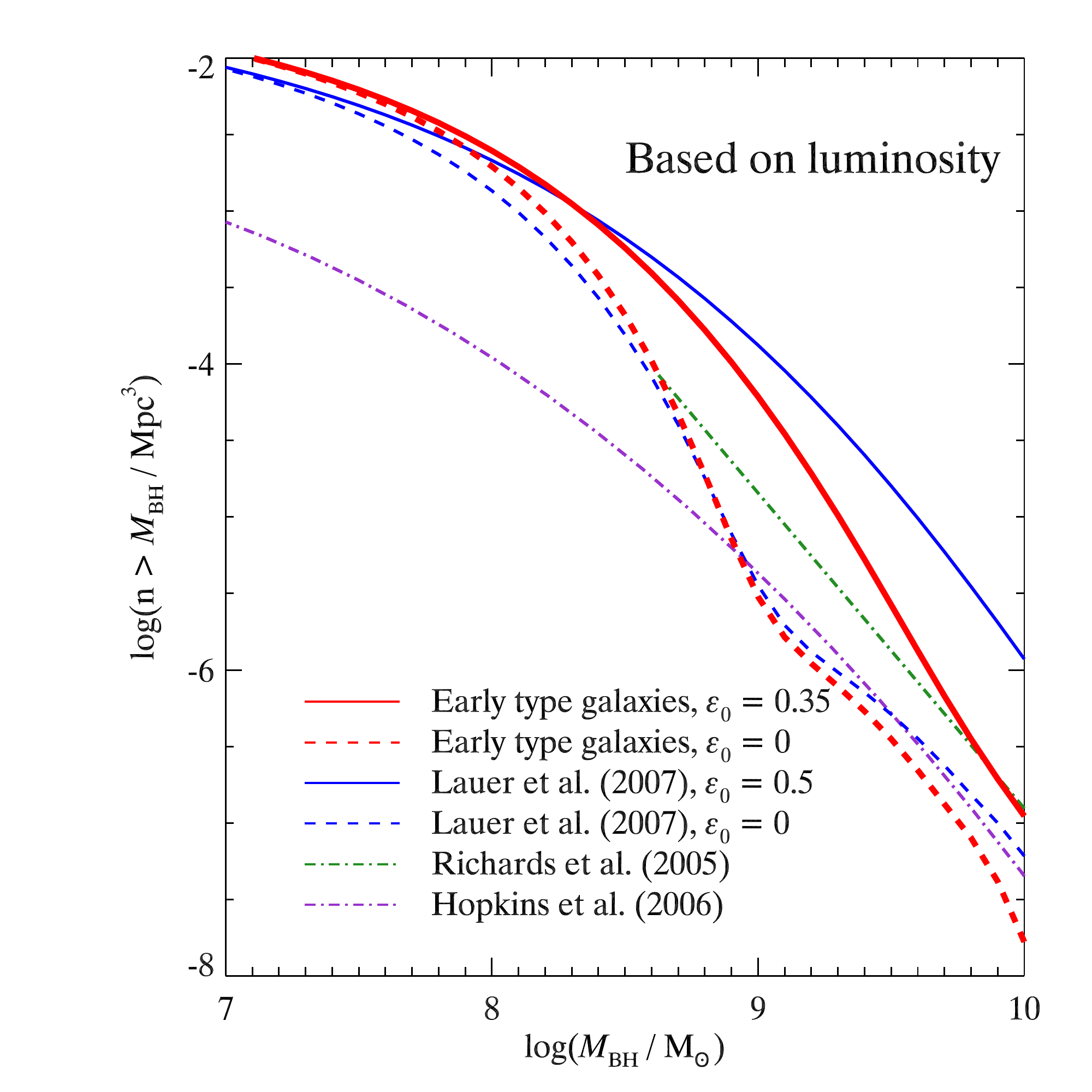}
\caption{BH space density derived from velocity dispersion
(\emph{left}) and luminosity (\emph{right}) of the host galaxy.  The
curves show logarithmic cumulative density for different assumptions.
The curves are the BH mass functions based on (\emph{left}) the SDSS
velocity dispersion function \protect{\citep{shethetal03}} and
(\emph{right}) a fit of SDSS data to the \protect{\citet{schecter76}}
luminosity function fit \protect{\citep{blantonetal03}} augmented by
the brightest cluster galaxies of \protect{\citet{pl95}}. Predictions
of BH mass based on velocity dispersion (\emph{left}) and luminosity
function (\emph{right}) are from fits to all galaxies in this paper
(\emph{orange}) and to elliptical galaxies in this paper (\emph{red}).
We also plot corresponding predictions from \protect{\citet{laueretal07}}
(\emph{blue}).  Dashed lines indicate mass functions assuming no
intrinsic scatter, and solid lines show mass functions assuming the
best-fit value of the intrinsic scatter.  For reference, both panels
show alternate density calculations.  The magenta dot-dashed line is
based on the \protect{\citet{2005MNRAS.360..839R}} luminosity function
from SDSS at a redshift of $z = 2.5$, assuming (1) a bolometric
correction of 6.5 \protect{\citep{elvisetal94,sos04}}, (2) that the
AGN are accreting at their Eddington limit, and (3) that the
AGN have a duty cycle of 0.03 \protect{\citep{steideletal02}}.  The
purple dot-dashed line is the quasar model due to
\protect{\citet{2006ApJS..163...50H}}.  The differences between the
solid and dashed curves of the same color show the large discrepancy
between considering and omitting intrinsic scatter in estimating the
density of the BHs.  The red, solid curves in both panels show the
best-fit results from this paper.}
\label{f:bhdf_sigma}
\label{f:bhdf_lum}
\bigskip
\end{figure*}

The intrinsic scatter in BH mass scaling relations affects estimates
of the space density of BHs based on velocity dispersion or luminosity
functions \citep{yt02,marconietal04,laueretal07}.  This is especially
true for the largest BHs.  The fundamental reason is that most of the
largest BHs will come from lower dispersion or lower luminosity hosts
with overmassive BHs.  We reproduce the cumulative density functions
based on our results in Figure~\ref{f:bhdf_sigma}.  In order to
calculate the density of BHs, a predictor for \mbh\ is needed, which
requires a \emph{regression} on $\sigma_e$ or $L_V$.  Our
maximum-likelihood method may be used to do a regression by simply
setting the uncertainties in $\sigma_e$ and $L_V$ to zero.  The
resulting fits differ very little from those presented above.

Figure~\ref{f:bhdf_sigma} shows that there is a marked difference in
the space density of BHs when assuming no scatter or our
best-fit scatter.  Regressions for both the entire sample and for
ellipticals-only are shown in the left panel of
Figure~\ref{f:bhdf_sigma}.  Because the largest BHs are found
in elliptical galaxies, the ellipticals-only curve with scatter best
reflects the velocity-dispersion-based calculation in this paper.
\citet{laueretal07} argued that the more fundamental scaling relation
for the largest BHs may be \ml.  If this is the case, then the curve
for early-type galaxies with scatter in the right-hand panel of
Figure~\ref{f:bhdf_sigma} is the appropriate function to use.  The
difference between the two curves at the high-mass end is
considerable, and it is a reflection of the inconsistency of the
\msigma\ and \ml\ relations in this regime.  Studies of the evolution
of \msigma\ are also biased due to the high scatter.

\section{Conclusions}
\label{concl}

In this work we have presented the latest results relating the mass of
a galaxy's central BH with the stellar velocity dispersion and
bulge luminosity of the galaxy.

(1) \emph{Most current \msigma\ fit.}  We compiled 49 BH mass
   measurements and fit a new \msigma\ relation using a
   maximum-likelihood method that includes an intrinsic scatter
   component.  We also include 18 upper limits to BH masses.  The best
   fit for the relation is
   \beq
   \log{\left(\frac{\mbh}{\msun}\right)} = (\msinterr) + (\msslopeerr) \log{\left(\frac{\sigma_e}{200~\kms}\right)},
   \label{e:concnoulmsigmafit}
   \eeq
   with an intrinsic rms Gaussian scatter of $\epsilon_0 = \msscaterr$.

(2) \emph{Characterization of the intrinsic scatter in \msigma.}
   Study of the distribution of BH masses in galaxies in a narrow
   range in velocity dispersion reveals that the intrinsic scatter in
   the \msigma\ relation is consistent with log-normal in BH mass and
   inconsistent with normal in BH mass.  Further, we find no evidence
   that adopting a log-quadratic, rather than a log-linear, relation
   between BH mass and velocity dispersion reduces the scatter.  

(3) \emph{Smaller scatter in the population of ellipticals.}  When we
   limit our sample to include elliptical galaxies, we find that the
   scatter decreases to $\epsilon_0 = 0.31 \pm 0.06$.  This is
   $\sim2\sigmaconf$ smaller than the intrinsic scatter in
   non-elliptical galaxies, $\epsilon_0 = 0.53 \pm 0.10$.  If real,
   this difference could either be because of larger systematic errors
   in BH mass determinations of late-type galaxies or because
   elliptical galaxies lie closer to the \msigma\ relation ridge line.

(4) \emph{Most current \ml\ fit.}  We also fit a relation
   between the mass of the central BH and the $V$-band bulge
   luminosity of the host.  The best-fit for the relation is
   \beq 
   \log{\left(\frac{M}{\msun}\right)} = (\mlinterr) + (\mlslopeerr) \log{\left(\frac{L_V}{10^{11} {\rm L}_{{\scriptscriptstyle \odot}, V}}\right)}
   \label{e:concnoulmlfit}
   \eeq
   with an intrinsic scatter of $\epsilon_0 = \mlscaterr$.

(5) \emph{Characterization of the scatter in \ml\ fit.}  We also
   find through study of the distribution of BH masses in
   galaxies in a narrow range of bulge luminosity that the scatter is
   adequately described by a log-normal intrinsic scatter in BH mass 
   and is inconsistent with a distribution normal in mass.

(6) \emph{Identification of bias in samples that use \rinfres\ as a
   selection criterion.}  Selecting a sample for fitting an \msigma\
   relation with a minimum value of \rinfres\ causes a bias that leads
   to an overestimate of the intercept, an overestimate estimate of the
   slope, and an incorrect estimate of the intrinsic scatter.  The bias
   arises because $R_{\mathrm{infl}} \approx G\mbh\sigma^{-2}$ and $\mbh
   \approx \sigma^4$ so that $R_{\mathrm{infl}} \approx \sigma^2$.  Thus,
   cuts in constant \rinfres\ systematically remove BH masses
   from the low-mass and low-velocity-dispersion portion of the
   \msigma\ plane.  For this reason, we exclude only those BH
   mass measurements that we believe to be unreliable.

(7) \emph{Implications for space density of BHs}.  Our findings that
   the intrinsic scatter in the \msigma\ for ellipticals and \ml\
   relations for early-type galaxies are $\epsilon_0 = 0.31 \pm 0.06$
   and $\epsilon_0 = \mlscaterr$, respectively, have an important
   influence on the determination of the space density of the most
   massive BHs.  We find that the density of BHs with $\mbh >
   3\times10^{9}~\msun$ is $8 \times 10^{-7}~\mathrm{Mpc^{-3}}$
   (based on the \msigma\ relation) and $\sim
   3\times10^{-6}~\mathrm{Mpc^{-3}}$ (based on the \ml\ relation).

\acknowledgements We thank Alessandra Beifiori for kindly providing
her data table and the anonymous referee for helpful comments,
especially for encouraging us to expand our discussion on issues
regarding resolution of sphere of influence.  KG thanks Marta Volonteri
and Monica Valluri for helpful discussions.  This work made use of the
NASA's Astrophysics Data System (ADS), and the NASA/IPAC Extragalactic
Database (NED), which is operated by the Jet Propulsion Laboratory,
California Institute of Technology, under contract with
NASA. Financial support was provided by NASA/{\it HST} grants GO-5999,
GO-6587, GO-6633, GO-7468, and GO-9107 from the Space Telescope
Science Institute, which is operated by AURA, Inc., under NASA
contract NAS 5-26555.

\appendix

\section{Fitting Method}
\label{analysis}

In this Appendix we describe our fitting method.  Previous methods
used for this problem include the \citet{ab96} extension of the 
least-squares estimator \citep{fm00} and a symmetric $\chi^2$ method
\citep{tremaineetal02}.  Neither of these methods, however, can
incorporate upper limits naturally. We use two general
methods for fitting the \msigma\ relation in this paper: (1) the
symmetric $\chi^2$ method of \citet{tremaineetal02}, and (2) a
generalized maximum-likelihood method developed for this work.  We
include the symmetric $\chi^2$ method for comparison to
\citet{tremaineetal02}, but the maximum likelihood method more
naturally includes an intrinsic scatter component and more naturally
incorporates upper limits.

We consider several functional forms for the measurement error
distribution and for the intrinsic scatter, both because it is far
from clear that the conventional log-normal distribution accurately
describes either distribution, and because distributions with fatter
tails than normal tend to handle outliers more robustly.  In
\S\S~\ref{xg}-\ref{ll} we describe our various assumptions about the
error distribution and scatter.

Ultimately, based on tests described in Appendix~\ref{difffits} below,
we find that the choice of error distribution does not significantly
change the fitted slope, intercept, or scatter; thus, we adopt
the results from the fit with Gaussian distributions as the most
straightforward.

We take the measurement error in logarithmic mass to be $0.5
[\log{(M_{\mathrm{high}})} - \log{(M_{\mathrm{low}})}]$, where
$M_\mathrm{high}$ and $M_\mathrm{low}$ are the published bounds to the
1$\sigma$ range in BH mass.  Even though some of the galaxies
in our sample have asymmetric errors (i.e., $\mbh - M_{\mathrm{low}}
\not= M_{\mathrm{high}} - \mbh$), we interpret the errors to be
symmetric in logarithmic mass.  This is a shortcoming of our methods,
which should be remedied in future work.  For the sake of consistency,
we interpret the errors to be the 68\% confidence intervals regardless
of the form of the error distribution, as described below.  In our
sample, IC~1459 and NGC~1399 each have two mass measurements that are
marginally inconsistent yet reliable.  To account for this we include
both measurements and weight each half as much.

Similar to the measurement errors in BH mass, we take the magnitude of
the intrinsic scatter to be the interval that contains 68\% of the
area under the curve, regardless of the shape of the intrinsic
scatter.  We further assume (1) that the shape of the intrinsic
scatter is symmetric above and below the ridge line, and (2) that the
magnitude of the intrinsic scatter is independent of velocity
dispersion.

\subsection{Maximum-Likelihood Method for Parameter Estimation}
\label{likelihood}
Let $P\left(\mu | s\right)$ be the probability that there is a BH of
logarithmic mass $\mu \equiv \log{\left(M/\msun\right)}$ given galaxy
properties $s$.  For our fits, $s$ is either the logarithm
of the velocity dispersion $s = \log{(\sigma/200~\kms)}$ or the
logarithm of the luminosity $s = \log{(L_V / 10^{11} {\rm
L}_{{\scriptscriptstyle \odot},V})}$. Thus, $P\left(\mu | s\right)$ is
either the \msigma\ or $M$--$L$ relation.  Given a set of observations
chosen on the basis of $s$ (or other properties not correlated with BH
mass), the likelihood of a set of observed points $\left\{ \mu_{i},
s_{i} \right\}$ is the product of the likelihood of each pair of
measurements $\ell_i = \ell\left(\mu_i, s_i\right)$:
\beq 
{\mathcal L} = \prod_{i} \ell_{i}.
\label{e:mainlike}
\eeq 
In the absence of measurement errors 
\beq 
\ell_{i} = P\left(\mu_{i} | s_{i}\right).
\label{e:elldef}
\eeq
For a given observation, the probability of measuring a mass between
$\mu_{\rm obs}$ and $\mu_{\rm obs}+d\mu_{\rm obs}$ in galaxy $i$,
given that the actual BH mass in this galaxy is $\mu$, is
$Q_i(\mu_{\rm obs}|\mu)d\mu_{\rm obs}$, normalized so that the
integral over all $\mu_{\rm obs}$ is unity.  Then
equation~(\ref{e:elldef}) becomes
\beq
\ell_{i} = \ell(\mu_i, s_i)= \int_{{\mathrm{all}\ } \mu} Q_{i}\left(\mu_i | \mu\right) P\left(\mu
| s_{i}\right) d\mu.
\label{e:genlike}
\eeq 
Note that equation~(\ref{e:genlike}) 
reduces to equation~(\ref{e:elldef}) if $Q_{i}(\mu_i | \mu) =
\delta(\mu-\mu_{i})$.  We also include observational upper limits to
BH masses in our fits.  In doing so, we must also allow for
the possibility that some galaxies have no BH; otherwise, a
single galaxy with no BH and a strong observational upper
limit could strongly bias our parameter fits.  To take this into
account, we modify equation~(\ref{e:genlike}) to be
\beq
\ell_i = P_\emptyset Q_i(\mu_i|-\infty) + (1-P_\emptyset) \int_{\mathrm{all}\ \mu} Q_{i}\left(\mu_i | \mu\right) P(\mu|s_i)d\mu,
\label{e:ulgenlike}
\eeq 
where $P_\emptyset$ is the probability of the galaxy having no BH.  In
all cases $P_\emptyset$ is consistent with zero, with typical
estimates of $P_\emptyset = 0.003 \pm 0.03$.

In the following sections we consider specific
forms for $Q_i(\mu_i | \mu)$ and $P(\mu | s_i)$.

\subsection{Gaussian Error Distribution with Gaussian Scatter}
\label{xg}
Let $G_{\epsilon}(x_1 - x_2) \equiv (1/\epsilon\sqrt{2\pi})\exp{[-(x_1 -
x_2)^2/2\epsilon^2]}$ be a normalized Gaussian with dispersion
$\epsilon$.  Then suppose
\beq
P\left(\mu | s_{i}\right) = G_{\epsilon_0}\left(\mu - f(s_i)\right),
\label{e:normscat}
\eeq
where $f$ is a ridge line through the joint distribution in $\mu$ and
$s$.  That is, $f$ is the \msigma\ or the \ml\ relation that we are
considering.  We refer to $\epsilon_0$ as the cosmic or intrinsic scatter
in the relation, and we assume it to be independent of all galaxy
properties.  Then
\beq
\ell_{i} = \int Q_i\left(\mu_i | \mu\right) G_{\epsilon_0}\left(\mu 
- f(s_i)\right) d\mu.
\eeq
If $Q_i$ is normally distributed about $\mu_i$ (i.e., the error in
measured logarithmic BH mass is Gaussian) so that
\beq
Q_i =  G_{\epsilon_i}\left(\mu_i - \mu\right),
\eeq
where $\epsilon_i$ is the measurement error, then
\beq
\ell_i = \int G_{\epsilon_i}\left(\mu_i - \mu\right) G_{\epsilon_0} 
\left(\mu - f(s_i)\right) d\mu.
\label{e:normallike1}
\eeq 
Equation~(\ref{e:normallike1}) is a convolution of two Gaussians ---
itself a Gaussian with variance equal to the sum of the two variances:
\beq
\ell_i = G_{\sqrt{{\epsilon_i}^2 + {\epsilon_0}^2}}\left(\mu_i - 
f(s_i)\right).
\label{e:normallike2}
\eeq
Equation~(\ref{e:normallike2}) is the justification for adding cosmic
scatter to the measurement error in \citet{tremaineetal02}.  We refer
to this model as Gaussian error distribution with Gaussian intrinsic
scatter (GG; see Table~\ref{t:abbrev} for a list of abbreviations for
models and samples).  This is the method we use for our final values
of the parameters in the \msigma\ and \ml\ relations.

\begin{deluxetable}{lrrl}
  \footnotesize
  \tablecaption{List of Abbreviations Used}
  \tablehead{
     \colhead{Abbrev.} &
     \colhead{$Q_i(\mu)$} & 
     \colhead{$P_i(\mu)$} & 
     \colhead{Short Description}
  }
  \startdata
GG & \protect{\ref{e:normallike2}} & \protect{\ref{e:normscat}} & Gaussian error distribution with Gaussian intrinsic scatter\\
CG & \protect{\ref{e:constproplike}} & \protect{\ref{e:normscat}} & Constant probability errors with Gaussian intrinsic scatter\\
DG & \protect{\ref{e:dblsidedexp}} & \protect{\ref{e:normscat}} & Double-sided exponential errors with Gaussian intrinsic  scatter\\
DD & \protect{\ref{e:dblsidedexp}} & \protect{\ref{e:dblsidedexpscat}} & Double-sided exponential errors\\ & & & with double-sided exponential intrinsic scatter\\
LG & \protect{\ref{e:lorentzian}} & \protect{\ref{e:normscat}} & Lorentzian errors with Gaussian intrinsic scatter\\
LL & \protect{\ref{e:lorentzian}} & \protect{\ref{e:lorentzianscat}} & Lorentzian errors with Lorentzian intrinsic scatter\\
SU  & & & Full sample with upper limits\\
S  & & & Sample without upper limits\\
RS  & & & Restricted sample, no upper limits\\
  \enddata
  \label{t:abbrev}

  \tablecomments{List of abbreviations.  Column~1 gives the
  abbreviation.  For the maximum-likelihood method abbreviations,
  Columns~2 and~3 give the equation numbers that describe
  the measurement error distribution $Q_i(\mu)$, and the intrinsic
  scatter distribution $P_i(\mu)$, respectively, while Column~4 gives 
  a short description.}
\end{deluxetable}

\subsection{Upper Limits}
\label{uls}
To include upper limits, we note that the probability that a galaxy
with properties $s$ has a logarithmic BH mass greater than
$\mu_u$ is
\beq
U(\mu_u | s) = (1 - P_\emptyset)\int_{\mu_u}^{\infty} P(\mu | s) d\mu.
\eeq
Suppose that an observation indicates the mass of the BH in a
given galaxy is less than $\mu_u$ at the $n\sigmaconf$ level.  This
means that the mass is greater than $\mu_u$ with probability
$\delta_n$ and less than $\mu_u$ with probability $1 - \delta_n$,
where $\delta_1 = 0.159$, $\delta_2 = 0.0228$, $\delta_3 = 0.00135$,
etc.  To include the upper limits in the maximum-likelihood method, we
must include the possibility that the observation has incorrectly
concluded that the mass is less than $\mu_u$, which will happen a
fraction $\delta_n$ of the time.  Then the likelihood of observing
an upper limit of mass $\mu_u$ at the $n\sigmaconf$ level in a galaxy
with property $s$ is
\beq
\ell_i = \delta_n U(\mu_u | s) + (1 - \delta_n) [1 - U(\mu_u | s)].
\eeq

\subsection{Constant Probability}
\label{cg}
We may also decide that a mass measurement means the mass is
restricted between $\mu_1$ and $\mu_2$, but that there is no preferred
mass in that range.  We may assume this, for example, because we
believe that the published error bars are a reasonable estimate of the
uncertainty in BH mass, but the error distribution is unknown.
In this case, $Q_i$ is a constant $C$, and
\beq
\ell_i = C \int_{\mu_1}^{\mu_2} G_{\epsilon_0} \left(\mu - f(s_i)\right) d\mu
\label{e:constproplike}
\eeq
with
\beq
C = \frac{1}{\mu_2 - \mu_1},
\eeq
where the normalization arises since the integral of $Q_i$ over all
$\mu_i$ should be unity.  In order to maintain the consistency of having
$\epsilon_i$ indicate the 68\% confidence limit, we take $\mu_{1,~2} =
\mu_i \mp \epsilon_i / 0.68$.  We refer to this method as CG.

\subsection{Double-Sided Exponential}
\label{dd}
We also use robust methods to estimate the parameters of $f(s_i)$ by
dropping the assumption of a Gaussian distribution of measurement
errors in $\mu$ and, optionally, the assumption of Gaussian intrinsic
scatter in $\mu$.  Robust methods are, in general, more tolerant of
outliers in distributions.  This may be especially appropriate for our
sample since the measurements come from several different groups using
several different measurement methods, and the systematic errors may
be larger than the intrinsic scatter.  There is no {\it a priori}
reason to assume that the intrinsic scatter is correctly described by
a Gaussian distribution.  We use two robust methods.  First, we use a
double-sided exponential, a standard robust method
\citep[e.g.,][]{numericalrecipes}:
\beq
Q_i(\mu_i | \mu) = \left(2 a_i\right)^{-1}\exp{\left(-\left| \mu - \mu_i \right|/a_i\right)},
\label{e:dblsidedexp}
\eeq
where we determine $a_i$ by assuming that the quoted 1$\sigmaconf$
measurement errors contain the 68\% confidence interval and thus
correspond to $-\ln{\left(0.32\right)}a_i$.  We use this form of measurement error
with either equation~(\ref{e:normscat}) (DG) or with a cosmic scatter
that is also described by a double-sided exponential (DD):
\beq
P_i(\mu) = \left(2 a_0\right)^{-1}\exp{\left(-\left| \mu_i - f(s_i) \right|/a_0\right)},
\label{e:dblsidedexpscat}
\eeq
where $a_0 = -\epsilon_0 / \ln{0.32}$.  

\subsection{Lorentzian}
\label{ll}
For our second robust method we use a Lorentzian to describe the
error distribution of the BH mass measurements.  Lorentzian
distributions have large tails and thus tend to be especially tolerant
of outliers:
\beq
Q_i(\mu_i | \mu) = \left(\frac{\Gamma_i}{\pi}\right)\frac{1}{\left(\mu - \mu_i \right)^2 + \Gamma^2_i},
\label{e:lorentzian}
\eeq
where we determine $\Gamma_i$ by assuming the 68\% confidence intervals correspond to $\tan{(0.68 \pi / 2)} \Gamma_i$.  We use this form
with either equation~(\ref{e:normscat}) (LG) or with a cosmic scatter
that is also described by a Lorentzian (LL):
\beq
P_i(\mu) = \left(\frac{\Gamma_0}{\pi}\right)\frac{1}{\left(\mu - f(s) \right)^2 + \Gamma^2_0},
\label{e:lorentzianscat}
\eeq
where $\Gamma_0 = \epsilon_0 / \tan{(0.68 \pi / 2)}$.  

\begin{figure*}
\centering
\includegraphics[width=0.4\textwidth]{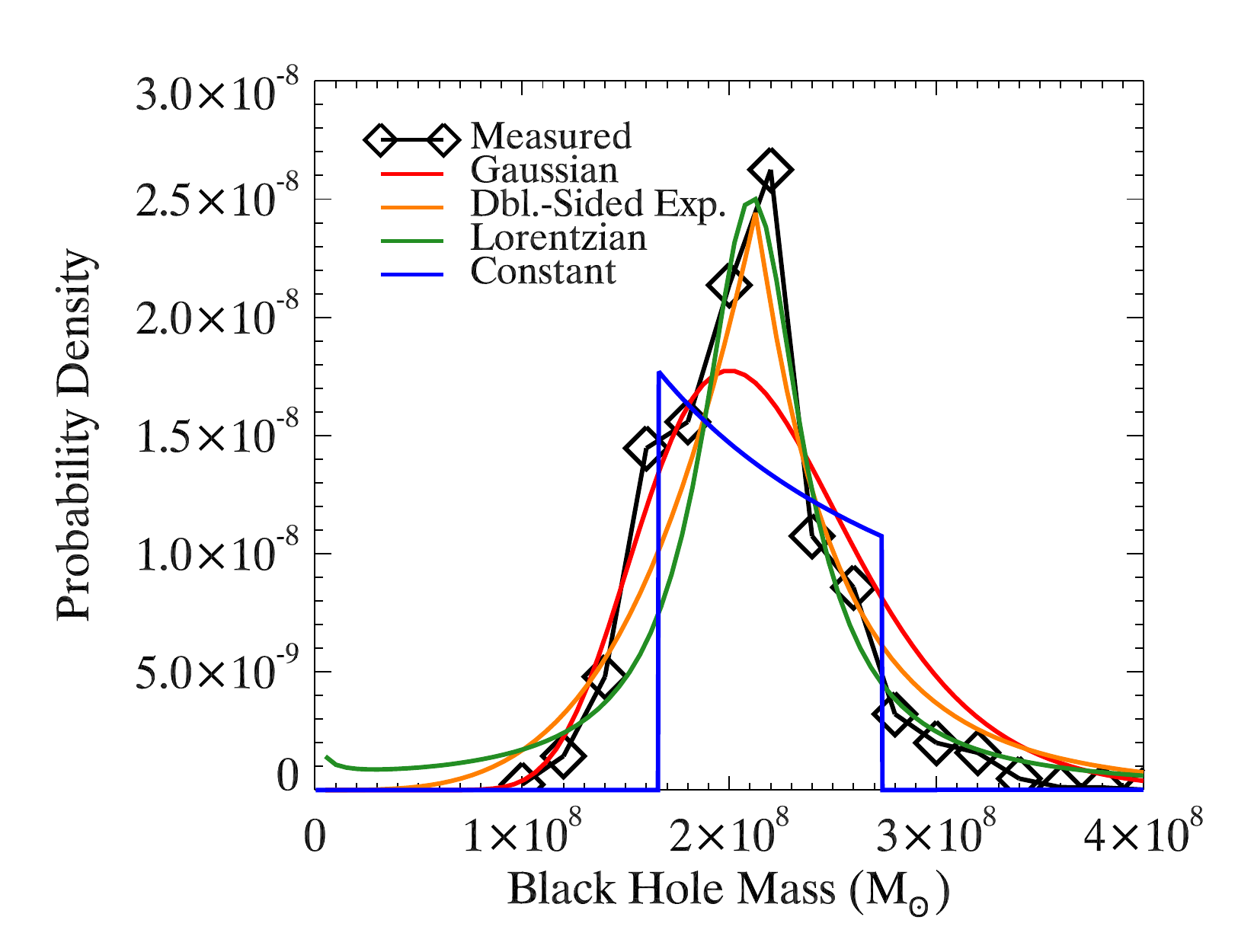}
\includegraphics[width=0.4\textwidth]{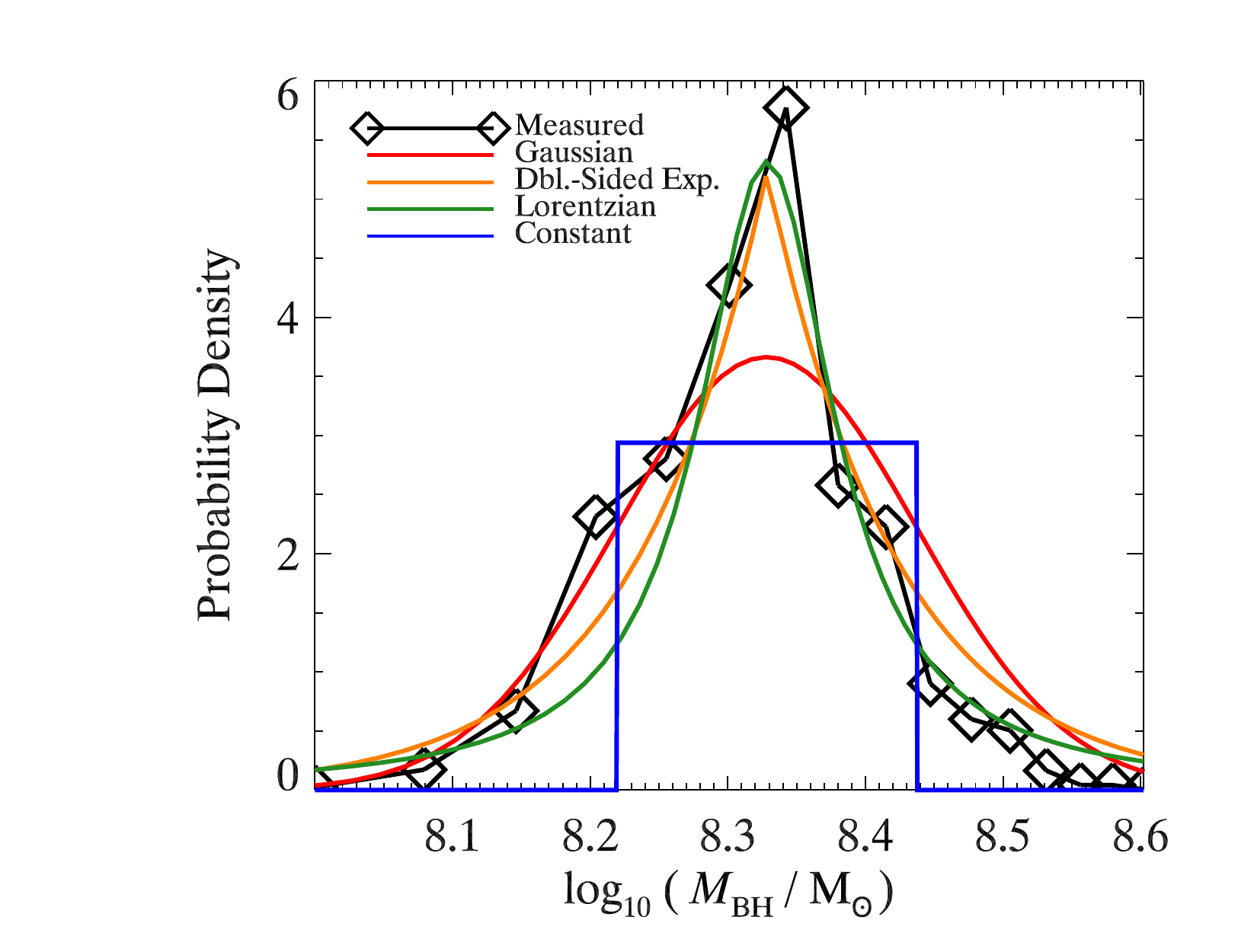}
\caption{Comparison of assumed error distributions with measured
probability distribution of NGC~4026 as a function of $\mbh$
(\emph{left}) and $\log{(\mbh/\msun)}$ (\emph{right}).  The measured
probability distribution comes from the $chi^2$ values from
\protect{\citet{Gultekin_etal_2008}}.  The
distributions are normalized so that the integral over
$\log{(\mbh/\msun)}$ is unity.  This figure illustrates the
differences among the assumed error distributions.}
\label{f:probprob}
\end{figure*}

The error distributions described above are compared to the measured
probability found for NGC~4026 by \citet{Gultekin_etal_2008} in
Figure~\ref{f:probprob}.

In practice, for all methods we minimize $-\ln{{\mathcal L}}$ by the
downhill simplex method \citep[e.g.,][]{numericalrecipes}.  To
calculate uncertainties in parameters, we change each parameter and
refit (allowing the other parameters to vary to maximize $\ln{{\mathcal L}}$)
until $\ln{{\mathcal L}}$ decreases by $0.5$.  We also run a Monte
Carlo bootstrap described in Appendix~\ref{bootstrap}.

\subsection{Errors in the Independent Variable}
\label{sigmaerrors}
Errors in the independent variables (velocity dispersion and bulge
luminosity) are incorporated by Monte Carlo sampling of the independent
variables.  For example, for determining the \msigma\ fit, we select
values of $\sigma$ according to the measured values and uncertainties.
For each fit $10^3$ realizations are done, and the small parameter
uncertainties from this are added in quadrature.

\subsection{Maximum-Likelihood Method:  Model Comparison}
\label{modelcomp}
To compare two models we calculate the odds ratio
\beq
{\mathcal R_{ab}} = \frac{\int{\mathcal L_a}(a_1, a_2, ..., a_m)
P_a(a_1, a_2, ..., a_m)da_1 da_2 ... da_m}
{\int{\mathcal L_b}(b_1, b_2, ..., b_n)
P_b(b_1, b_2, ..., b_n)db_1 db_2 ... db_n},
\label{e:oddsratio}
\eeq
where ${\mathcal L_a}$ is the likelihood of the data given model $a$
with $m$ parameters $a_i$, which have a prior probability distribution
$P_a$, and similarly for model $b$.  The models need not have the same
number of parameters.  We assume that the prior probability
distributions are uniform within the ranges $[4, 12]$ for intercepts
[$\mu(\sigma_e=200\kms)$ or $\mu(L_V = 10^{11} {\rm L}_{{\scriptscriptstyle 
\odot},V})$], $[-10,10]$ for slopes, and $[0, 2]$ for cosmic scatter, 
though any reasonable set of ranges
produces the same qualitative results.  The integrals are calculated
by the Vegas Monte Carlo method \citep{lepage78}.

\section{Results from Different Error Distributions, Scatter Forms, and Fit Samples}
\label{difffits}
\begin{figure*}
\includegraphics[width=0.70\textwidth,angle=90]{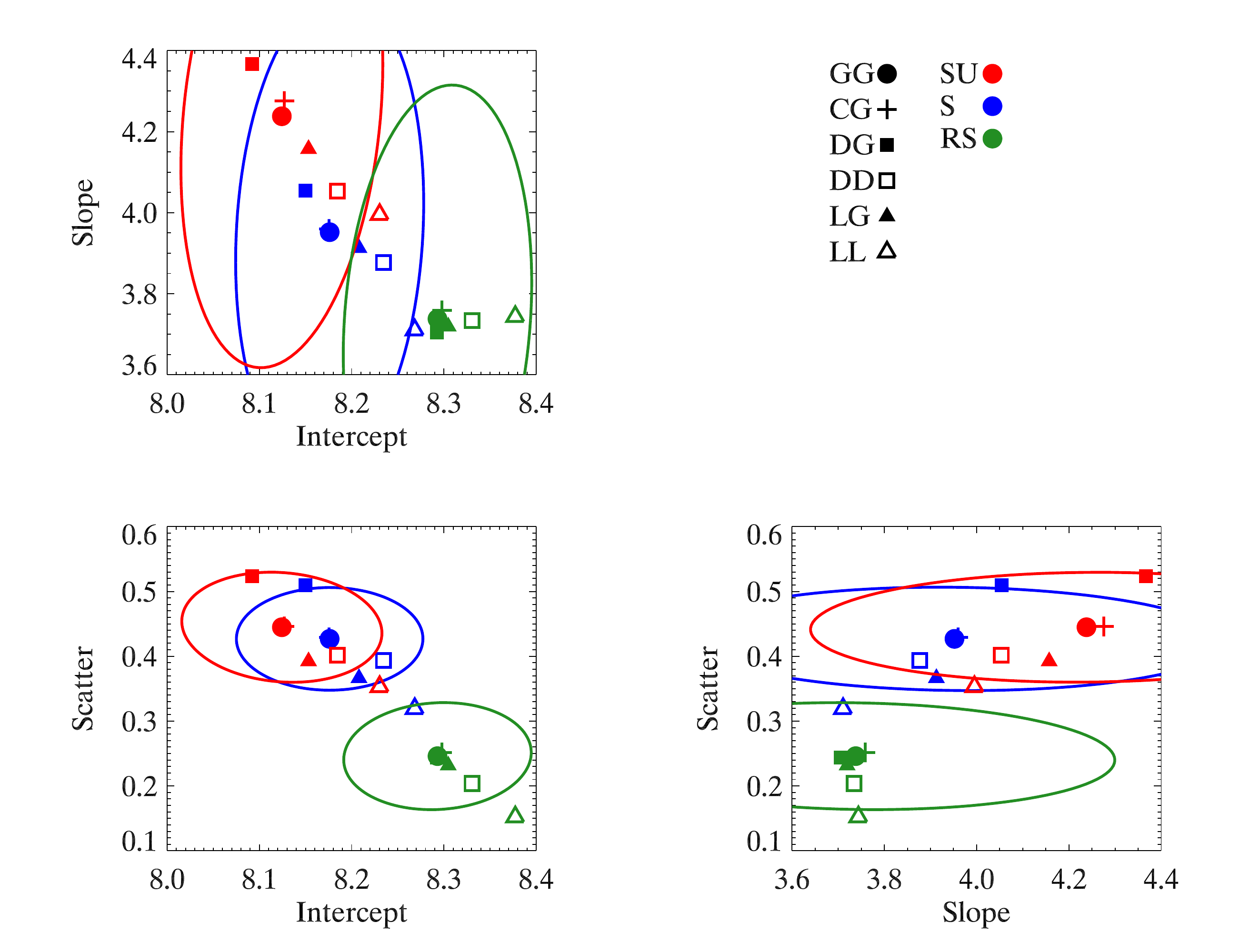}
\caption{Results of fits to the \msigma\ relation showing best-fit
intercepts ($\alpha$), best-fit slopes ($\beta$), and best-fit
intrinsic scatters ($\epsilon_0$) against each other. The adopted
fitting methods are plotted with the symbol from the legend, which
uses abbreviations from Table~\protect{\ref{t:abbrev}}.  For clarity,
we show representative error ellipses for method GG for each of the
samples.  The error ellipses are the 68\% confidence limits in the
joint distribution of the two parameters plotted.  The sample used is
indicated by the color of the parameter uncertainties: \emph{blue}
(sample without limits [S]), \emph{red} (full sample including upper
limits [SU]), and \emph{green} (restricted sample [RS]).  The
parameter values for the various adopted methods do not significantly
vary, as can be seen from the fact that most points from the same
sample (same color) fall inside the uncertainty box implied by the
error bars.  The values from different samples, however, do vary.  We
discuss the reasons for the differences in sample RS (which has a
cutoff in \rinfres) in \S~\protect{\ref{influence}} and for the
differences when including upper limits in
\S~\protect{\ref{upperlimits}}.}
\label{f:allfits}
\end{figure*}

The results from \msigma\ fits are summarized in Table~\ref{t:fits}
and plotted in Figure~\ref{f:allfits}.  

First, we analyze our dataset excluding upper limits (S) with the
method of \citet{tremaineetal02}, which consists of two distinct
variants. (1) Measurement errors and intrinsic scatter are combined
into a single number, the rms deviation of $\log(\mbh/\msun)$ from the
ridge line (denoted $\epsilon_0$, even though this is normally
reserved for the intrinsic scatter), which is assumed to be the same
for all galaxies and is determined by requiring that the reduced
$\chi^2$ of the fit be unity.  (2) The stated measurement errors and
the intrinsic scatter $\epsilon_0$ are added in quadrature, as in
Equation~(\ref{e:normallike2}).  When using the first method, we find
$\alpha = 8.19 \pm 0.06$ and $\beta = 3.99 \pm 0.37$ for $\epsilon_0 =
0.43$.  When using the second method, we find $\alpha = 8.19 \pm 0.06$
and $\beta = 4.06 \pm 0.37$ for $\epsilon_0 = 0.40$.  The slopes we
obtain here are consistent with \citet{tremaineetal02}, but the
scatters are larger (see \S~\ref{flatness}).

\begin{deluxetable}{llrrrr}
  \footnotesize
  \tablecaption{Parameter Estimates for $M$--$\sigma$ Relation}
  \tablehead{
     \colhead{Method} & 
     \colhead{Sample} & 
     \colhead{$\alpha$} & 
     \colhead{$\beta$} & 
     \colhead{$\epsilon_0$} &
     \colhead{$P_\emptyset$} 
  }
  \startdata

T02\phantom{eq} 
       & S & $8.19 \pm 0.063$ & $4.02 \pm 0.369$ & $0.41 \phantom{\pm 0.000}$ & \dots\\
T02eq & S & $8.19 \pm 0.063$ & $3.99 \pm 0.369$ & $0.43 \phantom{\pm 0.000}$ & \dots\\
T02ind & S & $8.19 \pm 0.064$ & $4.06 \pm 0.370$ & $0.40 \phantom{\pm 0.000}$ & \dots\\
\\
GG & SU & $8.12 \pm 0.080$ & $4.24 \pm 0.410$ & $0.44 \pm 0.059$ & $0.0004 \pm 0.018$\\
CG & SU & $8.13 \pm 0.085$ & $4.28 \pm 0.437$ & $0.45 \pm 0.063$ & $0.0006 \pm 0.016$\\
DG & SU & $8.09 \pm 0.088$ & $4.37 \pm 0.603$ & $0.52 \pm 0.064$ & $0.0002 \pm 0.015$\\
DD & SU & $8.18 \pm 0.075$ & $4.05 \pm 0.382$ & $0.40 \pm 0.069$ & $0.0001 \pm 0.015$\\
LG & SU & $8.15 \pm 0.079$ & $4.16 \pm 0.481$ & $0.39 \pm 0.068$ & $0.0003 \pm 0.021$\\
LL & SU & $8.23 \pm 0.077$ & $4.00 \pm 0.496$ & $0.35 \pm 0.105$ & $0.0002 \pm 0.020$\\
\\
GG &  S & $8.18 \pm 0.079$ & $3.95 \pm 0.423$ & $0.43 \pm 0.058$ & \dots\\
CG &  S & $8.18 \pm 0.079$ & $3.96 \pm 0.426$ & $0.43 \pm 0.058$ & \dots\\
DG &  S & $8.15 \pm 0.093$ & $4.05 \pm 0.507$ & $0.51 \pm 0.067$ & \dots\\
DD &  S & $8.23 \pm 0.073$ & $3.88 \pm 0.760$ & $0.39 \pm 0.082$ & \dots\\
LG &  S & $8.21 \pm 0.073$ & $3.91 \pm 0.676$ & $0.37 \pm 0.068$ & \dots\\
LL &  S & $8.27 \pm 0.072$ & $3.71 \pm 0.402$ & $0.32 \pm 0.094$ & \dots\\
\\
GG & RS & $8.29 \pm 0.078$ & $3.74 \pm 0.404$ & $0.25 \pm 0.059$ & \dots\\
CG & RS & $8.30 \pm 0.069$ & $3.76 \pm 0.369$ & $0.25 \pm 0.050$ & \dots\\
DG & RS & $8.29 \pm 0.073$ & $3.71 \pm 0.405$ & $0.24 \pm 0.059$ & \dots\\
DD & RS & $8.33 \pm 0.067$ & $3.73 \pm 0.342$ & $0.20 \pm 0.060$ & \dots\\
LG & RS & $8.30 \pm 0.083$ & $3.72 \pm 0.417$ & $0.23 \pm 0.070$ & \dots\\
LL & RS & $8.38 \pm 0.087$ & $3.74 \pm 0.708$ & $0.15 \pm 0.108$ & \dots

  \enddata
  \label{t:fits}
  \tablecomments{Results from fits.  The first three lines are the
  methods of \protect{\citet{tremaineetal02}}.  The first line (T02)
  is the average of two symmetric $\chi^2$ variants in which intrinsic
  scatter is increased until $\chi^2$ per degree of freedom is unity,
  with each mass measurement either given equal weight (T02eq) or with
  individual errors (T02ind).  Method and sample abbreviations are
  described in Table~\ref{t:abbrev}.}
  \end{deluxetable}

Then we use the maximum-likelihood method with the various assumed
forms for measurement errors and the intrinsic scatter.  Using the
different error distributions for the sample that includes upper
limits (SU), we find a range of intercepts from 8.09 to 8.23, which
is consistent compared to the largest parameter uncertainty estimate 
of 0.09.  Similarly, we find slopes in the range 4.00 to 4.37, a
smaller range than the $\sim 0.4$ uncertainty.  The estimates of the
intrinsic scatter range from 0.35 to 0.52 with uncertainties of
0.06--0.11.

For the sample that does not include upper limits (S), we find a
similar self-consistency among the different error and scatter
distributions.  The intercept is larger when excluding upper limits.
The slope is shallower in the sample without limits, though not at a
significant level, and the scatter is slightly (but not significantly)
smaller.  The consistency of the results among the different error
distributions for a given sample indicates that the choice of
distribution is not driving the values we obtain for parameters.  The
results for the restricted sample (RS) are similarly self-consistent,
though compared to the full sample SU they have a larger intercept,
shallower slope, and smaller intrinsic scatter, for reasons
discussed in \S~\ref{influence}.  Thus, assuming a different error
distribution does not have a strong effect on the results, but
selecting a different sample does.

\begin{figure*}
\includegraphics[width=0.45\textwidth]{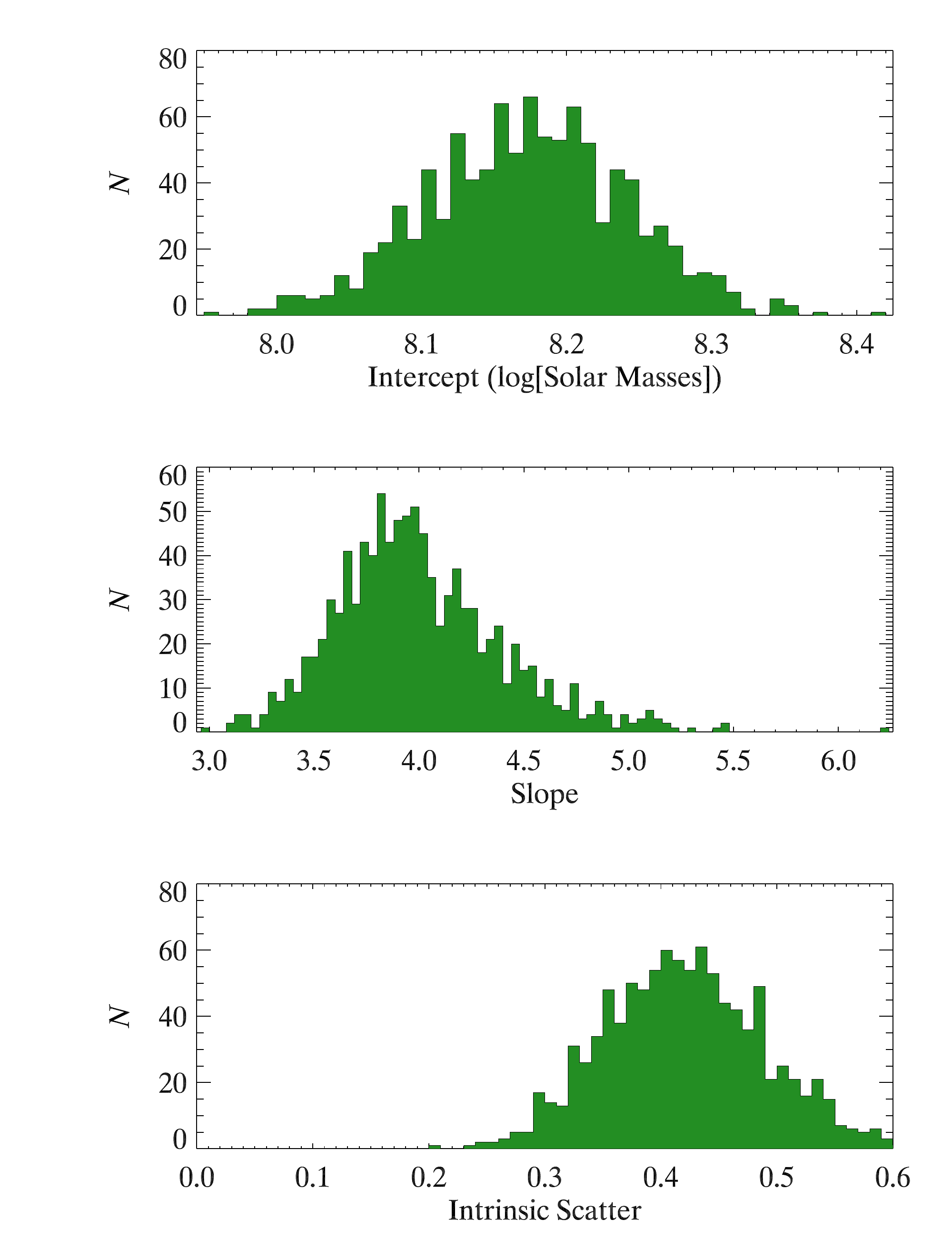}
\includegraphics[width=0.45\textwidth]{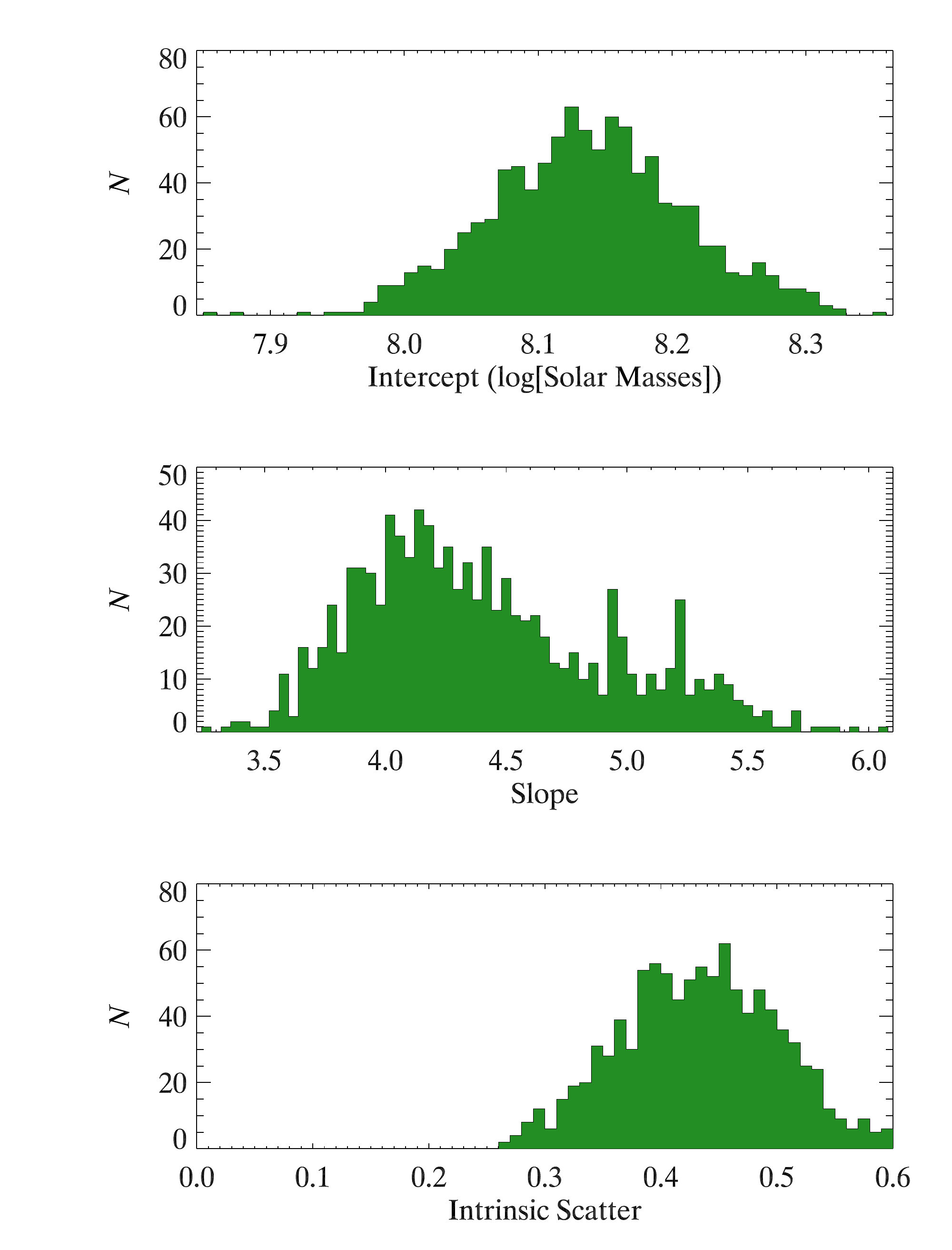}
\caption{Histograms of parameter estimates from the bootstrap samples
assuming Gaussian measurement errors and Gaussian intrinsic scatter.
Left panels are for the full sample, SU; right panels do not
include upper limits (sample S).  The distributions are unimodal, and
the medians and 68\% intervals are in good agreement with our best-fit
values and uncertainties.}
\label{f:bootstrapchinoul}
\label{f:bootstrapchiul}
\end{figure*}

\begin{figure*}
\includegraphics[width=0.45\textwidth]{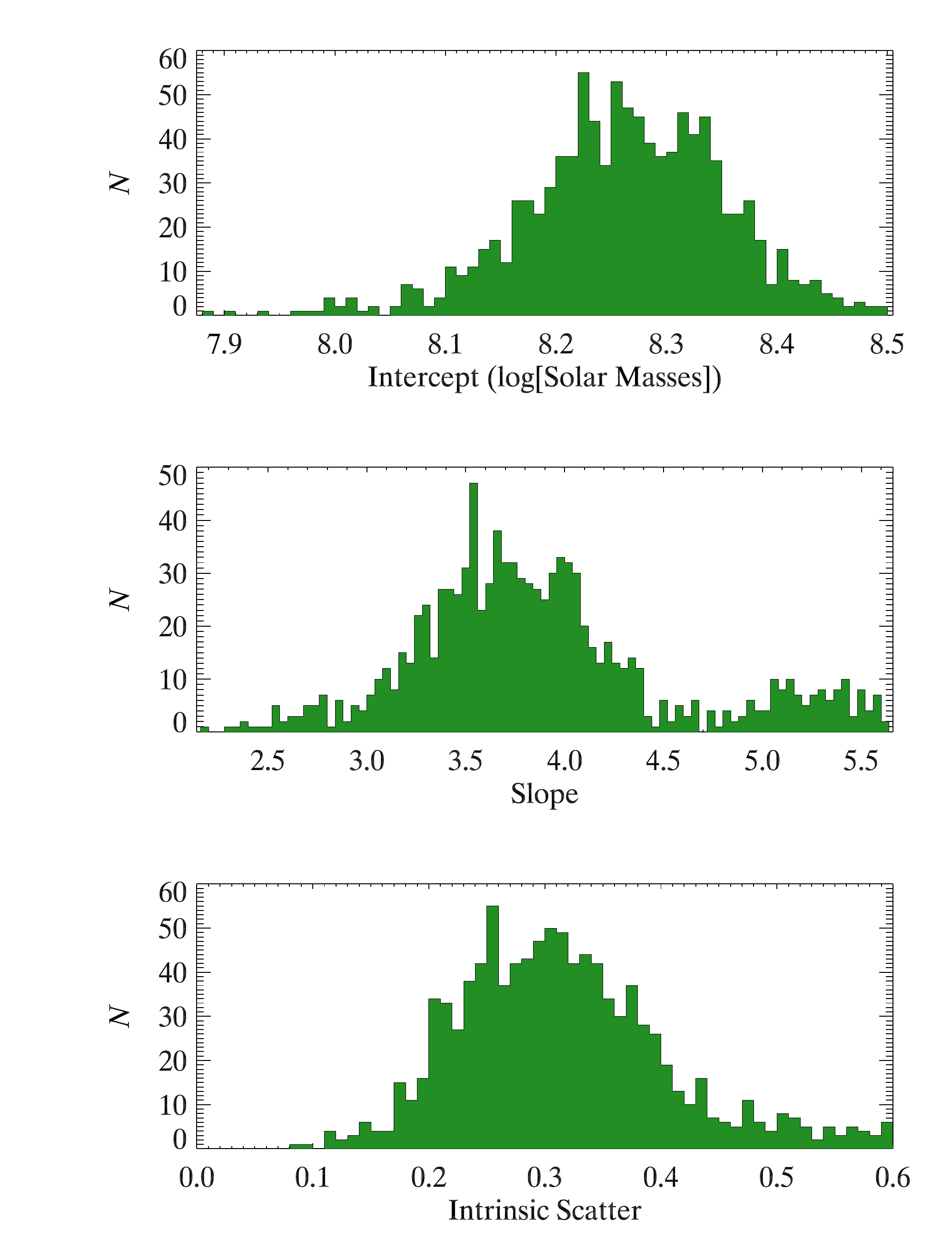}
\includegraphics[width=0.45\textwidth]{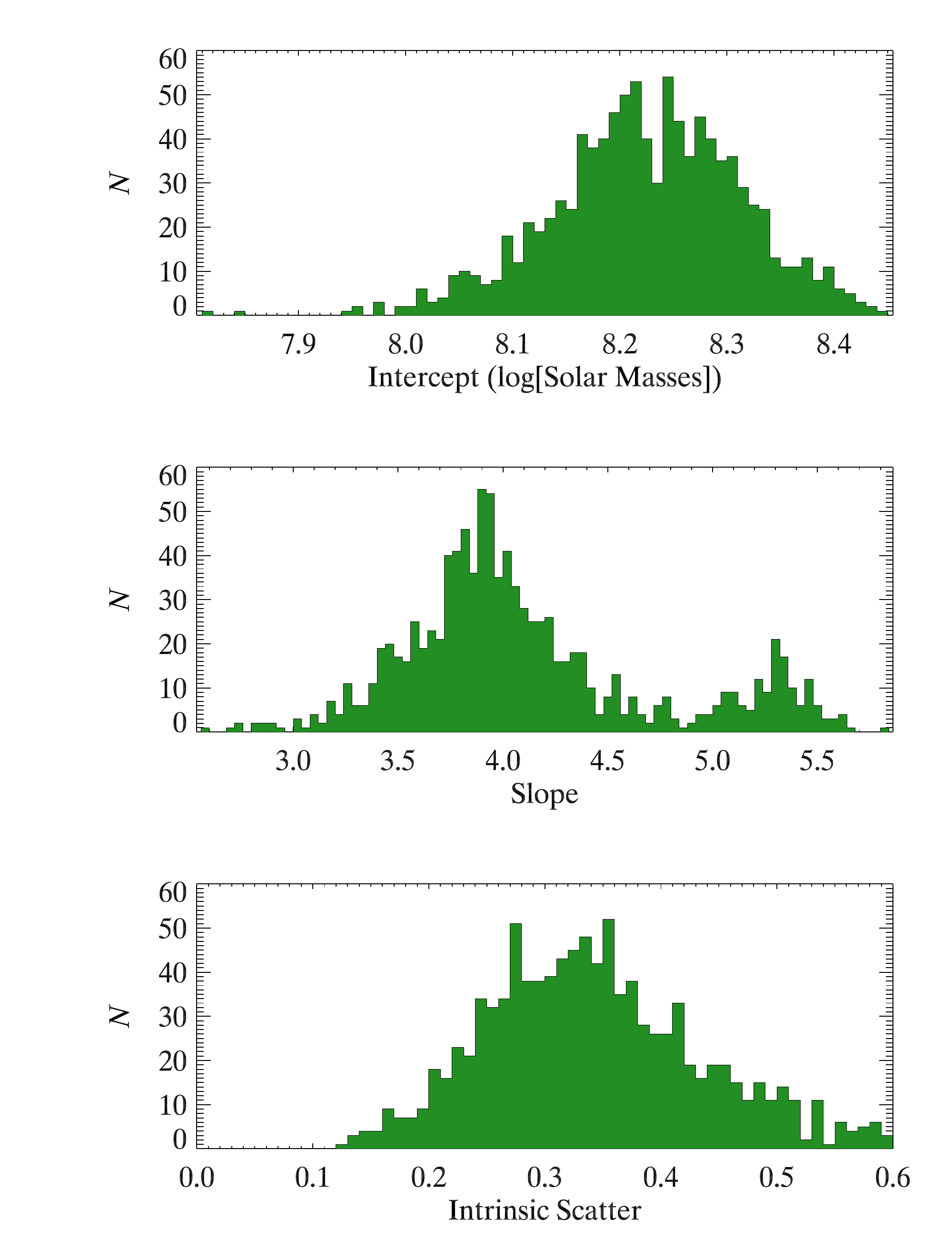}
\caption{Histograms of parameter estimates from the bootstrap samples
assuming Lorentzian measurement errors and Lorentzian intrinsic scatter.
Left panels are for the full sample, SU; right panels do not
include upper limits (sample S).  The distributions are unimodal, and
the medians and 68\% intervals are in good agreement with our best-fit
values and uncertainties.}
\label{f:bootstraplorentznoul}
\label{f:bootstraplorentzul}
\end{figure*}

\label{bootstrap}
As a check on the uncertainty estimation, we calculate a bootstrap
Monte Carlo of our sample.  For each combination of measurement error
and intrinsic scatter distributions, we randomly extract data points
from the original sample with replacement for $10^3$ realizations and
fit each sample.  The bootstrap method provides a powerful way of
examining whether any outlying measurements are driving the fit as
well as a consistency check on uncertainty estimations.  We show the
distribution of best-fit parameters for two combinations of
distributions: Gaussian measurement error with Gaussian intrinsic
scatter (Fig.~\ref{f:bootstrapchinoul}) and the Lorentzian measurement
error with Lorentzian intrinsic scatter
(Fig.~\ref{f:bootstraplorentznoul}), each with and without upper
limits.  Table~\ref{t:bootstrap} shows that in all cases the median
fit parameters with 68\% intervals of the distributions are very close
to the values we found above.  The distributions appear well
approximated by a Gaussian.

\begin{deluxetable}{llrrr}
  \footnotesize

  \tablecaption{Bootstrap Monte Carlo}
  \tablehead{
     \colhead{Method} &
     \colhead{Sample} &
     \colhead{$\alpha$} &
     \colhead{$\beta$} &
     \colhead{$\epsilon_0$}
  }
  \startdata
GG & SU & $8.14 \pm .07$ & $4.31 \pm .52$ & $0.44 \pm .07$\\
LL & SU & $8.23 \pm .08$ & $3.96 \pm .48$ & $0.33 \pm .09$
\\
GG & S & $8.17 \pm .07$ & $3.95 \pm .36$ & $0.42 \pm .07$\\
LL & S & $8.27 \pm .08$ & $3.78 \pm .52$ & $0.31 \pm .08$\\
  \enddata
  \label{t:bootstrap}

  \tablecomments{Results from bootstrap fits expressed as median value.
  The uncertainties quoted encompass 68\% of the values.  Methods are
  as given in Table~\protect{\ref{t:fits}}.  The values are in very
  close agreement with the corresponding values given in
  Table~\protect{\ref{t:fits}}.  This indicates that the uncertainties
  derived from our fitting method are accurate and are not strongly
  influenced by outliers.}
\end{deluxetable}


\phantom{text}


Given the consistency among the different assumptions and the
consistency between the different uncertainty estimation methods, we
conclude that the results are insensitive to the choice of
distribution of measurement error and intrinsic scatter (so long as
the ``width'' $\epsilon_0$ is consistently defined as the interval
containing 68\% of the distribution).  The best-fit parameters of the
\msigma\ and \ml\ relations, however, do depend on the sample choice.
Thus, we adopt the simplest combination of measurement error and
intrinsic scatter distributions, GG, for the estimates of the fit
parameters and their uncertainties that we quote in the abstract.

\eject
\bibliographystyle{apjads}
\bibliography{gultekin}

\newcommand{\noopsort}[1]{}
\begin{thebibliography}{133}
\expandafter\ifx\csname natexlab\endcsname\relax\def\natexlab#1{#1}\fi
\expandafter\ifx\csname href\endcsname\relax
  \def\href#1#2{}\fi
\expandafter\ifx\csname urllinklabel\endcsname\relax
  \def\urllinklabel{[LINK]}\fi
\expandafter\ifx\csname adsurllinklabel\endcsname\relax
  \def\adsurllinklabel{[ADS]}\fi

\bibitem[{{Adams} {et~al.}(2003){Adams}, {Graff}, {Mbonye}, \&
  {Richstone}}]{adamsetal03}
  \href{http://adsabs.harvard.edu/abs/2003ApJ...591..125A}{{Adams}, F.~C.,
  {Graff}, D.~S., {Mbonye}, M., \& {Richstone}, D.~O. 2003, \apj, 591, 125%
}%

\bibitem[{{Adams} {et~al.}(2001){Adams}, {Graff}, \& {Richstone}}]{agr01}
  \href{http://adsabs.harvard.edu/cgi-bin/nph-bib_query?bibcode=2001ApJ...551L%
..31A&db_key=AST}{{Adams}, F.~C., {Graff}, D.~S., \& {Richstone}, D.~O. 2001,
  \apjl, 551, L31%
}%

\bibitem[{{Akritas} \& {Bershady}(1996)}]{ab96}
  \href{http://adsabs.harvard.edu/abs/1996ApJ...470..706A}{{Akritas}, M.~G., \&
  {Bershady}, M.~A. 1996, \apj, 470, 706%
}%

\bibitem[{{Atkinson} {et~al.}(2005){Atkinson}, {Collett}, {Marconi}, {Axon},
  {Alonso-Herrero}, {Batcheldor}, {Binney}, {Capetti}, {Carollo}, {Dressel},
  {Ford}, {Gerssen}, {Hughes}, {Macchetto}, {Maciejewski}, {Merrifield},
  {Scarlata}, {Sparks}, {Stiavelli}, {Tsvetanov}, \& {van der
  Marel}}]{2005MNRAS.359..504A}
  \href{http://adsabs.harvard.edu/abs/2005MNRAS.359..504A}{{Atkinson}, J.~W.,
  {et~al.} 2005, \mnras, 359, 504%
}%

\bibitem[{{Barth} {et~al.}(2001){Barth}, {Sarzi}, {Rix}, {Ho}, {Filippenko}, \&
  {Sargent}}]{2001ApJ...555..685B}
  \href{http://adsabs.harvard.edu/abs/2001ApJ...555..685B}{{Barth}, A.~J.,
  {Sarzi}, M., {Rix}, H.-W., {Ho}, L.~C., {Filippenko}, A.~V., \& {Sargent},
  W.~L.~W. 2001, \apj, 555, 685%
}%

\bibitem[{{Beifiori} {et~al.}(2008){Beifiori}, {Sarzi}, {Corsini}, {Dalla
  Bont{\`a}}, {Pizzella}, {Coccato}, \& {Bertola}}]{beifiorietal08}
  \href{http://adsabs.harvard.edu/abs/2008arXiv0809.5103B}{{Beifiori}, A.,
  {Sarzi}, M., {Corsini}, E.~M., {Dalla Bont{\`a}}, E., {Pizzella}, A.,
  {Coccato}, L., \& {Bertola}, F. 2008,
  }\href{http://www.arXiv.org/abs/0809.5103}{{preprint (0809.5103)}%
}%

\bibitem[{{Bender} {et~al.}(2005){Bender}, {Kormendy}, {Bower}, {Green},
  {Thomas}, {Danks}, {Gull}, {Hutchings}, {Joseph}, {Kaiser}, {Lauer},
  {Nelson}, {Richstone}, {Weistrop}, \& {Woodgate}}]{benderetal05}
  \href{http://adsabs.harvard.edu/abs/2005ApJ...631..280B}{{Bender}, R.,
  {et~al.} 2005, \apj, 631, 280%
}%

\bibitem[{{Bentz} {et~al.}(2006){Bentz}, {Denney}, {Cackett}, {Dietrich},
  {Fogel}, {Ghosh}, {Horne}, {Kuehn}, {Minezaki}, {Onken}, {Peterson}, {Pogge},
  {Pronik}, {Richstone}, {Sergeev}, {Vestergaard}, {Walker}, \&
  {Yoshii}}]{bentzetal06b}
  \href{http://adsabs.harvard.edu/abs/2006ApJ...651..775B}{{Bentz}, M.~C.,
  {et~al.} 2006, \apj, 651, 775%
}%

\bibitem[{{Bernardi} {et~al.}(2006){Bernardi}, {Sheth}, {Nichol}, {Miller},
  {Schlegel}, {Frieman}, {Schneider}, {Subbarao}, {York}, \&
  {Brinkmann}}]{bernardietal06}
  \href{http://adsabs.harvard.edu/abs/2006AJ....131.2018B}{{Bernardi}, M.,
  {et~al.} 2006, \aj, 131, 2018%
}%

\bibitem[{{Blanton} {et~al.}(2003){Blanton}, {Hogg}, {Bahcall}, {Brinkmann},
  {Britton}, {Connolly}, {Csabai}, {Fukugita}, {Loveday}, {Meiksin}, {Munn},
  {Nichol}, {Okamura}, {Quinn}, {Schneider}, {Shimasaku}, {Strauss}, {Tegmark},
  {Vogeley}, \& {Weinberg}}]{blantonetal03}
  \href{http://adsabs.harvard.edu/abs/2003ApJ...592..819B}{{Blanton}, M.~R.,
  {et~al.} 2003, \apj, 592, 819%
}%

\bibitem[{{Bower} {et~al.}(1998){Bower}, {Green}, {Danks}, {Gull}, {Heap},
  {Hutchings}, {Joseph}, {Kaiser}, {Kimble}, {Kraemer}, {Weistrop}, {Woodgate},
  {Lindler}, {Hill}, {Malumuth}, {Baum}, {Sarajedini}, {Heckman}, {Wilson}, \&
  {Richstone}}]{1998ApJ...492L.111B}
  \href{http://adsabs.harvard.edu/abs/1998ApJ...492L.111B}{{Bower}, G.~A.,
  {et~al.} 1998, \apjl, 492, L111%
}%

\bibitem[{{Bower} {et~al.}(2001){Bower}, {Green}, {Bender}, {Gebhardt},
  {Lauer}, {Magorrian}, {Richstone}, {Danks}, {Gull}, {Hutchings}, {Joseph},
  {Kaiser}, {Weistrop}, {Woodgate}, {Nelson}, \&
  {Malumuth}}]{2001ApJ...550...75B}
  \href{http://adsabs.harvard.edu/abs/2001ApJ...550...75B}{---------. 2001,
  \apj, 550, 75%
}%

\bibitem[{{Burkert} \& {Silk}(2001)}]{bs01}
  \href{http://adsabs.harvard.edu/abs/2001ApJ...554L.151B}{{Burkert}, A., \&
  {Silk}, J. 2001, \apjl, 554, L151%
}%

\bibitem[{{Burstein}(1979)}]{1979ApJ...234..435B}
  \href{http://adsabs.harvard.edu/abs/1979ApJ...234..435B}{{Burstein}, D. 1979,
  \apj, 234, 435%
}%

\bibitem[{{Capaccioli} {et~al.}(1987){Capaccioli}, {Held}, \&
  {Nieto}}]{1987AJ.....94.1519C}
  \href{http://adsabs.harvard.edu/abs/1987AJ.....94.1519C}{{Capaccioli}, M.,
  {Held}, E.~V., \& {Nieto}, J.-L. 1987, \aj, 94, 1519%
}%

\bibitem[{{Capetti} {et~al.}(2005){Capetti}, {Marconi}, {Macchetto}, \&
  {Axon}}]{2005A&A...431..465C}
  \href{http://adsabs.harvard.edu/abs/2005A%26A...431..465C}{{Capetti}, A.,
  {Marconi}, A., {Macchetto}, D., \& {Axon}, D. 2005, \aap, 431, 465%
}%

\bibitem[{{Cappellari} {et~al.}(2008){Cappellari}, {Neumayer}, {Reunanen}, {van
  der Werf}, {de Zeeuw}, \& {Rix}}]{2008arXiv0812.1000C}
  \href{http://adsabs.harvard.edu/abs/2008arXiv0812.1000C}{{Cappellari}, M.,
  {Neumayer}, N., {Reunanen}, J., {van der Werf}, P.~P., {de Zeeuw}, P.~T., \&
  {Rix}, H.~. 2008, }\href{http://www.arXiv.org/abs/0812.1000}{{preprint
  (0812.1000)}%
}%

\bibitem[{{Cappellari} {et~al.}(2002){Cappellari}, {Verolme}, {van der Marel},
  {Kleijn}, {Illingworth}, {Franx}, {Carollo}, \& {de
  Zeeuw}}]{2002ApJ...578..787C}
  \href{http://adsabs.harvard.edu/abs/2002ApJ...578..787C}{{Cappellari}, M.,
  {Verolme}, E.~K., {van der Marel}, R.~P., {Kleijn}, G.~A.~V., {Illingworth},
  G.~D., {Franx}, M., {Carollo}, C.~M., \& {de Zeeuw}, P.~T. 2002, \apj, 578,
  787%
}%

\bibitem[{{Coccato} {et~al.}(2006){Coccato}, {Sarzi}, {Pizzella}, {Corsini},
  {Dalla Bont{\`a}}, \& {Bertola}}]{2006MNRAS.366.1050C}
  \href{http://adsabs.harvard.edu/abs/2006MNRAS.366.1050C}{{Coccato}, L.,
  {Sarzi}, M., {Pizzella}, A., {Corsini}, E.~M., {Dalla Bont{\`a}}, E., \&
  {Bertola}, F. 2006, \mnras, 366, 1050%
}%

\bibitem[{{Cretton} \& {van den Bosch}(1999)}]{1999ApJ...514..704C}
  \href{http://adsabs.harvard.edu/abs/1999ApJ...514..704C}{{Cretton}, N., \&
  {van den Bosch}, F.~C. 1999, \apj, 514, 704%
}%

\bibitem[{{Dalla Bont{\`a}} {et~al.}(2008){Dalla Bont{\`a}}, {Ferrarese},
  {Corsini}, {Miralda-Escud{\'e}}, {Coccato}, {Sarzi}, {Pizzella}, \&
  {Beifiori}}]{2008arXiv0809.0766D}
  \href{http://adsabs.harvard.edu/abs/2008arXiv0809.0766D}{{Dalla Bont{\`a}},
  E., {Ferrarese}, L., {Corsini}, E.~M., {Miralda-Escud{\'e}}, J., {Coccato},
  L., {Sarzi}, M., {Pizzella}, A., \& {Beifiori}, A. 2008,
  }\href{http://www.arXiv.org/abs/0809.0766}{{preprint (0809.0766)}%
}%

\bibitem[{{de Francesco} {et~al.}(2006){de Francesco}, {Capetti}, \&
  {Marconi}}]{2006AandA...460..439D}
  \href{http://adsabs.harvard.edu/abs/2006A%26A...460..439D}{{de Francesco},
  G., {Capetti}, A., \& {Marconi}, A. 2006, \aap, 460, 439%
}%

\bibitem[{{de Francesco} {et~al.}(2008){de Francesco}, {Capetti}, \&
  {Marconi}}]{2008arXiv0801.0064D}
  \href{http://adsabs.harvard.edu/abs/2008arXiv0801.0064D}{---------. 2008,
  }\href{http://www.arXiv.org/abs/0801.0064}{{preprint (0801.0064)}%
}%

\bibitem[{{de Vaucouleurs} {et~al.}(1991){de Vaucouleurs}, {de Vaucouleurs},
  {Corwin}, {Buta}, {Paturel}, \& {Fouque}}]{rc3}
  \href{http://adsabs.harvard.edu/abs/1991trcb.book.....D}{{de Vaucouleurs},
  G., {de Vaucouleurs}, A., {Corwin}, H.~G., Jr., {Buta}, R.~J., {Paturel}, G.,
  \& {Fouque}, P. 1991, {Third Reference Catalogue of Bright Galaxies}
  (Springer-Verlag Berlin Heidelberg New York)%
}%

\bibitem[{{Devereux} {et~al.}(2003){Devereux}, {Ford}, {Tsvetanov}, \&
  {Jacoby}}]{2003AJ....125.1226D}
  \href{http://adsabs.harvard.edu/abs/2003AJ....125.1226D}{{Devereux}, N.,
  {Ford}, H., {Tsvetanov}, Z., \& {Jacoby}, G. 2003, \aj, 125, 1226%
}%

\bibitem[{{Dressler}(1989)}]{dressler89}
  \href{http://adsabs.harvard.edu/abs/1989IAUS..134..217D}{{Dressler}, A. 1989,
  in IAU Symposium 134, Active Galactic Nuclei, ed. D.~E. {Osterbrock} \& J.~S.
  {Miller}, 217%
}%

\bibitem[{{Elvis} {et~al.}(1994){Elvis}, {Wilkes}, {McDowell}, {Green},
  {Bechtold}, {Willner}, {Oey}, {Polomski}, \& {Cutri}}]{elvisetal94}
  \href{http://adsabs.harvard.edu/abs/1994ApJS...95....1E}{{Elvis}, M.,
  {et~al.} 1994, \apjs, 95, 1%
}%

\bibitem[{{Emsellem} {et~al.}(1999){Emsellem}, {Dejonghe}, \&
  {Bacon}}]{1999MNRAS.303..495E}
  \href{http://adsabs.harvard.edu/abs/1999MNRAS.303..495E}{{Emsellem}, E.,
  {Dejonghe}, H., \& {Bacon}, R. 1999, \mnras, 303, 495%
}%

\bibitem[{{Erwin}(2004)}]{2004AandA...415..941E}
  \href{http://adsabs.harvard.edu/abs/2004A%26A...415..941E}{{Erwin}, P. 2004,
  \aap, 415, 941%
}%

\bibitem[{{Ferrarese}(2002)}]{ferrarese02}
  \href{http://adsabs.harvard.edu/abs/2002chee.conf....3F}{{Ferrarese}, L.
  2002, in Current high-energy emission around black holes, ed. C.-H. {Lee} \&
  H.-Y. {Chang}, 3%
}%

\bibitem[{{Ferrarese} \& {Ford}(1999)}]{1999ApJ...515..583F}
  \href{http://adsabs.harvard.edu/abs/1999ApJ...515..583F}{{Ferrarese}, L., \&
  {Ford}, H.~C. 1999, \apj, 515, 583%
}%

\bibitem[{{Ferrarese} {et~al.}(1996){Ferrarese}, {Ford}, \&
  {Jaffe}}]{1996ApJ...470..444F}
  \href{http://adsabs.harvard.edu/abs/1996ApJ...470..444F}{{Ferrarese}, L.,
  {Ford}, H.~C., \& {Jaffe}, W. 1996, \apj, 470, 444%
}%

\bibitem[{{Ferrarese} \& {Ford}(2005)}]{ff05}
  \href{http://adsabs.harvard.edu/abs/2005SSRv..116..523F}{{Ferrarese}, L., \&
  {Ford}, H. 2005, Space Science Reviews, 116, 523%
}%

\bibitem[{{Ferrarese} \& {Merritt}(2000)}]{fm00}
  \href{http://adsabs.harvard.edu/cgi-bin/nph-bib_query?bibcode=2000ApJ...539L%
...9F&db_key=AST}{{Ferrarese}, L., \& {Merritt}, D. 2000, \apjl, 539, L9%
}%

\bibitem[{{Gebhardt} \& {Thomas}(2009)}]{gt09} {{Gebhardt}, K., \& {Thomas}, J.
  2009, \apj, submitted, 0%
}%

\bibitem[{{Gebhardt} {et~al.}(2000{\natexlab{a}}){Gebhardt}, {Bender}, {Bower},
  {Dressler}, {Faber}, {Filippenko}, {Green}, {Grillmair}, {Ho}, {Kormendy},
  {Lauer}, {Magorrian}, {Pinkney}, {Richstone}, \&
  {Tremaine}}]{gebhardtetal00a}
  \href{http://adsabs.harvard.edu/cgi-bin/nph-bib_query?bibcode=2000ApJ...539L%
..13G&db_key=AST}{{Gebhardt}, K., {et~al.} 2000{\natexlab{a}}, \apjl, 539, L13%
}%

\bibitem[{{Gebhardt} {et~al.}(2000{\natexlab{b}}){Gebhardt}, {Kormendy}, {Ho},
  {Bender}, {Bower}, {Dressler}, {Faber}, {Filippenko}, {Green}, {Grillmair},
  {Lauer}, {Magorrian}, {Pinkney}, {Richstone}, \& {Tremaine}}]{gebhardtetal00}
  \href{http://adsabs.harvard.edu/cgi-bin/nph-bib_query?bibcode=2000ApJ...543L%
...5G&db_key=AST}{---------. 2000{\natexlab{b}}, \apjl, 543, L5%
}%

\bibitem[{{Gebhardt} {et~al.}(2000{\natexlab{c}}){Gebhardt}, {Richstone},
  {Kormendy}, {Lauer}, {Ajhar}, {Bender}, {Dressler}, {Faber}, {Grillmair},
  {Magorrian}, \& {Tremaine}}]{2000AJ....119.1157G}
  \href{http://adsabs.harvard.edu/abs/2000AJ....119.1157G}{---------.
  2000{\natexlab{c}}, \aj, 119, 1157%
}%

\bibitem[{{Gebhardt} {et~al.}(2001){Gebhardt}, {Lauer}, {Kormendy}, {Pinkney},
  {Bower}, {Green}, {Gull}, {Hutchings}, {Kaiser}, {Nelson}, {Richstone}, \&
  {Weistrop}}]{2001AJ....122.2469G}
  \href{http://adsabs.harvard.edu/abs/2001AJ....122.2469G}{---------. 2001,
  \aj, 122, 2469%
}%

\bibitem[{{Gebhardt} {et~al.}(2003){Gebhardt}, {Richstone}, {Tremaine},
  {Lauer}, {Bender}, {Bower}, {Dressler}, {Faber}, {Filippenko}, {Green},
  {Grillmair}, {Ho}, {Kormendy}, {Magorrian}, \&
  {Pinkney}}]{2003ApJ...583...92G}
  \href{http://adsabs.harvard.edu/abs/2003ApJ...583...92G}{---------. 2003,
  \apj, 583, 92%
}%

\bibitem[{{Gebhardt} {et~al.}(2007){Gebhardt}, {Lauer}, {Pinkney}, {Bender},
  {Richstone}, {Aller}, {Bower}, {Dressler}, {Faber}, {Filippenko}, {Green},
  {Ho}, {Kormendy}, {Siopis}, \& {Tremaine}}]{2007ApJ...671.1321G}
  \href{http://adsabs.harvard.edu/abs/2007ApJ...671.1321G}{---------. 2007,
  \apj, 671, 1321%
}%

\bibitem[{{Ghez} {et~al.}(2005){Ghez}, {Salim}, {Hornstein}, {Tanner}, {Lu},
  {Morris}, {Becklin}, \& {Duch{\^e}ne}}]{2005ApJ...620..744G}
  \href{http://adsabs.harvard.edu/abs/2005ApJ...620..744G}{{Ghez}, A.~M.,
  {Salim}, S., {Hornstein}, S.~D., {Tanner}, A., {Lu}, J.~R., {Morris}, M.,
  {Becklin}, E.~E., \& {Duch{\^e}ne}, G. 2005, \apj, 620, 744%
}%

\bibitem[{{Ghez} {et~al.}(2008){Ghez}, {Salim}, {Weinberg}, {Lu}, {Do}, {Dunn},
  {Matthews}, {Morris}, {Yelda}, {Becklin}, {Kremenek}, {Milosavljevic}, \&
  {Naiman}}]{ghezetal08}
  \href{http://adsabs.harvard.edu/abs/2008ApJ...689.1044G}{{Ghez}, A.~M.,
  {et~al.} 2008, \apj, 689, 1044%
}%

\bibitem[{{Gillessen} {et~al.}(2008){Gillessen}, {Eisenhauer}, {Trippe},
  {Alexander}, {Genzel}, {Martins}, \& {Ott}}]{2008arXiv0810.4674G}
  \href{http://adsabs.harvard.edu/abs/2008arXiv0810.4674G}{{Gillessen}, S.,
  {Eisenhauer}, F., {Trippe}, S., {Alexander}, T., {Genzel}, R., {Martins}, F.,
  \& {Ott}, T. 2008, }\href{http://www.arXiv.org/abs/0810.4674}{{preprint
  (0810.4674)}%
}%

\bibitem[{{Gnedin} {et~al.}(2004){Gnedin}, {Kravtsov}, {Klypin}, \&
  {Nagai}}]{gnedinetal04}
  \href{http://adsabs.harvard.edu/abs/2004ApJ...616...16G}{{Gnedin}, O.~Y.,
  {Kravtsov}, A.~V., {Klypin}, A.~A., \& {Nagai}, D. 2004, \apj, 616, 16%
}%

\bibitem[{{Graham}(2008)}]{graham08}
  \href{http://adsabs.harvard.edu/abs/2008ApJ...680..143G}{{Graham}, A.~W.
  2008, \apj, 680, 143%
}%

\bibitem[{{Graham} {et~al.}(2001){Graham}, {Erwin}, {Caon}, \&
  {Trujillo}}]{grahametal01}
  \href{http://adsabs.harvard.edu/abs/2001ApJ...563L..11G}{{Graham}, A.~W.,
  {Erwin}, P., {Caon}, N., \& {Trujillo}, I. 2001, \apjl, 563, L11%
}%

\bibitem[{{Greenhill} \& {Gwinn}(1997)}]{1997ApandSS.248..261G}
  \href{http://adsabs.harvard.edu/abs/1997Ap%26SS.248..261G}{{Greenhill},
  L.~J., \& {Gwinn}, C.~R. 1997, \apss, 248, 261%
}%

\bibitem[{{Greenhill} {et~al.}(1997){Greenhill}, {Moran}, \&
  {Herrnstein}}]{1997ApJ...481L..23G}
  \href{http://adsabs.harvard.edu/abs/1997ApJ...481L..23G}{{Greenhill}, L.~J.,
  {Moran}, J.~M., \& {Herrnstein}, J.~R. 1997, \apjl, 481, L23%
}%

\bibitem[{{Greenhill} {et~al.}(2003){Greenhill}, {Booth}, {Ellingsen},
  {Herrnstein}, {Jauncey}, {McCulloch}, {Moran}, {Norris}, {Reynolds}, \&
  {Tzioumis}}]{2003ApJ...590..162G}
  \href{http://adsabs.harvard.edu/abs/2003ApJ...590..162G}{{Greenhill}, L.~J.,
  {et~al.} 2003, \apj, 590, 162%
}%

\bibitem[{{G{\"u}ltekin} {et~al.}(2009){G{\"u}ltekin}, {Richstone}, {Gebhardt},
  {Lauer}, {Pinkney}, {Bender}, {Aller}, {Bower}, {Dressler}, {Faber},
  {Filippenko}, {Green}, {Ho}, {Kormendy}, \& {Siopis}}]{Gultekin_etal_2008}
  \href{http://adsabs.harvard.edu/abs/2009arXiv0901.4162G}{{G{\"u}ltekin}, K.,
  {et~al.} 2009, \apj, accepted, 0%
}%

\bibitem[{{H{\"a}ring-Neumayer} {et~al.}(2006){H{\"a}ring-Neumayer},
  {Cappellari}, {Rix}, {Hartung}, {Prieto}, {Meisenheimer}, \&
  {Lenzen}}]{2006ApJ...643..226H}
  \href{http://adsabs.harvard.edu/abs/2006ApJ...643..226H}{{H{\"a}ring-Neumaye%
r}, N., {Cappellari}, M., {Rix}, H.-W., {Hartung}, M., {Prieto}, M.~A.,
  {Meisenheimer}, K., \& {Lenzen}, R. 2006, \apj, 643, 226%
}%

\bibitem[{{Herrnstein} {et~al.}(2005){Herrnstein}, {Moran}, {Greenhill}, \&
  {Trotter}}]{2005ApJ...629..719H}
  \href{http://adsabs.harvard.edu/abs/2005ApJ...629..719H}{{Herrnstein}, J.~R.,
  {Moran}, J.~M., {Greenhill}, L.~J., \& {Trotter}, A.~S. 2005, \apj, 629, 719%
}%

\bibitem[{{Ho} {et~al.}(2002){Ho}, {Sarzi}, {Rix}, {Shields}, {Rudnick},
  {Filippenko}, \& {Barth}}]{hoetal02}
  \href{http://adsabs.harvard.edu/abs/2002PASP..114..137H}{{Ho}, L.~C.,
  {Sarzi}, M., {Rix}, H.-W., {Shields}, J.~C., {Rudnick}, G., {Filippenko},
  A.~V., \& {Barth}, A.~J. 2002, \pasp, 114, 137%
}%

\bibitem[{{Hopkins} {et~al.}(2006{\natexlab{a}}){Hopkins}, {Hernquist}, {Cox},
  {Robertson}, \& {Springel}}]{2006ApJS..163...50H}
  \href{http://adsabs.harvard.edu/abs/2006ApJS..163...50H}{{Hopkins}, P.~F.,
  {Hernquist}, L., {Cox}, T.~J., {Robertson}, B., \& {Springel}, V.
  2006{\natexlab{a}}, \apjs, 163, 50%
}%

\bibitem[{{Hopkins} {et~al.}(2006{\natexlab{b}}){Hopkins}, {Robertson},
  {Krause}, {Hernquist}, \& {Cox}}]{hopkinsetal06}
  \href{http://adsabs.harvard.edu/abs/2006ApJ...652..107H}{{Hopkins}, P.~F.,
  {Robertson}, B., {Krause}, E., {Hernquist}, L., \& {Cox}, T.~J.
  2006{\natexlab{b}}, \apj, 652, 107%
}%

\bibitem[{{Houghton} {et~al.}(2006){Houghton}, {Magorrian}, {Sarzi}, {Thatte},
  {Davies}, \& {Krajnovi{\'c}}}]{2006MNRAS.367....2H}
  \href{http://adsabs.harvard.edu/abs/2006MNRAS.367....2H}{{Houghton},
  R.~C.~W., {Magorrian}, J., {Sarzi}, M., {Thatte}, N., {Davies}, R.~L., \&
  {Krajnovi{\'c}}, D. 2006, \mnras, 367, 2%
}%

\bibitem[{{Hu}(2008)}]{hu08}
  \href{http://adsabs.harvard.edu/abs/2008MNRAS.386.2242H}{{Hu}, J. 2008,
  \mnras, 386, 2242%
}%

\bibitem[{{Knapen} {et~al.}(2006){Knapen}, {Mazzuca}, {B{\"o}ker}, {Shlosman},
  {Colina}, {Combes}, \& {Axon}}]{2006AandA...448..489K}
  \href{http://adsabs.harvard.edu/abs/2006A%26A...448..489K}{{Knapen}, J.~H.,
  {Mazzuca}, L.~M., {B{\"o}ker}, T., {Shlosman}, I., {Colina}, L., {Combes},
  F., \& {Axon}, D.~J. 2006, \aap, 448, 489%
}%

\bibitem[{{Kormendy}(1988)}]{1988ApJ...335...40K}
  \href{http://adsabs.harvard.edu/abs/1988ApJ...335...40K}{{Kormendy}, J. 1988,
  \apj, 335, 40%
}%

\bibitem[{{Kormendy}(1993)}]{kormendy93a}
  \href{http://adsabs.harvard.edu/abs/1993nag..conf..197K}{---------. 1993, in
  The Nearest Active Galaxies, ed. J.~{Beckman}, L.~{Colina}, \& H.~{Netzer}
  (Madrid: Consejo Superior de Investicaciones Cientificas), 197%
}%

\bibitem[{{Kormendy}(2004)}]{kormendy04}
  \href{http://adsabs.harvard.edu/abs/2004cbhg.symp....1K}{---------. 2004, in
  Coevolution of Black Holes and Galaxies, ed. L.~C. {Ho} (Cambridge: Cambridge
  Univ. Press), 1%
}%

\bibitem[{{Kormendy} {et~al.}(2008){Kormendy}, {Fisher}, {Cornell}, \&
  {Bender}}]{2008arXiv0810.1681K}
  \href{http://adsabs.harvard.edu/abs/2008arXiv0810.1681K}{{Kormendy}, J.,
  {Fisher}, D.~B., {Cornell}, M.~E., \& {Bender}, R. 2008,
  }\href{http://www.arXiv.org/abs/0810.1681}{{preprint (0810.1681)}%
}%

\bibitem[{{Kormendy} \& {Gebhardt}(2001)}]{kg01}
  \href{http://adsabs.harvard.edu/abs/2001AIPC..586..363K}{{Kormendy}, J., \&
  {Gebhardt}, K. 2001, in American Institute of Physics Conference Series 586,
  20th Texas Symposium on relativistic astrophysics, ed. J.~C. {Wheeler} \&
  H.~{Martel} (Melville, NY: AIP), 363%
}%

\bibitem[{{Kormendy} \& {Illingworth}(1982)}]{1982ApJ...256..460K}
  \href{http://adsabs.harvard.edu/abs/1982ApJ...256..460K}{{Kormendy}, J., \&
  {Illingworth}, G. 1982, \apj, 256, 460%
}%

\bibitem[{{Kormendy} \& {Illingworth}(1983)}]{1983ApJ...265..632K}
  \href{http://adsabs.harvard.edu/abs/1983ApJ...265..632K}{---------. 1983,
  \apj, 265, 632%
}%

\bibitem[{{Kormendy} \& {Kennicutt}(2004)}]{kk04}
  \href{http://adsabs.harvard.edu/abs/2004ARA%26A..42..603K}{{Kormendy}, J., \&
  {Kennicutt}, R.~C., Jr. 2004, \araa, 42, 603%
}%

\bibitem[{{Kormendy} \& {Richstone}(1995)}]{kr95}
  \href{http://adsabs.harvard.edu/cgi-bin/nph-bib_query?bibcode=1995ARA%26A..3%
3..581K&db_key=AST}{{Kormendy}, J., \& {Richstone}, D. 1995, \araa, 33, 581%
}%

\bibitem[{{Kormendy} {et~al.}(1997){Kormendy}, {Bender}, {Magorrian},
  {Tremaine}, {Gebhardt}, {Richstone}, {Dressler}, {Faber}, {Grillmair}, \&
  {Lauer}}]{kormendyetal97}
  \href{http://adsabs.harvard.edu/abs/1997ApJ...482L.139K}{{Kormendy}, J.,
  {et~al.} 1997, \apjl, 482, L139%
}%

\bibitem[{{Krajnovi{\'c}} {et~al.}(2007){Krajnovi{\'c}}, {Sharp}, \&
  {Thatte}}]{2007MNRAS.374..385K}
  \href{http://adsabs.harvard.edu/abs/2007MNRAS.374..385K}{{Krajnovi{\'c}}, D.,
  {Sharp}, R., \& {Thatte}, N. 2007, \mnras, 374, 385%
}%

\bibitem[{{Lauer} {et~al.}(2007{\natexlab{a}}){Lauer}, {Tremaine}, {Richstone},
  \& {Faber}}]{laueretal07c}
  \href{http://adsabs.harvard.edu/abs/2007ApJ...670..249L}{{Lauer}, T.~R.,
  {Tremaine}, S., {Richstone}, D., \& {Faber}, S.~M. 2007{\natexlab{a}}, \apj,
  670, 249%
}%

\bibitem[{{Lauer} {et~al.}(1996){Lauer}, {Tremaine}, {Ajhar}, {Bender},
  {Dressler}, {Faber}, {Gebhardt}, {Grillmair}, {Kormendy}, \&
  {Richstone}}]{laueretal96}
  \href{http://adsabs.harvard.edu/abs/1996ApJ...471L..79L}{{Lauer}, T.~R.,
  {et~al.} 1996, \apjl, 471, L79%
}%

\bibitem[{{Lauer} {et~al.}(2005){Lauer}, {Faber}, {Gebhardt}, {Richstone},
  {Tremaine}, {Ajhar}, {Aller}, {Bender}, {Dressler}, {Filippenko}, {Green},
  {Grillmair}, {Ho}, {Kormendy}, {Magorrian}, {Pinkney}, \&
  {Siopis}}]{laueretal05}
  \href{http://adsabs.harvard.edu/abs/2005AJ....129.2138L}{---------. 2005,
  \aj, 129, 2138%
}%

\bibitem[{{Lauer} {et~al.}(2007{\natexlab{b}}){Lauer}, {Faber}, {Richstone},
  {Gebhardt}, {Tremaine}, {Postman}, {Dressler}, {Aller}, {Filippenko},
  {Green}, {Ho}, {Kormendy}, {Magorrian}, \& {Pinkney}}]{laueretal07}
  \href{http://adsabs.harvard.edu/abs/2007ApJ...662..808L}{---------.
  2007{\natexlab{b}}, \apj, 662, 808%
}%

\bibitem[{{Lauer} {et~al.}(2007{\natexlab{c}}){Lauer}, {Gebhardt}, {Faber},
  {Richstone}, {Tremaine}, {Kormendy}, {Aller}, {Bender}, {Dressler},
  {Filippenko}, {Green}, \& {Ho}}]{laueretal07b}
  \href{http://adsabs.harvard.edu/abs/2007ApJ...664..226L}{---------.
  2007{\natexlab{c}}, \apj, 664, 226%
}%

\bibitem[{{Lepage}(1978)}]{lepage78}
  \href{http://adsabs.harvard.edu/abs/1978JCoPh..27..192L}{{Lepage}, G.~P.
  1978, Journal of Computational Physics, 27, 192%
}%

\bibitem[{{Lodato} \& {Bertin}(2003)}]{2003AandA...398..517L}
  \href{http://adsabs.harvard.edu/abs/2003A%26A...398..517L}{{Lodato}, G., \&
  {Bertin}, G. 2003, \aap, 398, 517%
}%

\bibitem[{{Macchetto} {et~al.}(1997){Macchetto}, {Marconi}, {Axon}, {Capetti},
  {Sparks}, \& {Crane}}]{1997ApJ...489..579M}
  \href{http://adsabs.harvard.edu/abs/1997ApJ...489..579M}{{Macchetto}, F.,
  {Marconi}, A., {Axon}, D.~J., {Capetti}, A., {Sparks}, W., \& {Crane}, P.
  1997, \apj, 489, 579%
}%

\bibitem[{{Maciejewski} \& {Binney}(2001)}]{mb01}
  \href{http://adsabs.harvard.edu/abs/2001MNRAS.323..831M}{{Maciejewski}, W.,
  \& {Binney}, J. 2001, \mnras, 323, 831%
}%

\bibitem[{{Magorrian}(2006)}]{magorrian06}
  \href{http://adsabs.harvard.edu/abs/2006MNRAS.373..425M}{{Magorrian}, J.
  2006, \mnras, 373, 425%
}%

\bibitem[{{Magorrian} {et~al.}(1998){Magorrian}, {Tremaine}, {Richstone},
  {Bender}, {Bower}, {Dressler}, {Faber}, {Gebhardt}, {Green}, {Grillmair},
  {Kormendy}, \& {Lauer}}]{magorrianetal98}
  \href{http://adsabs.harvard.edu/cgi-bin/nph-bib_query?bibcode=1998AJ....115.%
2285M&db_key=AST}{{Magorrian}, J., {et~al.} 1998, \aj, 115, 2285%
}%

\bibitem[{{Marconi} {et~al.}(2001){Marconi}, {Capetti}, {Axon}, {Koekemoer},
  {Macchetto}, \& {Schreier}}]{2001ApJ...549..915M}
  \href{http://adsabs.harvard.edu/abs/2001ApJ...549..915M}{{Marconi}, A.,
  {Capetti}, A., {Axon}, D.~J., {Koekemoer}, A., {Macchetto}, D., \&
  {Schreier}, E.~J. 2001, \apj, 549, 915%
}%

\bibitem[{{Marconi} \& {Hunt}(2003)}]{mh03}
  \href{http://adsabs.harvard.edu/abs/2003ApJ...589L..21M}{{Marconi}, A., \&
  {Hunt}, L.~K. 2003, \apjl, 589, L21%
}%

\bibitem[{{Marconi} {et~al.}(2006){Marconi}, {Pastorini}, {Pacini}, {Axon},
  {Capetti}, {Macchetto}, {Koekemoer}, \& {Schreier}}]{2006AandA...448..921M}
  \href{http://adsabs.harvard.edu/abs/2006A%26A...448..921M}{{Marconi}, A.,
  {Pastorini}, G., {Pacini}, F., {Axon}, D.~J., {Capetti}, A., {Macchetto}, D.,
  {Koekemoer}, A.~M., \& {Schreier}, E.~J. 2006, \aap, 448, 921%
}%

\bibitem[{{Marconi} {et~al.}(2004){Marconi}, {Risaliti}, {Gilli}, {Hunt},
  {Maiolino}, \& {Salvati}}]{marconietal04}
  \href{http://adsabs.harvard.edu/abs/2004MNRAS.351..169M}{{Marconi}, A.,
  {Risaliti}, G., {Gilli}, R., {Hunt}, L.~K., {Maiolino}, R., \& {Salvati}, M.
  2004, \mnras, 351, 169%
}%

\bibitem[{{Marconi} {et~al.}(2003){Marconi}, {Axon}, {Capetti}, {Maciejewski},
  {Atkinson}, {Batcheldor}, {Binney}, {Carollo}, {Dressel}, {Ford}, {Gerssen},
  {Hughes}, {Macchetto}, {Merrifield}, {Scarlata}, {Sparks}, {Stiavelli},
  {Tsvetanov}, \& {van der Marel}}]{2003ApJ...586..868M}
  \href{http://adsabs.harvard.edu/abs/2003ApJ...586..868M}{{Marconi}, A.,
  {et~al.} 2003, \apj, 586, 868%
}%

\bibitem[{{Martel} {et~al.}(2000){Martel}, {Turner}, {Sparks}, \&
  {Baum}}]{marteletal00}
  \href{http://adsabs.harvard.edu/abs/2000ApJS..130..267M}{{Martel}, A.~R.,
  {Turner}, N.~J., {Sparks}, W.~B., \& {Baum}, S.~A. 2000, \apjs, 130, 267%
}%

\bibitem[{{Merritt} \& {Ferrarese}(2001{\natexlab{a}})}]{mf01b}
  \href{http://adsabs.harvard.edu/abs/2001ASPC..249..335M}{{Merritt}, D., \&
  {Ferrarese}, L. 2001{\natexlab{a}}, in Astronomical Society of the Pacific
  Conference Series 249, The Central Kiloparsec of Starbursts and AGN: The La
  Palma Connection, ed. J.~H. {Knapen}, J.~E. {Beckman}, I.~{Shlosman}, \&
  T.~J. {Mahoney}, 335%
}%

\bibitem[{{Merritt} \& {Ferrarese}(2001{\natexlab{b}})}]{mf01a}
  \href{http://adsabs.harvard.edu/abs/2001ApJ...547..140M}{---------.
  2001{\natexlab{b}}, \apj, 547, 140%
}%

\bibitem[{{Merritt} {et~al.}(2001){Merritt}, {Ferrarese}, \&
  {Joseph}}]{2001Sci...293.1116M}
  \href{http://adsabs.harvard.edu/abs/2001Sci...293.1116M}{{Merritt}, D.,
  {Ferrarese}, L., \& {Joseph}, C.~L. 2001, Science, 293, 1116%
}%

\bibitem[{{Miyoshi} {et~al.}(1995){Miyoshi}, {Moran}, {Herrnstein},
  {Greenhill}, {Nakai}, {Diamond}, \& {Inoue}}]{1995Natur.373..127M}
  \href{http://adsabs.harvard.edu/abs/1995Natur.373..127M}{{Miyoshi}, M.,
  {Moran}, J., {Herrnstein}, J., {Greenhill}, L., {Nakai}, N., {Diamond}, P.,
  \& {Inoue}, M. 1995, \nat, 373, 127%
}%

\bibitem[{{Neumayer} {et~al.}(2007){Neumayer}, {Cappellari}, {Reunanen}, {Rix},
  {van der Werf}, {de Zeeuw}, \& {Davies}}]{2007ApJ...671.1329N}
  \href{http://adsabs.harvard.edu/abs/2007ApJ...671.1329N}{{Neumayer}, N.,
  {Cappellari}, M., {Reunanen}, J., {Rix}, H.-W., {van der Werf}, P.~P., {de
  Zeeuw}, P.~T., \& {Davies}, R.~I. 2007, \apj, 671, 1329%
}%

\bibitem[{{Novak} {et~al.}(2006){Novak}, {Faber}, \& {Dekel}}]{nfd06}
  \href{http://adsabs.harvard.edu/abs/2006ApJ...637...96N}{{Novak}, G.~S.,
  {Faber}, S.~M., \& {Dekel}, A. 2006, \apj, 637, 96%
}%

\bibitem[{{Nowak} {et~al.}(2007){Nowak}, {Saglia}, {Thomas}, {Bender},
  {Pannella}, {Gebhardt}, \& {Davies}}]{nowaketal07}
  \href{http://adsabs.harvard.edu/abs/2007MNRAS.379..909N}{{Nowak}, N.,
  {Saglia}, R.~P., {Thomas}, J., {Bender}, R., {Pannella}, M., {Gebhardt}, K.,
  \& {Davies}, R.~I. 2007, \mnras, 379, 909%
}%

\bibitem[{{Onken} {et~al.}(2004){Onken}, {Ferrarese}, {Merritt}, {Peterson},
  {Pogge}, {Vestergaard}, \& {Wandel}}]{onkenetal04}
  \href{http://adsabs.harvard.edu/abs/2004ApJ...615..645O}{{Onken}, C.~A.,
  {Ferrarese}, L., {Merritt}, D., {Peterson}, B.~M., {Pogge}, R.~W.,
  {Vestergaard}, M., \& {Wandel}, A. 2004, \apj, 615, 645%
}%

\bibitem[{{Onken} {et~al.}(2007){Onken}, {Valluri}, {Peterson}, {Pogge},
  {Bentz}, {Ferrarese}, {Vestergaard}, {Crenshaw}, {Sergeev}, {McHardy},
  {Merritt}, {Bower}, {Heckman}, \& {Wandel}}]{onkenetal07}
  \href{http://adsabs.harvard.edu/abs/2007ApJ...670..105O}{{Onken}, C.~A.,
  {et~al.} 2007, \apj, 670, 105%
}%

\bibitem[{{Pastorini} {et~al.}(2007){Pastorini}, {Marconi}, {Capetti}, {Axon},
  {Alonso-Herrero}, {Atkinson}, {Batcheldor}, {Carollo}, {Collett}, {Dressel},
  {Hughes}, {Macchetto}, {Maciejewski}, {Sparks}, \& {van der
  Marel}}]{2007AandA...469..405P}
  \href{http://adsabs.harvard.edu/abs/2007A%26A...469..405P}{{Pastorini}, G.,
  {et~al.} 2007, \aap, 469, 405%
}%

\bibitem[{{Peng} {et~al.}(2006){Peng}, {Impey}, {Rix}, {Kochanek}, {Keeton},
  {Falco}, {Leh{\'a}r}, \& {McLeod}}]{pengetal06}
  \href{http://adsabs.harvard.edu/abs/2006ApJ...649..616P}{{Peng}, C.~Y.,
  {Impey}, C.~D., {Rix}, H.-W., {Kochanek}, C.~S., {Keeton}, C.~R., {Falco},
  E.~E., {Leh{\'a}r}, J., \& {McLeod}, B.~A. 2006, \apj, 649, 616%
}%

\bibitem[{{Peterson} {et~al.}(2004){Peterson}, {Ferrarese}, {Gilbert}, {Kaspi},
  {Malkan}, {Maoz}, {Merritt}, {Netzer}, {Onken}, {Pogge}, {Vestergaard}, \&
  {Wandel}}]{petersonetal04}
  \href{http://adsabs.harvard.edu/abs/2004ApJ...613..682P}{{Peterson}, B.~M.,
  {et~al.} 2004, \apj, 613, 682%
}%

\bibitem[{{Postman} \& {Lauer}(1995)}]{pl95}
  \href{http://adsabs.harvard.edu/abs/1995ApJ...440...28P}{{Postman}, M., \&
  {Lauer}, T.~R. 1995, \apj, 440, 28%
}%

\bibitem[{{Press} {et~al.}(1986){Press}, {Flannery}, \&
  {Teukolsky}}]{numericalrecipes}
  \href{http://adsabs.harvard.edu/abs/1986nras.book.....P}{{Press}, W.~H.,
  {Flannery}, B.~P., \& {Teukolsky}, S.~A. 1986, {Numerical Recipes. The Art of
  Scientific Computing} (Cambridge: Cambridge University Press)%
}%

\bibitem[{{Richards} {et~al.}(2005){Richards}, {Croom}, {Anderson},
  {Bland-Hawthorn}, {Boyle}, {De Propris}, {Drinkwater}, {Fan}, {Gunn},
  {Ivezi{\'c}}, {Jester}, {Loveday}, {Meiksin}, {Miller}, {Myers}, {Nichol},
  {Outram}, {Pimbblet}, {Roseboom}, {Ross}, {Schneider}, {Shanks}, {Sharp},
  {Stoughton}, {Strauss}, {Szalay}, {Vanden Berk}, \&
  {York}}]{2005MNRAS.360..839R}
  \href{http://adsabs.harvard.edu/abs/2005MNRAS.360..839R}{{Richards}, G.~T.,
  {et~al.} 2005, \mnras, 360, 839%
}%

\bibitem[{{Richstone}(2004)}]{richstone04}
  \href{http://adsabs.harvard.edu/abs/2004cbhg.symp..280R}{{Richstone}, D.
  2004, in Coevolution of Black Holes and Galaxies, ed. L.~C. {Ho} (Cambridge:
  Univ. Chicago Press), 280%
}%

\bibitem[{{Richstone} {et~al.}(1998){Richstone}, {Ajhar}, {Bender}, {Bower},
  {Dressler}, {Faber}, {Filippenko}, {Gebhardt}, {Green}, {Ho}, {Kormendy},
  {Lauer}, {Magorrian}, \& {Tremaine}}]{richstoneetal98}
  \href{http://adsabs.harvard.edu/cgi-bin/nph-bib_query?bibcode=1998Natur.395A%
..14R&db_key=AST}{{Richstone}, D., {et~al.} 1998, \nat, 395, A14%
}%

\bibitem[{{Sarzi} {et~al.}(2001){Sarzi}, {Rix}, {Shields}, {Rudnick}, {Ho},
  {McIntosh}, {Filippenko}, \& {Sargent}}]{2001ApJ...550...65S}
  \href{http://adsabs.harvard.edu/abs/2001ApJ...550...65S}{{Sarzi}, M., {Rix},
  H.-W., {Shields}, J.~C., {Rudnick}, G., {Ho}, L.~C., {McIntosh}, D.~H.,
  {Filippenko}, A.~V., \& {Sargent}, W.~L.~W. 2001, \apj, 550, 65%
}%

\bibitem[{{Sarzi} {et~al.}(2002){Sarzi}, {Rix}, {Shields}, {McIntosh}, {Ho},
  {Rudnick}, {Filippenko}, {Sargent}, \& {Barth}}]{2002ApJ...567..237S}
  \href{http://adsabs.harvard.edu/abs/2002ApJ...567..237S}{{Sarzi}, M.,
  {et~al.} 2002, \apj, 567, 237%
}%

\bibitem[{{Sazonov} {et~al.}(2004){Sazonov}, {Ostriker}, \& {Sunyaev}}]{sos04}
  \href{http://adsabs.harvard.edu/abs/2004MNRAS.347..144S}{{Sazonov}, S.~Y.,
  {Ostriker}, J.~P., \& {Sunyaev}, R.~A. 2004, \mnras, 347, 144%
}%

\bibitem[{{Schechter}(1976)}]{schecter76}
  \href{http://adsabs.harvard.edu/abs/1976ApJ...203..297S}{{Schechter}, P.
  1976, \apj, 203, 297%
}%

\bibitem[{{Shen} {et~al.}(2008){Shen}, {Vanden Berk}, {Schneider}, \&
  {Hall}}]{shenetal08}
  \href{http://adsabs.harvard.edu/abs/2008AJ....135..928S}{{Shen}, J., {Vanden
  Berk}, D.~E., {Schneider}, D.~P., \& {Hall}, P.~B. 2008, \aj, 135, 928%
}%

\bibitem[{{Shen} {et~al.}(2007){Shen}, {Strauss}, {Oguri}, {Hennawi}, {Fan},
  {Richards}, {Hall}, {Gunn}, {Schneider}, {Szalay}, {Thakar}, {Vanden Berk},
  {Anderson}, {Bahcall}, {Connolly}, \& {Knapp}}]{shenetal07}
  \href{http://adsabs.harvard.edu/abs/2007AJ....133.2222S}{{Shen}, Y., {et~al.}
  2007, \aj, 133, 2222%
}%

\bibitem[{{Sheth} {et~al.}(2003){Sheth}, {Bernardi}, {Schechter}, {Burles},
  {Eisenstein}, {Finkbeiner}, {Frieman}, {Lupton}, {Schlegel}, {Subbarao},
  {Shimasaku}, {Bahcall}, {Brinkmann}, \& {Ivezi{\'c}}}]{shethetal03}
  \href{http://adsabs.harvard.edu/abs/2003ApJ...594..225S}{{Sheth}, R.~K.,
  {et~al.} 2003, \apj, 594, 225%
}%

\bibitem[{{Silge} {et~al.}(2005){Silge}, {Gebhardt}, {Bergmann}, \&
  {Richstone}}]{2005AJ....130..406S}
  \href{http://adsabs.harvard.edu/abs/2005AJ....130..406S}{{Silge}, J.~D.,
  {Gebhardt}, K., {Bergmann}, M., \& {Richstone}, D. 2005, \aj, 130, 406%
}%

\bibitem[{{Silk} \& {Rees}(1998)}]{sr98}
  \href{http://adsabs.harvard.edu/abs/1998A%26A...331L...1S}{{Silk}, J., \&
  {Rees}, M.~J. 1998, \aap, 331, L1%
}%

\bibitem[{{Siopis} {et~al.}(2008){Siopis}, {Gebhardt}, {Lauer}, {Kormendy},
  {Pinkney}, {Richstone}, {Faber}, {Tremaine}, {Aller}, {Bender}, {Bower},
  {Dressler}, {Filippenko}, {Green}, {Ho}, \& {Magorrian}}]{siopisetal08}
  \href{http://adsabs.harvard.edu/abs/2008arXiv0808.4001S}{{Siopis}, C.,
  {et~al.} 2008, }\href{http://www.arXiv.org/abs/0808.4001}{{preprint
  (0808.4001)}%
}%

\bibitem[{{Steidel} {et~al.}(2002){Steidel}, {Hunt}, {Shapley}, {Adelberger},
  {Pettini}, {Dickinson}, \& {Giavalisco}}]{steideletal02}
  \href{http://adsabs.harvard.edu/abs/2002ApJ...576..653S}{{Steidel}, C.~C.,
  {Hunt}, M.~P., {Shapley}, A.~E., {Adelberger}, K.~L., {Pettini}, M.,
  {Dickinson}, M., \& {Giavalisco}, M. 2002, \apj, 576, 653%
}%

\bibitem[{Stephens(1974)}]{stephens74} {Stephens, M. 1974, Journal of the
  American Statistical Association, 69, 730%
}%

\bibitem[{{Tadhunter} {et~al.}(2003){Tadhunter}, {Marconi}, {Axon}, {Wills},
  {Robinson}, \& {Jackson}}]{2003MNRAS.342..861T}
  \href{http://adsabs.harvard.edu/abs/2003MNRAS.342..861T}{{Tadhunter}, C.,
  {Marconi}, A., {Axon}, D., {Wills}, K., {Robinson}, T.~G., \& {Jackson}, N.
  2003, \mnras, 342, 861%
}%

\bibitem[{{Tonry} {et~al.}(2001){Tonry}, {Dressler}, {Blakeslee}, {Ajhar},
  {Fletcher}, {Luppino}, {Metzger}, \& {Moore}}]{tonryetal01}
  \href{http://adsabs.harvard.edu/abs/2001ApJ...546..681T}{{Tonry}, J.~L.,
  {Dressler}, A., {Blakeslee}, J.~P., {Ajhar}, E.~A., {Fletcher}, A.~B.,
  {Luppino}, G.~A., {Metzger}, M.~R., \& {Moore}, C.~B. 2001, \apj, 546, 681%
}%

\bibitem[{{Tremaine}(1995)}]{tremaine95}
  \href{http://adsabs.harvard.edu/abs/1995AJ....110..628T}{{Tremaine}, S. 1995,
  \aj, 110, 628%
}%

\bibitem[{{Tremaine} {et~al.}(2002){Tremaine}, {Gebhardt}, {Bender}, {Bower},
  {Dressler}, {Faber}, {Filippenko}, {Green}, {Grillmair}, {Ho}, {Kormendy},
  {Lauer}, {Magorrian}, {Pinkney}, \& {Richstone}}]{tremaineetal02}
  \href{http://adsabs.harvard.edu/abs/2002ApJ...574..740T}{{Tremaine}, S.,
  {et~al.} 2002, \apj, 574, 740%
}%

\bibitem[{{Treu} {et~al.}(2004){Treu}, {Malkan}, \& {Blandford}}]{tmb04}
  \href{http://adsabs.harvard.edu/abs/2004ApJ...615L..97T}{{Treu}, T.,
  {Malkan}, M.~A., \& {Blandford}, R.~D. 2004, \apjl, 615, L97%
}%

\bibitem[{{Treu} {et~al.}(2007){Treu}, {Woo}, {Malkan}, \&
  {Blandford}}]{treuetal07}
  \href{http://adsabs.harvard.edu/abs/2007ApJ...667..117T}{{Treu}, T., {Woo},
  J.-H., {Malkan}, M.~A., \& {Blandford}, R.~D. 2007, \apj, 667, 117%
}%

\bibitem[{{Valluri} {et~al.}(2005){Valluri}, {Ferrarese}, {Merritt}, \&
  {Joseph}}]{vallurietal05}
  \href{http://adsabs.harvard.edu/abs/2005ApJ...628..137V}{{Valluri}, M.,
  {Ferrarese}, L., {Merritt}, D., \& {Joseph}, C.~L. 2005, \apj, 628, 137%
}%

\bibitem[{{Valluri} {et~al.}(2004){Valluri}, {Merritt}, \& {Emsellem}}]{vme04}
  \href{http://adsabs.harvard.edu/abs/2004ApJ...602...66V}{{Valluri}, M.,
  {Merritt}, D., \& {Emsellem}, E. 2004, \apj, 602, 66%
}%

\bibitem[{{van den Bosch} {et~al.}(1998){van den Bosch}, {Jaffe}, \& {van der
  Marel}}]{1998MNRAS.293..343V}
  \href{http://adsabs.harvard.edu/abs/1998MNRAS.293..343V}{{van den Bosch},
  F.~C., {Jaffe}, W., \& {van der Marel}, R.~P. 1998, \mnras, 293, 343%
}%

\bibitem[{{van der Marel} \& {van den Bosch}(1998)}]{1998AJ....116.2220V}
  \href{http://adsabs.harvard.edu/abs/1998AJ....116.2220V}{{van der Marel},
  R.~P., \& {van den Bosch}, F.~C. 1998, \aj, 116, 2220%
}%

\bibitem[{{Verolme} {et~al.}(2002){Verolme}, {Cappellari}, {Copin}, {van der
  Marel}, {Bacon}, {Bureau}, {Davies}, {Miller}, \& {de
  Zeeuw}}]{2002MNRAS.335..517V}
  \href{http://adsabs.harvard.edu/abs/2002MNRAS.335..517V}{{Verolme}, E.~K.,
  {et~al.} 2002, \mnras, 335, 517%
}%

\bibitem[{{Vestergaard} {et~al.}(2008){Vestergaard}, {Fan}, {Tremonti},
  {Osmer}, \& {Richards}}]{vestergaardetal08}
  \href{http://adsabs.harvard.edu/abs/2008ApJ...674L...1V}{{Vestergaard}, M.,
  {Fan}, X., {Tremonti}, C.~A., {Osmer}, P.~S., \& {Richards}, G.~T. 2008,
  \apjl, 674, L1%
}%

\bibitem[{{Volonteri}(2007)}]{volonteri07}
  \href{http://adsabs.harvard.edu/abs/2007ApJ...663L...5V}{{Volonteri}, M.
  2007, \apjl, 663, L5%
}%

\bibitem[{{Wold} {et~al.}(2006){Wold}, {Lacy}, {K{\"a}ufl}, \&
  {Siebenmorgen}}]{2006AandA...460..449W}
  \href{http://adsabs.harvard.edu/abs/2006A%26A...460..449W}{{Wold}, M.,
  {Lacy}, M., {K{\"a}ufl}, H.~U., \& {Siebenmorgen}, R. 2006, \aap, 460, 449%
}%

\bibitem[{{Wyithe}(2006{\natexlab{a}})}]{wyithe06}
  \href{http://adsabs.harvard.edu/abs/2006MNRAS.365.1082W}{{Wyithe}, J.~S.~B.
  2006{\natexlab{a}}, \mnras, 365, 1082%
}%

\bibitem[{{Wyithe}(2006{\natexlab{b}})}]{wyithe06erratum}
  \href{http://adsabs.harvard.edu/abs/2006MNRAS.371.1536W}{---------.
  2006{\natexlab{b}}, \mnras, 371, 1536%
}%

\bibitem[{{Yu} \& {Tremaine}(2002)}]{yt02}
  \href{http://adsabs.harvard.edu/abs/2002MNRAS.335..965Y}{{Yu}, Q., \&
  {Tremaine}, S. 2002, \mnras, 335, 965%
}%

\end{thebibliography}

\clearpage
\LongTables
\begin{deluxetable}{llrrrrrr@{$\pm$}lrrrr}
\tabletypesize{\scriptsize}
  \tablecaption{Sample of Dynamically Detected Black Hole Masses}
  \tablewidth{0pt}
  \tablehead{
    & & \colhead{Dist.} & \colhead{$M_{\mathrm{BH}}$} & \colhead{$M_{\mathrm{low}}$} & \colhead{$M_{\mathrm{high}}$} & \colhead{Method,} & \multicolumn{2}{c}{$\sigma_e$} & & & & \\
     \colhead{Galaxy} &
     \colhead{Type%
\tablenotemark{a} 
} &
     \colhead{Mpc} &
     \colhead{\msun}  &
     \colhead{\msun}  &
     \colhead{\msun}  &
     \colhead{Ref.} &
     \multicolumn{2}{c}{$\mathrm{km~s^{-1}}$} &
     \colhead{$M_{V,T}^0$} &
     \colhead{$M_{V,{\mathrm{bulge}}}^0$%
\tablenotemark{b} 
} &
     \colhead{${R_{\mathrm{infl}}/d_{\mathrm{res}}}$} &
     \colhead{Samp.} 
  }
  \startdata
Circinus\tablenotemark{c}\tablenotemark{d}       &        Sb &    4.0 & $1.7\times10^{6}$ & $1.4\times10^{6}$ & $2.1\times10^{6}$ & maser, ~1 & $158$&$18$\tablenotemark{d} & $-17.36$ &  . . . &   6.06 & ~S\\
IC1459\tablenotemark{e}       &        E4 &   30.9 & $2.8\times10^{9}$ & $1.6\times10^{9}$ & $3.9\times10^{9}$ & stars, ~2 & $340$&$17$ & $-22.57$ & $-22.57 \pm 0.15$ &   0.56 & ~S\\
MW\tablenotemark{f}\tablenotemark{g}       &       Sbc &  0.008 & $4.1\times10^{6}$ & $3.5\times10^{6}$ & $4.7\times10^{6}$ & stars, ~3 & $105$&$20$ &  . . . &  . . . &  20622 & ~S\\
 N0221 ~~~ M32 &        E2 &   0.86 & $3.1\times10^{6}$ & $2.5\times10^{6}$ & $3.7\times10^{6}$ & stars, ~4 & $ 75$&$ 3$ & $-16.83$ & $-16.83 \pm 0.05$ &   12.2 & RS\\
 N0224 ~~~ M31 &        Sb &   0.80 & $1.5\times10^{8}$ & $1.2\times10^{8}$ & $2.4\times10^{8}$ & stars, ~5 & $160$&$ 8$ & $-21.84$ &  . . . &    113 & ~S\\
N0821\tablenotemark{h}       &        E4 &   25.5 & $4.2\times10^{7}$ & $3.4\times10^{7}$ & $7.0\times10^{7}$ & stars, ~6 & $209$&$10$ & $-21.24$ & $-21.24 \pm 0.13$ &   0.33 & ~S\\
 N1023       &       SB0 &   12.1 & $4.6\times10^{7}$ & $4.1\times10^{7}$ & $5.1\times10^{7}$ & stars, ~7 & $205$&$10$ & $-21.26$ & $-20.61 \pm 0.28$ &   0.81 & ~S\\
N1068\tablenotemark{g}\tablenotemark{i} ~~~ M77 &        Sb &   15.4 & $8.6\times10^{6}$ & $8.3\times10^{6}$ & $8.9\times10^{6}$ & maser, ~8 & $151$&$ 7$ & $-22.17$ &  . . . &   22.5 & ~S\\
N1300\tablenotemark{g}       &  SB(rs)bc &   20.1 & $7.1\times10^{7}$ & $3.6\times10^{7}$ & $1.4\times10^{8}$ &   gas, ~9 & $218$&$10$ & $-21.34$ &  . . . &   0.65 & ~S\\
N1399\tablenotemark{j}       &        E1 &   21.1 & $5.1\times10^{8}$ & $4.4\times10^{8}$ & $5.8\times10^{8}$ & stars, 10 & $337$&$16$ & $-22.13$ & $-22.13 \pm 0.10$ &   1.82 & ~S\\
N1399\tablenotemark{j}       &        E1 &   21.1 & $1.3\times10^{9}$ & $6.4\times10^{8}$ & $1.8\times10^{9}$ & stars, 11 & $337$&$16$ & $-22.13$ & $-22.13 \pm 0.10$ &   3.02 & ~S\\
N2748\tablenotemark{g}       &        Sc &   24.9 & $4.7\times10^{7}$ & $8.6\times10^{6}$ & $8.5\times10^{7}$ &   gas, ~9 & $115$&$ 5$ & $-20.97$ &  . . . &   1.27 & ~S\\
N2778\tablenotemark{h}       &        E2 &   24.2 & $1.6\times10^{7}$ & $5.8\times10^{6}$ & $2.5\times10^{7}$ & stars, ~6 & $175$&$ 8$ & $-19.62$ & $-19.62 \pm 0.13$ &   0.45 & ~S\\
N2787\tablenotemark{g}\tablenotemark{k}       &       SB0 &    7.9 & $4.3\times10^{7}$ & $3.8\times10^{7}$ & $4.7\times10^{7}$ &   gas, 12 & $189$&$ 9$ & $-18.90$ &  . . . &   1.09 & RS\\
 N3031 ~~~ M81 &        Sb &    4.1 & $8.0\times10^{7}$ & $6.9\times10^{7}$ & $1.0\times10^{8}$ &   gas, 13 & $143$&$ 7$ & $-21.51$ &  . . . &   6.61 & ~S\\
 N3115       &        S0 &   10.2 & $9.6\times10^{8}$ & $6.7\times10^{8}$ & $1.5\times10^{9}$ & stars, 14 & $230$&$11$ & $-21.25$ & $-21.18 \pm 0.05$ &   13.1 & ~S\\
N3227\tablenotemark{d}\tablenotemark{g}       &       SBa &   17.0 & $1.5\times10^{7}$ & $7.0\times10^{6}$ & $2.0\times10^{7}$ & stars, 15 & $133$&$12$\tablenotemark{d} & $-20.73$ &  . . . &   0.52 & ~S\\
N3245\tablenotemark{g}       &        S0 &   22.1 & $2.2\times10^{8}$ & $1.7\times10^{8}$ & $2.7\times10^{8}$ &   gas, 15 & $205$&$10$ & $-20.96$ &  . . . &   1.01 & RS\\
N3377\tablenotemark{h}       &        E6 &   11.7 & $1.1\times10^{8}$ & $1.0\times10^{8}$ & $2.2\times10^{8}$ & stars, ~6 & $145$&$ 7$ & $-20.11$ & $-20.11 \pm 0.10$ &   4.49 & RS\\
N3379\tablenotemark{h}       &        E0 &   11.7 & $1.2\times10^{8}$ & $6.2\times10^{7}$ & $2.0\times10^{8}$ & stars, 16 & $206$&$10$ & $-21.10$ & $-21.10 \pm 0.03$ &   2.18 & ~S\\
N3384\tablenotemark{h}\tablenotemark{g}       &       SB0 &   11.7 & $1.8\times10^{7}$ & $1.5\times10^{7}$ & $1.9\times10^{7}$ & stars, ~6 & $143$&$ 7$ & $-20.50$ & $-19.93 \pm 0.22$ &   0.60 & ~S\\
 N3585       &        S0 &   21.2 & $3.4\times10^{8}$ & $2.8\times10^{8}$ & $4.9\times10^{8}$ & stars, 17 & $213$&$10$ & $-21.88$ & $-21.80 \pm 0.20$ &   6.69 & RS\\
 N3607       &        E1 &   19.9 & $1.2\times10^{8}$ & $7.9\times10^{7}$ & $1.6\times10^{8}$ & stars, 17 & $229$&$11$ & $-21.62$ & $-21.62 \pm 0.10$ &   2.12 & RS\\
N3608\tablenotemark{h}       &        E1 &   23.0 & $2.1\times10^{8}$ & $1.4\times10^{8}$ & $3.2\times10^{8}$ & stars, ~6 & $182$&$ 9$ & $-21.05$ & $-21.05 \pm 0.10$ &   2.17 & RS\\
 N3998       &        S0 &   14.9 & $2.4\times10^{8}$ & $6.2\times10^{7}$ & $4.5\times10^{8}$ &   gas, 18 & $305$&$15$ & $-20.32$ &  . . . &   1.52 & ~S\\
 N4026       &        S0 &   15.6 & $2.1\times10^{8}$ & $1.7\times10^{8}$ & $2.8\times10^{8}$ & stars, 17 & $180$&$ 9$ & $-20.28$ & $-19.83 \pm 0.20$ &   7.72 & RS\\
 N4258       &     SABbc &    7.2 & $3.78\times10^{7}$ & $3.77\times10^{7}$ & $3.79\times10^{7}$ & maser, 19 & $115$&$10$ & $-21.31$ &  . . . &   64.5 & RS\\
 N4261       &        E2 &   33.4 & $5.5\times10^{8}$ & $4.3\times10^{8}$ & $6.6\times10^{8}$ &   gas, 20 & $315$&$15$ & $-22.72$ & $-22.72 \pm 0.06$ &   0.76 & RS\\
N4291\tablenotemark{h}       &        E2 &   25.0 & $3.2\times10^{8}$ & $8.3\times10^{7}$ & $4.1\times10^{8}$ & stars, ~6 & $242$&$12$ & $-20.67$ & $-20.67 \pm 0.13$ &   1.41 & RS\\
N4342\tablenotemark{g}       &        S0 &   18.0 & $3.6\times10^{8}$ & $2.4\times10^{8}$ & $5.6\times10^{8}$ & stars, 21 & $225$&$11$ & $-18.84$ &  . . . &   0.35 & RS\\
 N4374\tablenotemark{l} ~~~ M84 &        E1 &   17.0 & $1.5\times10^{9}$ & $9.0\times10^{8}$ & $2.6\times10^{9}$ &   gas, 22 & $296$&$14$ & $-22.45$ & $-22.45 \pm 0.05$ &   9.89 & ~S\\
 N4459\tablenotemark{k}       &        E2 &   17.0 & $7.4\times10^{7}$ & $6.0\times10^{7}$ & $8.8\times10^{7}$ &   gas, 12 & $167$&$ 8$ & $-21.06$ & $-21.06 \pm 0.04$ &   1.34 & ~S\\
N4473\tablenotemark{h}       &        E4 &   17.0 & $1.3\times10^{8}$ & $3.6\times10^{7}$ & $1.8\times10^{8}$ & stars, ~6 & $190$&$ 9$ & $-21.14$ & $-21.14 \pm 0.04$ &   2.11 & RS\\
 N4486 ~~~ M87 &        E1 &   17.0 & $3.6\times10^{9}$ & $2.6\times10^{9}$ & $4.6\times10^{9}$ &   gas, 23 & $375$&$18$ & $-22.92$ & $-22.92 \pm 0.04$ &   6.43 & RS\\
N4486A       &        E2 &   17.0 & $1.3\times10^{7}$ & $9.0\times10^{6}$ & $1.8\times10^{7}$ & stars, 24 & $111$&$ 5$ & $-18.70$ & $-18.70 \pm 0.05$ &   3.74 & ~S\\
N4564\tablenotemark{h}       &        S0 &   17.0 & $6.9\times10^{7}$ & $5.9\times10^{7}$ & $7.3\times10^{7}$ & stars, ~6 & $162$&$ 8$ & $-20.10$ & $-19.60 \pm 0.32$ &   1.46 & RS\\
 N4594       &        Sa &   10.3 & $5.7\times10^{8}$ & $1.7\times10^{8}$ & $1.1\times10^{9}$ & stars, 25 & $240$&$12$ & $-22.52$ & $-22.44 \pm 0.15$ &   8.48 & ~S\\
 N4596\tablenotemark{k}       &       SB0 &   18.0 & $8.4\times10^{7}$ & $5.9\times10^{7}$ & $1.2\times10^{8}$ &   gas, 12 & $136$&$ 6$ & $-20.70$ &  . . . &   1.86 & RS\\
N4649\tablenotemark{h} ~~~ M60 &        E2 &   16.5 & $2.1\times10^{9}$ & $1.5\times10^{9}$ & $2.6\times10^{9}$ & stars, ~6 & $385$&$19$ & $-22.65$ & $-22.65 \pm 0.05$ &   10.2 & RS\\
N4697\tablenotemark{h}       &        E6 &   12.4 & $2.0\times10^{8}$ & $1.8\times10^{8}$ & $2.2\times10^{8}$ & stars, ~6 & $177$&$ 8$ & $-21.29$ & $-21.29 \pm 0.11$ &   4.63 & RS\\
 N5077       &        E3 &   44.9 & $8.0\times10^{8}$ & $4.7\times10^{8}$ & $1.3\times10^{9}$ &   gas, 26 & $222$&$11$ & $-22.04$ & $-22.04 \pm 0.13$ &   2.45 & RS\\
 N5128\tablenotemark{j}\tablenotemark{m}       &      S0/E &    4.4 & $3.0\times10^{8}$ & $2.8\times10^{8}$ & $3.4\times10^{8}$ & stars, 27 & $150$&$ 7$ & $-21.82$ & $-21.82 \pm 0.08$ &   42.4 & ~S\\
 N5128\tablenotemark{j}\tablenotemark{m}       &      S0/E &    4.4 & $7.0\times10^{7}$ & $3.2\times10^{7}$ & $8.3\times10^{7}$ & stars, 32 & $150$&$ 7$ & $-21.82$ & $-21.82 \pm 0.08$ &   42.4 & ~S\\
 N5576       &        E3 &   27.1 & $1.8\times10^{8}$ & $1.4\times10^{8}$ & $2.1\times10^{8}$ & stars, 17 & $183$&$ 9$ & $-21.26$ & $-21.26 \pm 0.13$ &   3.55 & ~S\\
N5845\tablenotemark{h}       &        E3 &   28.7 & $2.9\times10^{8}$ & $1.2\times10^{8}$ & $3.4\times10^{8}$ & stars, ~6 & $234$&$11$ & $-19.77$ & $-19.77 \pm 0.13$ &   1.65 & ~S\\
 N6251       &        E1 &  106.0 & $6.0\times10^{8}$ & $4.0\times10^{8}$ & $8.0\times10^{8}$ &   gas, 28 & $290$&$14$ &  . . . &  . . . &   0.52 & ~S\\
 N7052       &        E3 &   70.9 & $4.0\times10^{8}$ & $2.4\times10^{8}$ & $6.8\times10^{8}$ &   gas, 29 & $266$&$13$ &  . . . &  . . . &   0.65 & ~S\\
N7457\tablenotemark{h}       &        S0 &   14.0 & $4.1\times10^{6}$ & $2.4\times10^{6}$ & $5.3\times10^{6}$ & stars, ~6 & $ 67$&$ 3$ & $-19.80$ & $-18.72 \pm 0.11$ &   0.53 & ~S\\
N7582\tablenotemark{d}\tablenotemark{g}       &      SBab &   22.3 & $5.5\times10^{7}$ & $4.4\times10^{7}$ & $7.1\times10^{7}$ &   gas, 30 & $156$&$19$\tablenotemark{d} & $-21.51$ &  . . . &   0.22 & ~S\\
A1836-BCG\tablenotemark{d}       &         E &  157.5 & $3.9\times10^{9}$ & $3.3\times10^{9}$ & $4.3\times10^{9}$ &   gas, 31 & $288$&$14$\tablenotemark{d} & $-23.31$ & $-23.31 \pm 0.15$ &   2.63 & ~S\\
A3565-BCG\tablenotemark{d}       &         E &   54.4 & $5.2\times10^{8}$ & $4.4\times10^{8}$ & $6.0\times10^{8}$ &   gas, 31 & $322$&$16$\tablenotemark{d} & $-23.27$ & $-23.27 \pm 0.15$ &   0.81 & ~S
\enddata
  \label{t:bigtable}
  \tablecomments{$M_{\mathrm{low}}$ and $M_{\mathrm{hi}}$ are the
lower and upper limits of the allowed ranges in the BH mass
measurement at the 1$\sigmaconf$ level.  ``Method'' column indicates
method of BH mass determination: stellar dynamics (``stars''), gas
dynamics (``gas''), or maser dynamics (``maser'').  Effective velocity
dispersion $\sigma_e$ is given as defined by
equation~(\ref{e:sigmae}).  Dereddened magnitudes are given for the
entire galaxy ($M_{V,T}^0$) and the bulge
($M_{V,{\mathrm{bulge}}}^0$).  The final column gives the most
restricted sample of which each galaxy is a member.  All members of
the restricted sample (RS) are members of the sample without limits
(S), which are all members of the full sample with upper limits (SU).
Galaxies that contain pseudobulges are marked by the superscript ``g''.}
\tablenotetext{%
a%
}{Galaxy types are taken from \citet{rc3} with the following exceptions:  NGC~3607, NGC~4459, and NGC~4564, which all come from \protect{\citet{2008arXiv0810.1681K}}.}
\tablenotetext{%
b%
}{Errors in bulge-disk decomposition are estimated from the range of decomposition values given in the literature and are propagated to an error in bulge luminosity.  Bulge-disk decomposition is taken from \protect{\citet{laueretal07}} with the following exceptions:  NGC~1023 \protect{\citep{1983ApJ...265..632K}}, NGC~3115 \protect{\citep{1987AJ.....94.1519C}}, and NGC~3384 \protect{\citep{1979ApJ...234..435B}}.}
%
\tablenotetext{%
c%
}{The Circinus galaxy is in the plane of the Milky
Way and has a BH mass measurement from masers
\protect{\citep{2003ApJ...590..162G}}.  \protect{\citet{ff05}} list
the BH mass of Circinus as possibly in error because the inclination
of the maser disk is not constrained, yet the velocities obtained are
clearly Keplerian, and the disk is unlikely to be far from edge-on.
If the location of the maser detections indicates the extent of the
disk, as has been assumed for NGC~4258
\protect{\citep{1995Natur.373..127M}}, then the inclination of the
disk is probably $\sim75^\circ$, which would increase the mass by
about 4\% from the value \protect{\citet{2003ApJ...590..162G}} find
assuming edge-on inclination.  We conclude that the presence of a BH
is almost certain and that the mass and uncertainties obtained by
\protect{\citet{2003ApJ...590..162G}} are reliable estimates.  If we
omit Circinus from our sample, our best-fit parameter
estimates are $\alpha = 8.17 \pm 0.07$, $\beta = 4.12 \pm 0.37$,
$\epsilon_0 = 0.38 \pm 0.06$.}
\tablenotetext{%
d%
}{Central dispersion $\sigma_c$ is given rather than effective $\sigma_e$; see eq.~(\ref{e:sigmae}).}
\tablenotetext{%
e%
}{
We use only the stellar dynamical measurement of the black hole in
IC~1459.  The gas dynamical measurement may be in error because the
gas kinematics are disturbed (M.\ Cappellari, private communication).}
\tablenotetext{%
f%
}{The uncertainty in the mass of the Galaxy's black
hole includes uncertainty in the distance to the Galactic center.}
\tablenotetext{%
g%
}{The galaxy has a pseudobulge.  If the classification is
not obvious from the images available in NED or from the high
bulge-to-disk ratio, we cite the reference for pseudobulge
classification here: NGC~224 \protect{\citep{1982ApJ...256..460K}}, NGC~1300 \protect{\citep{2006AandA...448..489K}}, NGC~2748 \protect{\citep{kk04}}, NGC~2787 \protect{\citep{2004AandA...415..941E}}, NGC~3384 \protect{\citep{2004AandA...415..941E}}, NGC~4342 \protect{\citep{1998MNRAS.293..343V}}.}
\tablenotetext{%
h%
}{Mass is 9\% larger than originally published due to
a numerical error.}  
\tablenotetext{%
i%
}{
The BH in NGC~1068 was originally measured to have $\mbh =
1.5_{0.5}^{1.6}\times10^{7} \msun$ by \citet{1997ApandSS.248..261G}
under the assumption of Keplerian rotation in the disk despite the
fact that they find that velocities fall off more slowly than
Keplerian.  Using the same data, \citet{2003AandA...398..517L} find a
formally better fit assuming a massive gas disk, and, thus, we use
their smaller value for \mbh.  If we use the larger value, we obtain
$\alpha = 8.13 \pm 0.08$, $\beta = 4.21 \pm 0.40$, and $\epsilon_0 =
0.44 \pm 0.06$.}
\tablenotetext{%
j%
}{Both NGC~1399 and NGC~5128 have two different mass
measurements that are at least marginally inconsistent with each other but 
appear to be individually reliable.  Rather than averaging these
values we use both and weight each measurement half as much, as
explained in Appendix~\ref{analysis}.}
\tablenotetext{%
k%
}{
The galaxies NGC~2787, NGC~4459, and NGC~4596 have masses reported by
\citet{2001ApJ...550...65S} for both unconstrained disk inclination
and for the best-fit disk inclination.  We use the latter as the
regions of the gas disk probed by the observations are likely to be
aligned in the plane of the axisymmetric bulge
\citep[e.g.][]{marteletal00}.  These galaxies are all low velocity
dispersion galaxies and thus unlikely to be triaxial.}
\tablenotetext{%
l%
}{
\citet{mb01} interpreted the double-peaked velocity profile in
NGC~4374 as coming from a single component rather than from two
separate components as \citet{1998ApJ...492L.111B} did.  The velocity
profile is clearly doubly peaked as can be seen in figure~3 of
\citet{1998ApJ...492L.111B}, and the second peak is likely to come
from a more slowly rotating separate component, thus we include it in
our sample.  If the interpretation of \citet{mb01} is correct, the
mass is a factor of 4 smaller than found by
\citet{1998ApJ...492L.111B}.  Omitting this galaxy, our best fit
parameters change to $\alpha = 8.12 \pm 0.08$, $\beta = 4.19 \pm
0.41$, and $\epsilon_0 = 0.45 \pm 0.06$.}
\tablenotetext{%
m%
}{
NGC~5128 has multiple \mbh\ measurements with ionized gas
\citep{2001ApJ...549..915M,2006ApJ...643..226H,2006AandA...448..921M},
two-dimensional neutral gas velocity measurements
\citep{2007MNRAS.374..385K,2007ApJ...671.1329N}, and stellar dynamical
measurements \citep{2005AJ....130..406S,2008arXiv0812.1000C}.
\citet{2007ApJ...671.1329N} argue that the ionized-gas measurements
may be contaminated by the galaxy's jet.  Three of the remaining
measurements
\citep{2007MNRAS.374..385K,2007ApJ...671.1329N,2008arXiv0812.1000C}
are in good agreement with each other, which we present as one value
in addition to the other measurement \citep{2005AJ....130..406S}.}

\tablerefs{
(1) \cite{2003ApJ...590..162G}, 
(2) \cite{2002ApJ...578..787C}, 
(3) \cite{ghezetal08} and \cite{2008arXiv0810.4674G}, 
(4) \cite{2002MNRAS.335..517V}, 
(5) \cite{benderetal05}, 
(6) \cite{2003ApJ...583...92G}, 
(7) \cite{2001ApJ...550...75B}, 
(8) \cite{2003AandA...398..517L}, 
(9) \cite{2005MNRAS.359..504A}, 
(10) \cite{2007ApJ...671.1321G}, 
(11) \cite{2006MNRAS.367....2H}, 
(12) \cite{2001ApJ...550...65S}, 
(13) \cite{2003AJ....125.1226D}, 
(14) \cite{1999MNRAS.303..495E}, 
(15) \cite{2001ApJ...555..685B}, 
(16) \cite{2000AJ....119.1157G}, 
(17) \cite{Gultekin_etal_2008}, 
(18) \cite{2006AandA...460..439D}, 
(19) \cite{2005ApJ...629..719H}, 
(20) \cite{1996ApJ...470..444F}, 
(21) \cite{1999ApJ...514..704C}, 
(22) \cite{1998ApJ...492L.111B}, 
(23) \cite{1997ApJ...489..579M}, 
(24) \cite{nowaketal07}, 
(25) \cite{1988ApJ...335...40K}, 
(26) \cite{2008arXiv0801.0064D}, 
(27) \cite{2005AJ....130..406S}, 
(28) \cite{1999ApJ...515..583F}, 
(29) \cite{1998AJ....116.2220V}, 
(30) \cite{2006AandA...460..449W}, 
(31) \cite{2008arXiv0809.0766D},
(32) \cite{2008arXiv0812.1000C}.
}
\end{deluxetable}

\clearpage
\begin{deluxetable}{llrrrrr@{$\pm$}lrrrr}
  \tabletypesize{\scriptsize}
  \tablecaption{Upper limits to black hole masses}
  \tablewidth{0pt}
  \tablehead{
    & & \colhead{Dist.} & \colhead{$M_{u}$} & & \colhead{Method,} & \multicolumn{2}{c}{$\sigma_e$} & & & \\
     \colhead{Galaxy} &
     \colhead{Type} &
     \colhead{Mpc} &
     \colhead{\msun}  &
     \colhead{Confidence}  &
     \colhead{Ref.} & 
    \multicolumn{2}{c}{$\mathrm{km~s^{-1}}$} &
     \colhead{$M_{V,T}^0$} &
     \colhead{$M_{V,{\mathrm{bulge}}}^0$%
\tablenotemark{a} 
} &
     \colhead{${R_{\mathrm{infl}}/d_{\mathrm{res}}}$} &
     \colhead{Sample} 
  }
  \startdata
 N3310 &   SB(r)bc &   17.4 & $4.2\times10^{7}$ & 2$\sigmaconf$ &   gas, ~1 & $ 83$&$ 4$ & $-20.56$ &  . . . &   2.39 & SU\\
N3351\tablenotemark{b} &       SBb &    8.7 & $8.6\times10^{6}$ & 1$\sigmaconf$ &   gas, ~2 & $ 93$&$ 4$ & $-20.15$ &  . . . &   0.90 & SU\\
 N3368 &      SBab &   11.0 & $3.7\times10^{7}$ & 1$\sigmaconf$ &   gas, ~2 & $114$&$ 5$ & $-21.19$ &  . . . &   1.80 & SU\\
 N3982 &      SBb: &   18.2 & $8.0\times10^{7}$ & 1$\sigmaconf$ &   gas, ~2 & $ 78$&$ 3$ &  . . . &  . . . &   8.75 & SU\\
 N3992 &      SBbc &   18.2 & $5.7\times10^{7}$ & 1$\sigmaconf$ &   gas, ~2 & $119$&$ 5$ & $-21.73$ &  . . . &   1.95 & SU\\
 N4041 &   S(rs)bc &   20.9 & $6.4\times10^{6}$ & 3$\sigmaconf$ &   gas, ~4 & $ 88$&$ 4$ & $-20.40$ &  . . . &   0.35 & SU\\
 N4143 &       SB0 &   16.8 & $1.4\times10^{8}$ & 1$\sigmaconf$ &   gas, ~2 & $271$&$13$ &  . . . &  . . . &   1.59 & SU\\
 N4203 &       SB0 &   16.0 & $3.8\times10^{7}$ & 1$\sigmaconf$ &   gas, ~2 & $110$&$ 5$ & $-20.34$ &  . . . &   0.80 & SU\\
N4321\tablenotemark{b} &      SBbc &   18.0 & $2.7\times10^{7}$ & 1$\sigmaconf$ &   gas, ~2 & $ 74$&$ 3$ & $-21.95$ &  . . . &   1.79 & SU\\
 N4435 &       SB0 &   17.0 & $8.0\times10^{6}$ & 3$\sigmaconf$ & stars, ~5 & $150$&$ 7$ & $-20.45$ &  . . . &   0.17 & SU\\
 N4450 &       Sab &   18.0 & $1.2\times10^{8}$ & 1$\sigmaconf$ &   gas, ~2 & $121$&$ 6$ & $-21.29$ &  . . . &   3.46 & SU\\
 N4477 &     SB0:? &   18.0 & $8.4\times10^{7}$ & 1$\sigmaconf$ &   gas, ~2 & $134$&$ 6$ & $-20.92$ &  . . . &   1.19 & SU\\
N4486B\tablenotemark{c} &        E1 &   17.0 & $1.1\times10^{9}$ & 1$\sigmaconf$ & stars, ~6 & $185$&$ 9$ & $-17.80$ & $-17.80 \pm 0.04$ &   10.1 & SU\\
 N4501 &        Sb &   18.0 & $7.9\times10^{7}$ & 1$\sigmaconf$ &   gas, ~2 & $136$&$ 6$ & $-22.02$ &  . . . &   1.51 & SU\\
 N4548 &       SBb &   20.3 & $3.4\times10^{7}$ & 1$\sigmaconf$ &   gas, ~2 & $154$&$ 7$ & $-21.51$ &  . . . &   0.71 & SU\\
 N4698 &       Sab &   18.0 & $7.6\times10^{7}$ & 1$\sigmaconf$ &   gas, ~2 & $116$&$ 5$ & $-20.87$ &  . . . &   2.13 & SU\\
 N4800 &        Sb &   16.3 & $2.1\times10^{7}$ & 1$\sigmaconf$ &   gas, ~2 & $112$&$ 5$ &  . . . &  . . . &   1.00 & SU\\
A2052-BCG\tablenotemark{d} &         E &  151.1 & $4.9\times10^{9}$ & 1$\sigmaconf$ &   gas, ~7 & $233$&$11$\tablenotemark{d} & $-24.21$ & $-24.21 \pm 0.15$ &   5.33 & SU\\
  \enddata
  \label{t:ultable}
  \tablecomments{$M_{u}$ is the upper limit to the BH mass at
confidence level given by the following column where 1$\sigmaconf = 84.2\%$, 2$\sigmaconf = 97.8\%$, and 3$\sigmaconf = 99.9\%$.  ``Method'' column indicates
method of BH mass determination: stellar dynamics (``stars''), gas dynamics (``gas'').  All galaxies in this table are
members of the full sample with upper limits (SU).}
\tablenotetext{%
a%
}{Bulge-disk decomposition is taken from \protect{\citet{laueretal07}}.
}
\tablenotetext{%
b%
}{The galaxy is a pseudobulge.}
\tablenotetext{%
c%
}{NGC~4486B hosts an asymmetric double nucleus \protect{\citep{laueretal96}}.
\protect{\citet{kormendyetal97}} used ground-based spectroscopic
observations of NGC~4486B, along with spherical, isotropic, dynamical
models, to
find a central dark object of mass between $3 \times 10^8$ and $3
\times 10^{9} \msun$.  \protect{\citet{magorrianetal98}} used
two-integral, axisymmetric dynamical models to find a mass of
$M=9.2^{+1.2}_{-0.7}\times10^8 \msun$.  The presence of an asymmetric double
nucleus could indicate an eccentric disk or torus of stars near the
center, which requires a massive BH to be stable
\protect{\citep{tremaine95}}.  
The exact mass of the BH , however,
may not be sufficiently well determined to warrant using the mass found with
isotropic models because (1) the three-integral axisymmetric models do
not significantly rule out the absence of a BH
\protect{\citep{kormendyetal97}} and (2) the presence of a double
nucleus may present problems for isotropic models.  To
account for this, we conservatively list it as an upper limit at the
1$\sigmaconf$ upper error estimate.}
\tablenotetext{%
d%
}{Central dispersion $\sigma_c$ is given rather than effective $\sigma_e$; see eq.~(\ref{e:sigmae}).}
\tablerefs{
(1) \cite{2007AandA...469..405P}, 
(2) \cite{2002ApJ...567..237S}, 
(3) \cite{Gultekin_etal_2008}, 
(4) \cite{2003ApJ...586..868M}, 
(5) \cite{2006MNRAS.366.1050C}, 
(6) \cite{kormendyetal97}, 
(7) \cite{2008arXiv0809.0766D}.
}
\end{deluxetable}

\clearpage
\begin{deluxetable}{lllrrrrrrrrr}
  \tabletypesize{\scriptsize}
  \tablecaption{Black Hole Masses Omitted from Fits}
  \tablewidth{0pt}
  \tablehead{
    \colhead{Galaxy} &
      &
     \colhead{Type} &
     \colhead{Dist.} &
     \colhead{$M_{\mathrm{BH}}$} &
     \colhead{$M_{\mathrm{low}}$} &
     \colhead{$M_{\mathrm{high}}$} &
     \colhead{Method,} &
     \colhead{$\sigma_e$} &
     \colhead{$M_{V,T}^0$} &
     \colhead{$M_{V,{\mathrm{bulge}}}^0$} &
     \colhead{${R_{\mathrm{infl}}/d_{\mathrm{res}}}$} \\
    & & & \colhead{Mpc} & \colhead{\msun} & \colhead{\msun} & \colhead{\msun} & \colhead{Ref.} & \colhead{$\mathrm{km~s^{-1}}$} & & 
  }
  \startdata
Cygnus~A\tablenotemark{a} &    &        E  &  257.1 & $2.7\times10^{9}$  & $1.9\times10^{9}$ & $3.4\times10^{9}$ &    gas, ~1 & 270    & $-$21.27 & $-$21.27 & 1.27 \\
N0205\tablenotemark{b} &  M101 &        Sph &   0.74 & . . . & . . .             & $3.8\times10^{4}$             &  stars, ~2 &  39    & $-$16.38 & $-$16.38 & 0.03 \\
N0598\tablenotemark{c} &   M33 &        Sc &   0.80 & . . . & . . .             & $3.0\times10^{3}$             &  stars, ~3 &  24    & $-$18.77 & . . .    & 0.06 \\
N3945\tablenotemark{d} & &      SB0+ &   19.9 & \dots & \dots & $5.1\times10^{7}$ &  stars, ~4 & $192$ & $-21.06$ & $-20.09$ &   1.50\\
N4151\tablenotemark{e} &    &   SAB(rs)ab: &   13.9 & $4.5\times10^{7}$  & $4.0\times10^{7}$ & $5.0\times10^{7}$ &  stars, ~5 &  93    & $-20.68$ &    . . . & 0.44 \\
N4303\tablenotemark{f} &   M61 &     SABbc &   17.9 & $4.5\times10^{6}$  & $2.8\times10^{6}$ & $1.4\times10^{7}$ &    gas, ~6 &  84    & $-$21.65 &    . . . & 0.31 \\
N4742\tablenotemark{g} &       &        E4 &   16.4 & $1.5\times10^{7}$  & $9.5\times10^{6}$ & $1.9\times10^{7}$ &  stars, ~7 &  90    & $-$19.91 & $-$19.91 & 0.99 \\
N4945\tablenotemark{h} &       &        Sc &    3.7 & $1.4\times10^{6}$  & $9.0\times10^{5}$ & $2.1\times10^{6}$ & masers, ~8 & 134    &    . . . &    . . . & 4.67 \\
N5252\tablenotemark{i} &       &        S0 &  103.7 & $1.0\times10^{9}$  & $5.4\times10^{8}$ & $2.6\times10^{9}$ &    gas, 9 & 190    &    . . . &    . . . & 2.42 \\

  \enddata
  \label{t:insecure}
\tablecomments{We list galaxies with dynamically determined BH masses
  that we do not include in our main fits.  We also do not include
  (but do not list above) the 105 upper limits due to
  \protect{\citet{beifiorietal08}}.  These upper limits come from
  \emph{HST} spectroscopy of ionized gas at the center of the
  galaxies, assuming isotropic inclinations of gas disks.  If we
  include all of the upper limits from
  \protect{\citet{beifiorietal08}} with our full sample, we find a
  best fit of $\alpha = 7.97 \pm 0.08$, $\beta = 4.41 \pm 0.39$,
  $\epsilon_0 = 0.45 \pm 0.06$.  In the table footnote for each galaxy
  we list estimates for the parameters of the \msigma\ relation
  (eq.~\ref{e:msigmaform}) when that galaxy is included with the rest
  of our full sample.  For almost all of the exclusions, there is no
  significant change.}

\tablerefs{
(1) \phantom{\;} \cite{2003MNRAS.342..861T},
(2) \cite{vallurietal05}
(3) \cite{2001AJ....122.2469G} and \cite{2001Sci...293.1116M},
(4) \cite{Gultekin_etal_2008},
(5) \cite{onkenetal07},
(6) \cite{2007AandA...469..405P}, 
(7) listed as in preparation in \cite{tremaineetal02} but never published,
(8) \cite{1997ApJ...481L..23G},
(9) \cite{2005A&A...431..465C}.
}

\newcounter{bhexpl}
\setcounter{bhexpl}{1}


\tablenotetext{a}{The gas velocity measurement in Cygnus~A
\citep{2003MNRAS.342..861T} gives strong evidence of the presence of a
BH, but the model was not able to explain consistently both their
infrared Pa$\alpha$ and optical [O~III] data, giving no quantitative
analysis of the goodness of fit for their models; $\alpha = 8.07 \pm
0.10$, $\beta = 4.13 \pm 0.53$, $\epsilon_0 = 0.54 \pm 0.08$.}

\tablenotetext{b}{The upper limit to NGC~0205 is considerably below
our best-fit ridge line.  We omit it from our sample because
determination of the mass-to-light ratio at the center of this galaxy
may be hampered by a very blue nuclear star cluster.  It is also a
diffuse spheroidal, and we exclude galaxies without either a classical
bulge or a pseudobulge.  Including this upper limit in our fits with
the given velocity dispersion of the nucleus, however, does not
significantly alter any fit parameters;  $\alpha = 8.12 \pm 0.08$,
$\beta = 4.37 \pm 0.43$, $\epsilon_0 = 0.45 \pm 0.06$.}

\tablenotetext{c}{NGC~0598 is a bulgeless disk galaxy
\protect{\citep{kk04}}.  The \msigma\ and \ml\ relations apply only to
elliptical galaxies or galaxies with a spheroid component.
Additionally, it is not clear whether to represent the bulge
dispersion by the stellar velocity dispersion of the nucleus (as given
in this table) or some function of the circular velocity of the disk;
$\alpha = 8.12 \pm 0.08$, $\beta = 4.38 \pm 0.45$, $\epsilon_0 = 0.45
\pm 0.06$.}

\tablenotetext{d}{The mass from NGC~3945, which is a psuedobulge, is
 derived from an axisymmetric code \citep{Gultekin_etal_2008}, but the
 galaxy is a double-barred system; $\alpha = \pm $, $\beta = \pm $,
 $\epsilon_0 = \pm $.}

\tablenotetext{e}{NGC~4151 has been measured by reverberation mapping
\protect{\citep{bentzetal06b}} to have $M = 4.57^{+0.57}_{-0.47} \times
10^7~\msun$ (after applying a scale factor that calibrates
reverberation mapping virial products to the \msigma\ relation), in
close agreement with the mass derived from the inclined model of
\citet{onkenetal07} by stellar dynamical measurement.  The models by
\citet{onkenetal07} that assume an edge-on inclination, however,
cannot rule out $M = 0$, and the authors cite the noise in their
$\Delta\chi^2$ contours as cause to label it a ``tentative estimate.''
Hence, we do not include this galaxy in our sample; $\alpha = 8.13 \pm
0.08$, $\beta = 4.06 \pm 0.40$, $\epsilon_0 = 0.46 \pm 0.06$.}

\tablenotetext{f}{NGC~4303 has an
unresolved point source at the center, which is likely an AGN but may
also have a significant amount of mass in stars
\protect{\citep{2007AandA...469..405P}}, especially if
nuclear stellar clusters are prevalent.  The velocity profile is also
irregular, and the authors do not consider the BH mass
estimate completely reliable \protect{\citep{2007AandA...469..405P}};
$\alpha = 8.13 \pm 0.08$, $\beta = 4.21 \pm 0.40$, $\epsilon_0 = 0.44
\pm 0.06$.}


\tablenotetext{g}{NGC~4742 appeared in the sample of
\protect{\citet{tremaineetal02}} listed as ``in preparation,'' but it
has still not appeared in a refereed publication, and thus we do not
include it in our sample;  $\alpha = 8.13 \pm 0.08$, $\beta = 4.14 \pm
0.39$, $\epsilon_0 = 0.44 \pm 0.06$.}

\tablenotetext{h}{NGC~4945 has a velocity profile that is asymmetric
from one side of the galaxy to the other and no constraint on the
disk inclination;  $\alpha = 8.10 \pm 0.08$, $\beta = 4.36 \pm 0.43$,
$\epsilon_0 = 0.47 \pm 0.06$.}

\tablenotetext{i}{The best-fit parameter set in the mass modeling
produces a reduced $\chi^2$ of 16.5 \citep{2005A&A...431..465C},
indicating a poor fit to the data; $\alpha = 8.14 \pm 0.08$, $\beta =
4.25 \pm 0.41$, $\epsilon_0 = 0.45 \pm 0.06$.}

\end{deluxetable}

\phantom{helpme}

\label{lastpage}
\end{document}